\documentclass[epsfig,useAMS,usenatbib]{mn2e}

\input{epsf}

\usepackage{graphicx}

\def\gsim{ \lower .75ex \hbox{$\sim$} \llap{\raise .27ex \hbox{$>$}} }
\def\lsim{ \lower .75ex \hbox{$\sim$} \llap{\raise .27ex \hbox{$<$}} }

\title[Effects of ram pressure on the LMC's disk]{Effects of ram pressure on the gas distribution and star formation in the Large Magellanic Cloud.}

\author[Mastropietro et al.]
{Chiara Mastropietro $^{1}$ \thanks{E-mail: chiara.mastropietro@obspm.fr}, 
Andreas Burkert $^{2}$, Ben Moore $^{3}$\\  
$^{1}$LERMA, Observatoire de Paris, UPMC, CNRS, 61 Av. de l' Observatoire, 75014 Paris, France\\
$^{2}$Universit\"ats Sternwarte M\"unchen, Scheinerstr.1, D-81679 M\"unchen,Germany\\
$^{3}$Institute for Theoretical Physics, University of Z\"urich, CH-8057 Z\"urich, Switzerland}

\begin{document}


\pagerange{\pageref{firstpage}--\pageref{lastpage}} \pubyear{00} 

\maketitle

\label{firstpage}

\begin{abstract}
We use high resolution N-body/SPH simulations to study the hydrodynamical interaction between the Large Magellanic Cloud (LMC) and the hot halo of the Milky Way. We investigate whether ram-pressure acting on the satellite's ISM  can explain the peculiarities observed in the HI distribution and the location of the recent star formation activity.

Due to the present nearly edge-on orientation of the disk with respect to the orbital motion, compression at the leading edge can explain the high density region observed in HI at the south-east border. In the case of a face-on disk (according to Mastropietro et al. 2008 the LMC was moving almost face-on before the last perigalactic passage), ram-pressure directed perpendicularly to the disk produces a clumpy structure characterized by voids and high density filaments that resemble those observed by the Parkes HI survey.
As a consequence of the very recent edge-on  motion, the H$\alpha$ emission is mainly concentrated on the eastern side where 30 Doradus and most of the supergiant shells are located, although some H$\alpha$ complexes form a patchy distribution on the entire disk. In this scenario only the youngest stellar complexes show a progression in age along the leading border of the disk.

\end{abstract}


\section{Introduction}

The Large Magellanic Cloud (LMC) has revealed a very complex structure both in the stellar and in the gaseous component. The elongation of the stellar disk  in the direction of the Galactic center, its substantial vertical thickness, the warp and the strong asymmetric bar are naturally predicted by numerical simulations
as a result of the gravitational interaction between the LMC and the Galaxy
 (Bekki \& Chiba 2005, Mastropietro et al. 2005, hereafter M05). 
The
old stellar distribution appears to be quite smooth in the outer
parts of the disk, with no signs of spiral structures out to a radius of 10
kpc \citep{vanderMarel01}. Within the same radius the HI large scale structure
reveals the presence of several asymmetric features that  have no
equivalent in the old stellar disk.   
The gaseous disk is characterized by the presence
of an elongated region located at the south-east of the galaxy and aligned
with the border of the optical disk, where the column density distribution
shows a steep increase \citep{StaveleySmithetal03, Putmanetal03}. 
Since the LMC proper motion vector is directed to the east, it appears natural to associate this high density region with  ram-pressure acting on
the leading edge of the disk due to the orbital motion of the LMC and its 
consequent interaction with the diffuse hot gas in the halo of the 
Milky Way (MW). 

The presence of an extended hot halo surrounding galaxies and
in hydrostatic equilibrium within the dark matter potential is expected by
current models of hierarchical structure formation \citep{Maller&Bullock04}.
In the MW, X-ray absorption lines produced by hot ($T \sim 10^6$ K) gas are detected
in the spectra of several bright AGN \citep{Williamsetal05, Fangetal06}. Some ionization features discovered in the Magellanic Stream and  high velocity Clouds indicate that this distribution of hot gas  extends well beyond the Galactic disk ( $>70$ kpc). 
Constraints from dynamical and thermal
arguments fix its density in a range between $10^{-5}$ and
$10^{-4}$ cm $^{-3}$ at the LMC distance from the Galactic center (but  Kaufmann et al. 2009 suggest a value ten times higher).

\citet{Cionietal06} have performed a detailed analysis of the LMC
global star formation rate using asymptotic giant branch stars.
They find an irregular and patchy distribution in age, with the
youngest carbon-rich systems located at the south-east of the disk. 
The present star formation activity is rather clumpy and concentrated in 
stellar complexes  
characterized by intense HII emission and associated with bright H$\alpha$ 
filamentary bubbles. 
Most of these very young structures lie on the south-east of the disk, in the 
proximity of 30 Doradus, the largest star forming region of the LMC, 
some are located in the bar and the remainder form an asymmetric
pattern that covers the entire disk with no apparent relation to the global 
geometry of the satellite.

It is not clear  which is the overall physical mechanism responsible for
triggering star formation with the observed asymmetric pattern
and different models have been proposed in the past. 
The stochastic self-propagating star formation (SSPSF) model  \citep{Gerola&Seiden78} predicts a clear age gradient in the LMC's stellar 
complexes, with the edges being younger with respect to the center, in
contradiction with observations \citep{Braunetal97}.
\citet{deBoeretal98} proposed a scenario where the bow shock originated by the
motion of the LMC through the hot galactic halo compresses the leading edge of
the disk and induces star formation. 
The pressure at the south-eastern edge of the LMC is indeed 10 times higher than the average in the 
rest of the LMC \citep{Blondiauetal97}.
This model, which assumes the orbital motion vector  lying in the plane of the disk, predicts increasing ages of the 
stellar complexes in the direction of the rotation, due to the fact that the
material compressed at the front side of the disk moves, in time, away to the
side. The youngest systems would indeed lie at the south-east border of the 
disk, where the relative velocity between the corotating interstellar medium 
and the external diffuse gas is maximum.
Several giant structures along the outer east and north edge of the LMC 
actually show a progression in age in a clockwise direction: moving from 
south-east to the north LMC 2, 30 Doradus and LMC 3, LMC 4, NGC1818.
In particular the difference in age between 30 Doradus and LMC4 is exactly their distance along the border of the disk divided by the satellite's rotational velocity (Harris, private communication).
\citet{Grebel&Brandner98} studied the recent star formation history of the LMC
using Cepheids and other supergiant stars and found that although the
majority of the star formation events in the last 30 Myr are concentrated on
the east border, others are distributed across the entire disk in partial
contrast with the bow shock induced star formation model, that can not
explain them.

In this work we use high resolution SPH simulations to study
the effects of the interaction between the LMC interstellar medium and the
diffuse hot halo of the MW.  We want to investigate whether the ram-pressure
acting on the leading edge of the LMC disk is responsible for the increase 
in density observed in the south-east and for triggering star
formation.   
The analytic model of \citet{deBoeretal98} assumes
a pure edge-on model, but according to \citet{vanderMareletal02} 
the present angle
between the LMC disk and the orbital motion is nearly $30^{\circ}$. Even in the
absence of precession and nutation, this angle is subjected to large 
variations during the orbital period in such a way that compression produced by the external hot gas can affect in time both edge-on and face-on.

Moreover, the ram-pressure felt by the LMC is not constant and
has a maximum when the satellite approaches the  perigalacticon.
The motion of the LMC through the hot halo of the MW during the last 1 Gyr is modeled using 
``test wind tunnel'' simulations with increasing ram-pressure
 values.

The paper is structured as follows.
Section 2 describes the models and the star formation criteria adopted,
Section 3 illustrates the results of simulations without star formation, focussed
on the investigation of pure effects of compression on the LMC interstellar
medium while Section 4 describes the runs where star formation is
activated. Several simulations have been performed, assuming different star formation
 models, disk inclinations and hot halo densities.

\section{Simulations}

\subsection{Galaxy model}

The initial conditions of the simulations are constructed using the technique
described by  \citet{Hernquist}. Our disk galaxy model is a multi-component
system with a stellar and gaseous disk embedded in a spherical NFW  \citep{Navarroetal97}  dark matter
halo. 
The density profile of the dark matter halo is adiabatically contracted in response to baryonic infall \citep{Blumenthaletal86}. The stellar disk follows an exponential surface density profile of the form:
\begin{equation} \label{stars}
\Sigma(R)=\frac{M_d}{2\pi {R_d}^2}\,\textrm{exp}\,(-R/R_d)\, ,
\end{equation}
where $M_d$ and $R_d$ are the disk mass and radial scale length (in cylindrical coordinates), respectively, while the thin vertical structure has a scale length $z_d \sim \frac{1}{5}R_d$:
\begin{equation} \label{diskh} 
\rho_d(R,z) = \frac{\Sigma(R)}{2z_d}\,\textrm{sech}^2\,(z/z_d)\, .
\end{equation}
The gaseous disk is characterized by an exponential profile with the same
radial and vertical scale length as the stellar component and by a constant
density layer which extends up to 8$R_d$. 

The structural parameters of the disk and the halo are chosen so that the
resulting rotation curve resembles that of a typical bulgeless late-type
(Sc/Sd) disk galaxy  \citep{Courteau97, Persic&Salucci97}. 
They are similar to those adopted in M05 for the initial LMC model and reproduce quite well the peak of the rotation curve inferred by \citet{vanderMareletal02} (Fig. 1). As seen in M05, the interaction with the MW does not affect significantly the stellar and dark matter mass in the inner $8-9$ kpc of the LMC and consequently the global rotation curve within this radial range.
The choice of an extended gaseous component for the initial LMC model is
motivated by the fact that spiral galaxies in the local universe are commonly observed to be embedded in
extended disks of neutral hydrogen significantly larger than their stellar
component \citep{Hunter&Gallagher85, Broeils&Woerden94}. 
As seen in M05, the combined effect of tidal interactions and ram-pressure stripping can remove a significant fraction of gas from a LMC disk orbiting within the hot halo of the MW, with a ram-pressure stripping radius which is a factor of three smaller than the initial radius of the gaseous disk.  
Also in the case of a LMC with orbital velocities significantly higher \citep{Mastropietro08} hydrodynamic and gravitational forces together are effective in resizing and reshaping the extended gaseous disk of the satellite beyond 8 kpc. 
In the present work we neglect the presence of gravitational forces 
focusing on the effects of pure ram-pressure. Therefore we do not expect to see a 
significant decrease in the radius of the gas distribution.
However, in order to take in account the loss of cold gas from the disk of the
satellite and the star formation events, we assumed an initial amount 
of gas in the disk which is about $3$ times larger than the HI mass in 
the LMC ($2.9 \times 10^8$ $M_{\odot}$ according to Putman et al. 2003).

\begin{figure}
\epsfxsize=8truecm \epsfbox{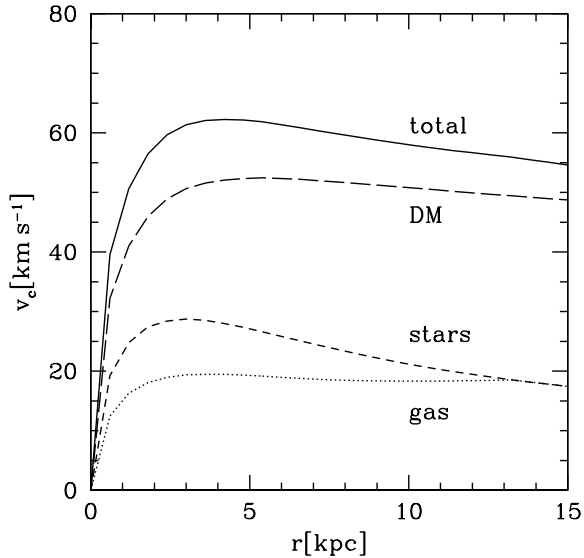} 
\caption{Galaxy model rotation curves.}
\label{rotcurve}
\end{figure}

The mass within the virial radius is set equal to $2.18 \times 10^{10}$ M$_\odot$   and the
fraction of mass in the disk is $\sim 10\%$,  equally distributed between the gaseous and stellar component. The contribution of the
different components to the global rotation curve, assuming a disk scale
length $R_d= 1.7$ kpc and a dark halo concentration $c = 9.5$ (where $c$ is defined as
$c=r_{vir}/r_{s}$, with $r_{vir}$ and $r_s$ virial and scale
radius of the NFW halo, respectively) is plotted in Fig. \ref{rotcurve}. The halo spin parameter, which sets the disk scale length in our modeling, is $\lambda = 0.074$, where $\lambda$ relates the
angular momentum $J$ and the total energy $E$ of a system with virial mass
$M_{vir}$ through the relation $\lambda=J|E|^{1/2}G^{-1}M_{vir}^{-5/2}$ 

\begin{figure}
\epsfxsize=8truecm \epsfbox{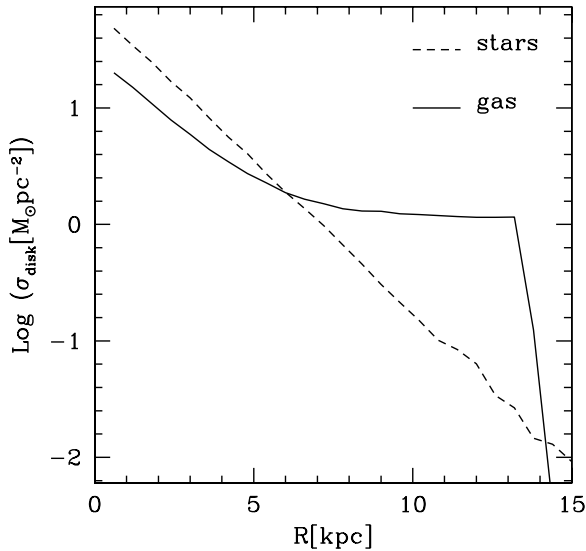} 
\caption{Gaseous and stellar disk surface density profiles (cylindrical coordinates).}
\label{diskprofile}
\end{figure}

The initial stellar disk of the satellite galaxy has, within its scale radius $R_d$, 
a central mass surface density of 
$\sim 35\, \textrm{M}_{\odot} \textrm{pc}^{-2}$ (Fig. \ref{diskprofile}), 
that corresponds to a B-band surface brightness of $\sim 24\, \textrm{mag
arcsec}^{-2}$, assuming a mass to light ratio $\simeq 2$.
The central gas surface density is only $\sim 16 \, \textrm{M}_{\odot}
\textrm{pc}^{-2}$ since a significant fraction of gas
is distributed in the external disk. Assuming $72\%$ HI abundance this 
value corresponds to an hydrogen column density of $\sim 1.5 \times
10^{21}$ within $R_d$, comparable with the values observed by 
\citet{StaveleySmithetal03} with the LMC Parkes multibeam HI survey.

\subsection {Stability criterium}

In order to obtain a strongly stable disk against bar formation even in
the presence of significant gas stripping and consequent perturbation of the
satellite potential, the thickness of the stellar component is set such that
the Toomre's \citep{Toomre64} stability criterion is largely satisfied. 
In particular the Toomre's parameter for the stellar disk: 

\begin{equation}  
Q_{s}(R)= \frac{\sigma_r(R) \kappa(R)}{3.36 G \Sigma_s(R)},
\label{Toomres} 
\end{equation} 
where $\sigma_r(R)$ is the radial velocity dispersion,
$k(R)$ is the local epicyclic frequency and $\Sigma_s(R)$ the unperturbed 
stellar surface
density, has a minimum at the disk scale length with $Q_s(R_d)
\sim 4$.   
For a gaseous disk the stability of the disk  is expressed  in terms of the gas sound
speed $v_s$ and surface density $\Sigma_g(r)$ through the relation: 
\begin{equation} 
Q_{g}(R)= \frac{v_s \kappa(R)}{\pi\, G \Sigma_g(R)}.
\label{Toomresgas} 
\end{equation}   
The gaseous disk has initially a constant temperature of 10000 K, which implies $Q_g(R)>
3$ and $Q(R_d)=3.2$.

\begin{figure}
\epsfxsize=8truecm \epsfbox{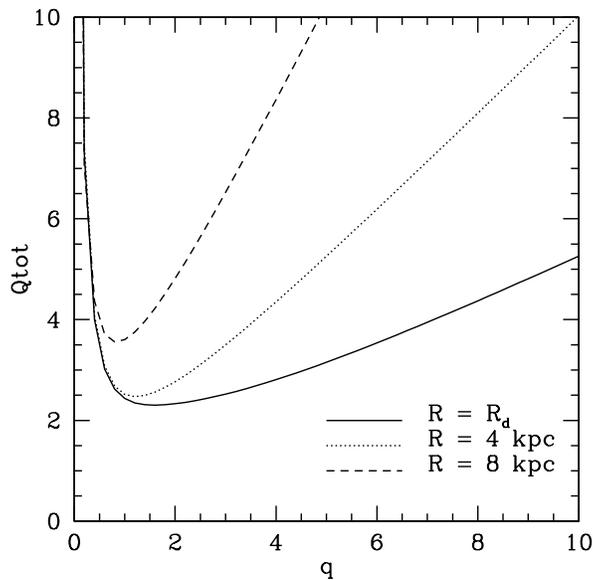} 
\caption{Total stability parameter $Q_{tot}$ of the disk (see Equation \ref{Toomretotal})
  versus the dimensionless wavenumber $q=k\sigma_s/\kappa$ of the
  perturbation. $Q_{tot}$ is calculated for three different radii of the disk.}
\label{qparameter}
\end{figure}

According to \citet{Jog&Solomon84} and \citet{Rafikov01}, the stability of a
multicomponent disk is not guaranteed by the individual stability of its 
single constituents, due to the mutual gravitational interaction between gas
and stars. Stars are characterized by velocity dispersions 3-4
times larger than the typical sound velocities in the cold gaseous disk and
even relatively small variations of the gaseous component parameters
can significantly affect the stability of the whole disk
\citep{Jog&Solomon84}. Therefore we choose a large value of $ Q$ to
contrast the effects of ram-pressure.
In the case of a two components - gaseous and stellar - disk, the stability
condition is expressed by  \citep{Jog&Solomon84}
\begin{equation}
Q_{tot}= \Big[\frac{2}{Q_s}\frac{\pi}{3.36}\frac{q}{1+q^2}+\frac{2}{Q_g}x\frac{q}{1+q^2x^2}\Big]^{-1}>1,
\label{Toomretotal}
\end{equation}
where $q=k\sigma_s/\kappa$ and $x=c_g/\sigma_s$. 
Fig. \ref{qparameter} illustrates the dependence of $Q_{tot}$ on the
dimensionless wavenumber of the perturbation $q$ within three different
regions of the disk: at the disk scale length $R_d$, at $R=5$ kpc and in the
external region ($R=8$ kpc), where the gas component predominates. The \citet{Jog&Solomon84} criterium is always satisfied and the disk is stable against
axisymmetric perturbations, independently of their wavelength.
In order to check stability the disk was initially evolved in isolation for 1 Gyr.

\subsection{Star formation recipes}


All the simulations we now discuss were 
carried out using GASOLINE, a parallel tree-code with multi-stepping
\citep{Wadsley04} which is an extension of the pure N-body gravity code
PKDGRAV developed by \citet{Stadel01}. The code uses a spline kernel with
compact support \citep{Monaghan&Lattanzio85} where the interaction distance
for a particle $i$ is set equal to two times the smoothing length $h_i$, defined
as the $k$-th neighbour distance from the particle. In this paper $k=32$. The
internal energy of the gas is integrated using the antisymmetric formulation
of \citet{Evrard88} that conserves entropy closely. Dissipation in shocks is
modeled using the quadratic term of the standard \citet{Monaghan92} artificial
viscosity. The Balsara \citep{Balsara95} correction term is used to suppress
the viscosity in non-shocking, shearing environments. 
The code includes radiative cooling for a primordial mixture of hydrogen and 
helium in collisional equilibrium. At temperatures below $10^4$ K the gas is 
entirely neutral and due to the lack of molecular cooling and metals, the 
efficiency of the cooling function drops rapidly to zero.
We used a star formation recipe that includes density and
temperature criteria, while converging flow criterium is not required in most of the
simulations.
Gas particles are eligible to form stars only if the density of the
star formation region has a minimum physical density corresponding to 0.1 
hydrogen atoms per $\textrm{cm}^3$ \citep{Summers93, Katz96} and an
overdensity $\rho_{gas}/{\bar\rho_{gas}}>55.7$ (Katz et al. 1996), 
which basically
restricts star formation to collapsed, virialized regions.
The physical density threshold describes the steep drop in star formation rate
observed in disk galaxies when the gas surface density is much lower than a
critical value $\Sigma_c$ \citep{Kennicutt89}. The density threshold $\rho_c =
0.1 $ cm$^{-3}$ is compatible with observational results.   
According to \citet{Katz92}, the star forming region has to be part of a
converging flow that implies a local negative divergence of the SPH velocity
field. However, the converging flow criterium was introduced to describe star formation
in cosmological simulations, where the geometry of the collapsing regions is
approximatively spherical. In the case of star formation regions like 30
Doradus, localized at the periphery of the LMC disk where gas particles
relatively close in distance can have significantly different kinematics, 
this criterium leads to underestimate of
the star formation rate. 
Therefore the converging flow is not required in most of the simulations.
A single run including the converging flow requirement has been performed for comparison.

The star formation rate is assumed to be proportional to $\rho_{gas}^{3/2}$
\citep{Silk87}, where $\rho_{gas}$ represents the volume density of the cold
gas, and is given by the expression \citep{Katz96}
\begin{equation}
\frac{\textrm{d}\rho_{\star}}{\textrm{d}t} = \frac{c_{\star}\rho_{gas}}{t_{form}},
\label{starformeq}
\end{equation}
where the star formation timescale $t_{form}$ is the maximum between the local
gas dynamical collapse time $t_{dyn}= (4\pi G \rho_{gas})^{-1/2}$ and the local cooling
time. If the gas is already cool enough to form stars i.e $T<T_{max}$, then
$t_{dyn}$ is used. We assumed $T_{max}= 30000$K. 
The constant star formation rate parameter $c_{\star}$ is chosen such that we
reproduce the global LMC  star formation rate \citep{Sandage86}. 
Once a gas particle satisfies the above criteria, it spawns stars according to
a probability distribution function. In particular, the probability $p$ that a
gas particle forms stars in a time $\Delta t$ is modeled as
\begin{equation}
p=(1-e^{-c_{\star}\Delta t/t_{form}}).
\label{starformeq2}
\end{equation}
A random number is then drawn to determine whether the gas particle forms
stars during $\Delta t$. For all the simulations in this paper $\Delta t = 4 $ Myr. 
The newly created collisionless particle has the same position, velocity and
softening length as the original gas particle while its mass is a fixed
fraction $\epsilon$ of the parent gas particle, whose mass is reduced
accordingly. Following \citet{Katz92} we assumed for our favorite models a
star formation efficiency $\epsilon = 1/3$. 
Up to six particles are then created for each gas particle in the disk. After
its mass has decreased below $10\% $ of its initial value the gas particle is 
removed and its mass is re-allocated among the neighbouring particles.

\subsection{Test wind tunnels}

In order to study the influence of pure ram pressure on a galaxy model
orbiting in a Milky Way halo, we performed ``wind tunnel'' simulations where 
the ram-pressure value varies with time. 
 
We represent the hot gas as a flux of particles moving along the major axis of
an oblong of base equal to the diameter of the dark matter halo of the satellite
and height $h=vt$, where $v$ is the velocity of the LMC at the perigalacticon
and $t$ is the time scale of the simulation.
The hot particles have an initial random distribution and a
temperature $T=10^6 \textrm{K}$. 
The box has periodic boundary conditions in order to restore the flow of hot
gas that leaves the oblong.
The  galaxy model is at rest at the center of the oblong.

Kinematical data \citep{vanderMareletal02} indicate that
the LMC, presently located at $\sim 50$ kpc from the Galactic center, 
is just past a perigalactic passage and has an orbital velocity of about 300 km s$^{-1}$.
Recent proper motion measurements 
by \citet{Kallivayaliletal06} and \citet{ Piatek07} suggest that the velocity of the satellite is substantially
higher (almost 100 km s$^{-1}$) than previously estimated and consistent with the hypothesis 
of a first passage about the MW \citep{Besla07}.
In both scenarios the Cloud is affected by the largest ram-pressure values during the last 
million years of its orbital evolution. 
Indeed, while in the models proposed by \citet{Besla07} the LMC does not enter the halo of the MW 
earlier than 1 Gyr ago (slightly different orbits are found in Mastropietro 2008), in M05 we 
have shown that the change in the orbital parameters due to dynamical friction strongly affects 
the ram-pressure stripping rate. Even in the case of a ``low velocity''  model the largest 
ram-pressure on the satellite is expected during the last orbital semi-period (about 1 Gyr) due to the 
increasing velocity and external gas density.

We followed ram-pressure acting on the LMC's IGM during the past 1 Gyr.
The density of the hot external gas increases with time, in such a way that in our low velocity model 
the external pressure experienced by the cold disk varies from  
$P_{min}=\rho_{min} v_{min}^2= 5 \times 10^{-15}
\textrm{dyn}\, \textrm{cm}^{-2}$ to 
$P_{max}= 1.52 \times 10^{-13} \textrm{dyn}\, \textrm{cm}^{-2}$ at the time corresponding to the  
pericentric passage.
This is equivalent of assuming 
$v_{min}= 170 \,\textrm{km}\, \textrm{s}^{-1}$ and $v_{max}\sim 300\, \textrm{km}\, \textrm{s}^{-1}$, 
and a number density of the external gas that increases from  $\sim 10^{-5}
\textrm{cm}^{-3}$ to $10^{-4}\textrm{cm}^{-3}$ at $\sim 50$ kpc from the
Galactic center.
These density values are comparable with those provided by
M05, who modeled the MW hot halo assuming a spherical 
distribution of gas that traces the dark matter profile, with a mean number 
density of $2 \times 10^{-5}$ within 150 kpc.
We also consider the eventuality of a less dense Galactic halo and performed 
runs where the gas density is a factor 10 lower.
Models with higher velocities \citep{Kallivayaliletal06} and orbital parameters similar 
to those suggested by \citet{Besla07} are also explored. 
In details, $P_{min}$ is the same as in the low velocity models since the higher orbital velocity at 
the beginning of the simulation (about 250 km s$^{-1}$) is compensated by a lower external density 
($\sim 5 \times 10^{-6}\textrm{cm}^{-3}$, according to Mastropietro 2008. Indeed 1 Gyr ago the satellite has just passed through the virial radius 
of the MW). 
The maximum pressure felt by the disk is 
$P_{max}= 2.67 \times 10^{-13} \textrm{dyn}\, \textrm{cm}^{-2}$, that corresponds to a Cloud moving with 
$v_{max}\sim 400\, \textrm{km}\, \textrm{s}^{-1}$ through an external hot medium of density  
$10^{-4}\textrm{cm}^{-3}$.

Each galaxy model is simulated using 750000 particles, of which $6 \times
10^5$ are in the dark halo and $1.5 \times 10^5 $ in the disk ($10^5$ 
collisional  and $5 \times 10^4$ collisionless). 
The hot gas in the ``wind tunnel'' has $2 \times 10^6$ particles, in such a
way that the mass ratio $m_h/m_{disk}$ between hot particles and particles in
the disk is close to the unity even when the halo density is the largest. This
choice permits to avoid the presence of scattering and numerical holes which 
artificially change the shape of the front edge and influence the morphology
of the disk (M05).
The gravitational spline softening is set equal to 0.5 kpc
for the dark halo and the hot gas in the oblong, while it is 0.2 kpc for stars
and gas in the disk.

\section{Cooling simulations}

\begin{figure}
\includegraphics[%
  scale=0.4]{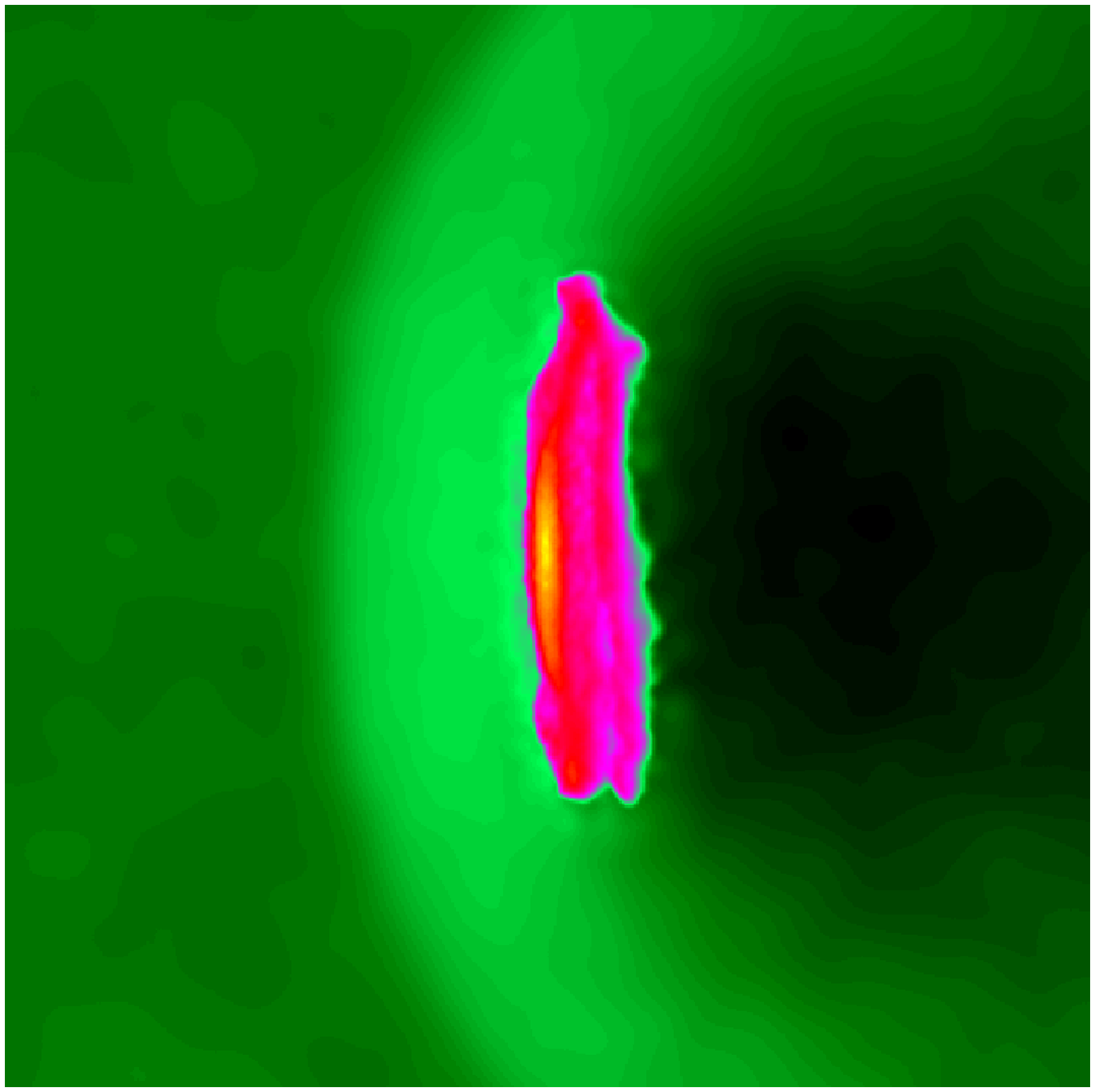}
\includegraphics[%
  scale=0.4]{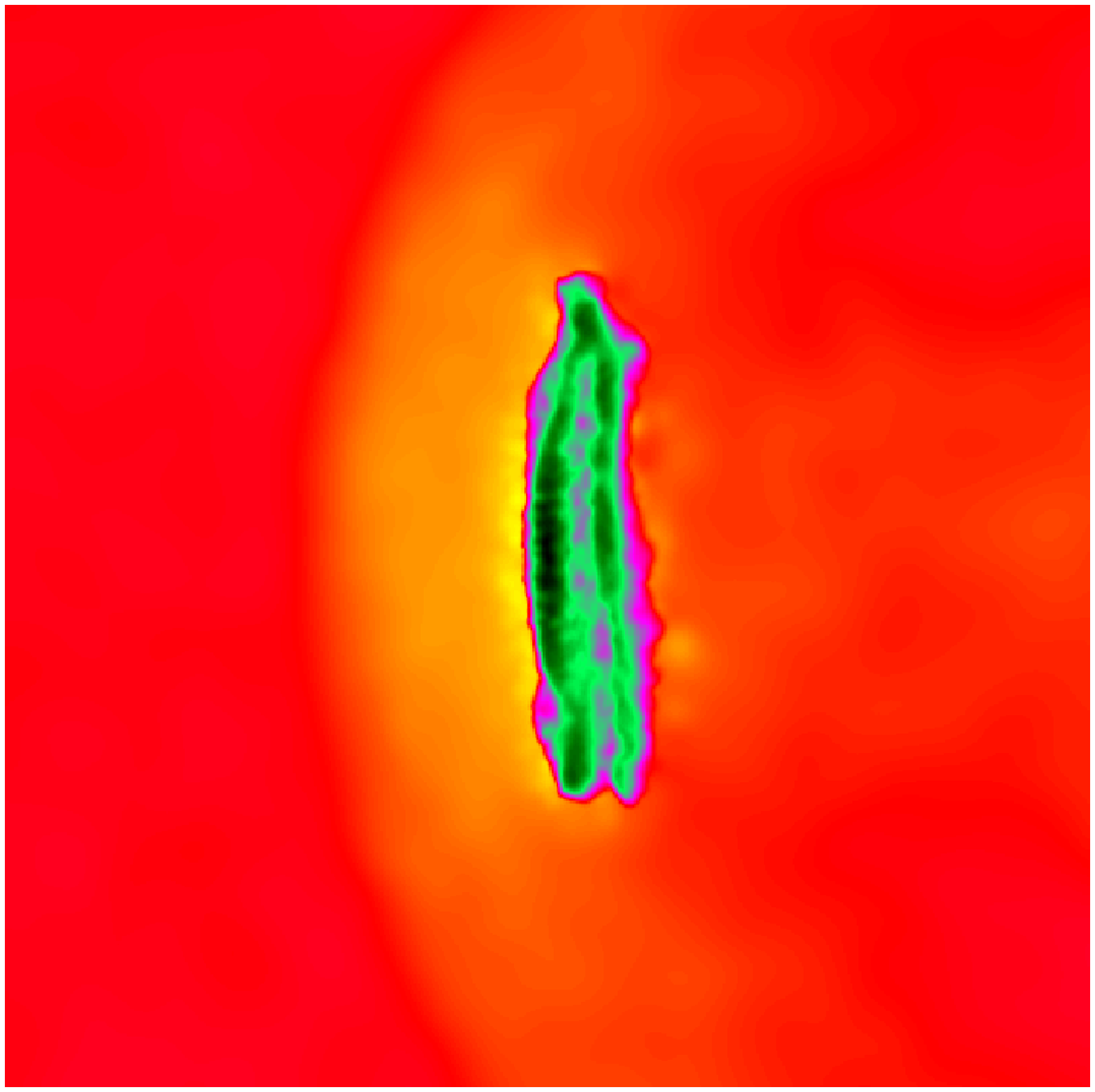}
\includegraphics[%
  scale=0.3]{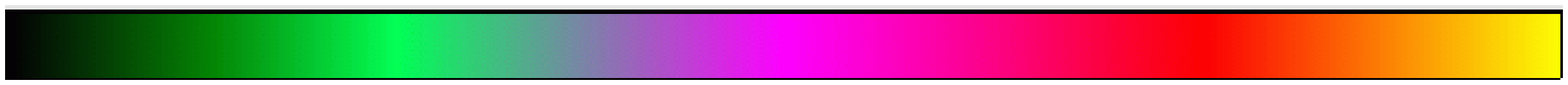}
\caption{Density (upper panel) and temperature (bottom) map of a 50x50
  kpc$^2$ region around the satellite, which is moving towards the 
left of the page nearly face-on (inclination angle $i=10^{\circ}$,
  cool10). 
 Logarithmic color scaling is indicated by the key at the bottom of
  the figure.  
 The surface density map ranges between $\sim 2 \times 10^3$ (black) and
  $\sim 2 \times 10^9$ M$_{\odot}$ kpc$^{-2}$ (yellow), while
  temperatures vary from $10^4$ (black) to $10^7$ K (yellow). 
}
\label{shockimage}
\end{figure}

In order to study the effect of pure compression on the density distribution
of cold gas in the LMC disk, we have run a first set of simulations where the gas
cools radiatively but star formation was not activated.
 
According to \citet{vanderMareletal02} and \citet{Kallivayaliletal06}, the present 
angle between the LMC's disk and its proper motion vector is roughly
$30^{\circ}$. Even neglecting the effects of precession and nutation on the
disk plane of the satellite \citep{vanderMareletal02} this angle is expected to
vary significantly during an orbital period, especially in the proximity of a
pericentric passage due to  rapid changes in the velocity vector.
Different relative orientations of the disk with respect to the orbital motion 
are therefore investigated. 
The inclination angle $i$ is defined as the angle between the angular momentum vector
of the disk and the flux of hot particles in the wind tube, so that a galaxy
moving edge-on through the external medium is characterized by $ i= 90
^{\circ}$, while the observed LMC disk would have $i \sim 60^{\circ}$.    
We explored cases with inclination angle $i$ of 
90, 45 and $10^{\circ}$ (runs cool90, cool45 and cool10, respectively).
  
With a hot halo temperature of $10^6$ K the relative velocity between the satellite
and the external medium is supersonic (sound speed $\sim 135$ km s$^{-1}$ and 
Mach number ${\cal M} = 2.2 $ and $3$ at the pericenter of the low and high velocity orbit, respectively) 
and a bow shock forms in front of the disk (Fig. \ref{shockimage}).
Since the cooling time of the post-shock gas is $\sim 16$ Gyr, 
the shock can be considered adiabatic and hydrodynamical
quantities at the two sides of the shock front are in first approximation
related by the Rankine-Hugoniot equations for a stationary normal shock.

\begin{figure}
\epsfxsize=8truecm \epsfbox{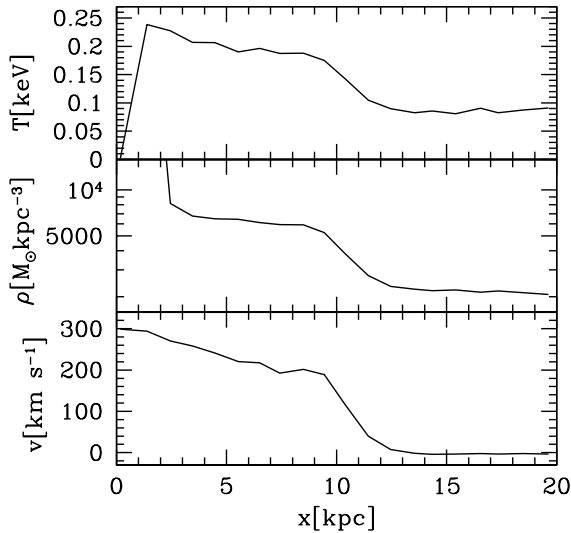} 
\caption{Gas temperature (top panel), density (middle) and one
  dimensional velocity (bottom) profiles across the shock discontinuity.}
\label{shock}
\end{figure}

For ${\cal M}>1$ the jump conditions give $\rho_1/\rho_2 = v_2/v_1< 1$ and
$T_1< T_2$, where subscripts 1 and 2 denote upstream and downstream
quantities.
The ram-pressure $P_2= \rho_2 v_2^2$ actually felt by the galaxy
behind the shock front is therefore smaller than that it would suffer
due to the upstream flux of hot particles, but conservation of
momentum flux across the shock discontinuity implies that the
reduction in dynamical pressure has to be balanced by an increase in
thermal pressure (see also Rasmussen et al. 2006).

Fig. \ref{shock} illustrates the behavior of hydrodynamical quantities across
the shock discontinuity for a snapshot corresponding to the
perigalacticon of a low velocity orbit. The disk inclination is $i=10^{\circ}$.
The horizontal axis is centered on the LMC stellar disk and oriented
perpendicularly to the bow shock nose, with the shock located at $x \sim
10$ kpc and the satellite moving towards increasing
values of $x$.
The Mach number derived by the temperature jump is ${\cal M}=2.1$, in
good agreement with the theoretical value for a normal shock.
Only hot halo particles are considered in computations but, due to the SPH nature of the simulations, close to the border of the disk we observe a further density increase and a sharp drop in temperature. 
The $x$-velocity profile is plotted in the system of
reference where the pre-shock gas is at rest.

\begin{figure}
\epsfxsize=8truecm \epsfbox{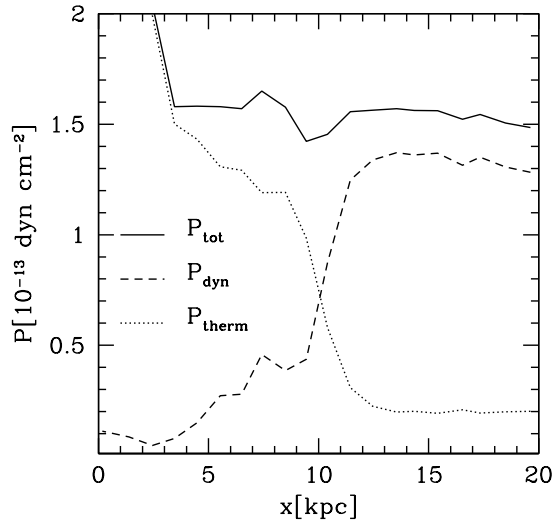} 
\caption{Dynamical ($P_{dyn}$), thermal ($P_{therm}$) and total ($P_{tot}$)
  pressure across the shock.}
\label{shockpressure}
\end{figure}

Pressure profiles across the shock are plotted in
Fig. \ref{shockpressure}, where $P_{dyn}$ is defined as $\rho v^2$. The
total pressure remains roughly constant until the edge of the disk. 
The steep increase at $x=2.5$ kpc is due the rapid
growth in density at the border of the disk, which is not immediately
followed by a decrease in temperature.

For $i>0^{\circ}$ the shock wave is inclined with respect to the
initial flow velocity and the Rankine-Hugoniot conditions apply to the
normal components of the velocity across the shock discontinuity,
while the component parallel to the shock front remain unchanged. 
The flow is therefore deflected toward an oblique shock wave and the
jump at the shock discontinuity is smaller.

An edge-on ($i=90^{\circ}$) disk behaves like a wedge moving
supersonically with the vertex facing upstream. If the wedge angle is smaller
than or equal to the maximum flow deflection angle, the oblique shock becomes
attached to the vertex of the wedge and the flow is deflected so
that the streamlines are parallel to the surfaces of the wedge.
The shock standoff
distance thus depends on the external density profile of the collisional
edge-on disk and on the Mach number of the incident flow.
In both low and high velocity edge-on models the shock results almost attached to the disk.

   \begin{figure*}
\includegraphics[%
  scale=0.32]{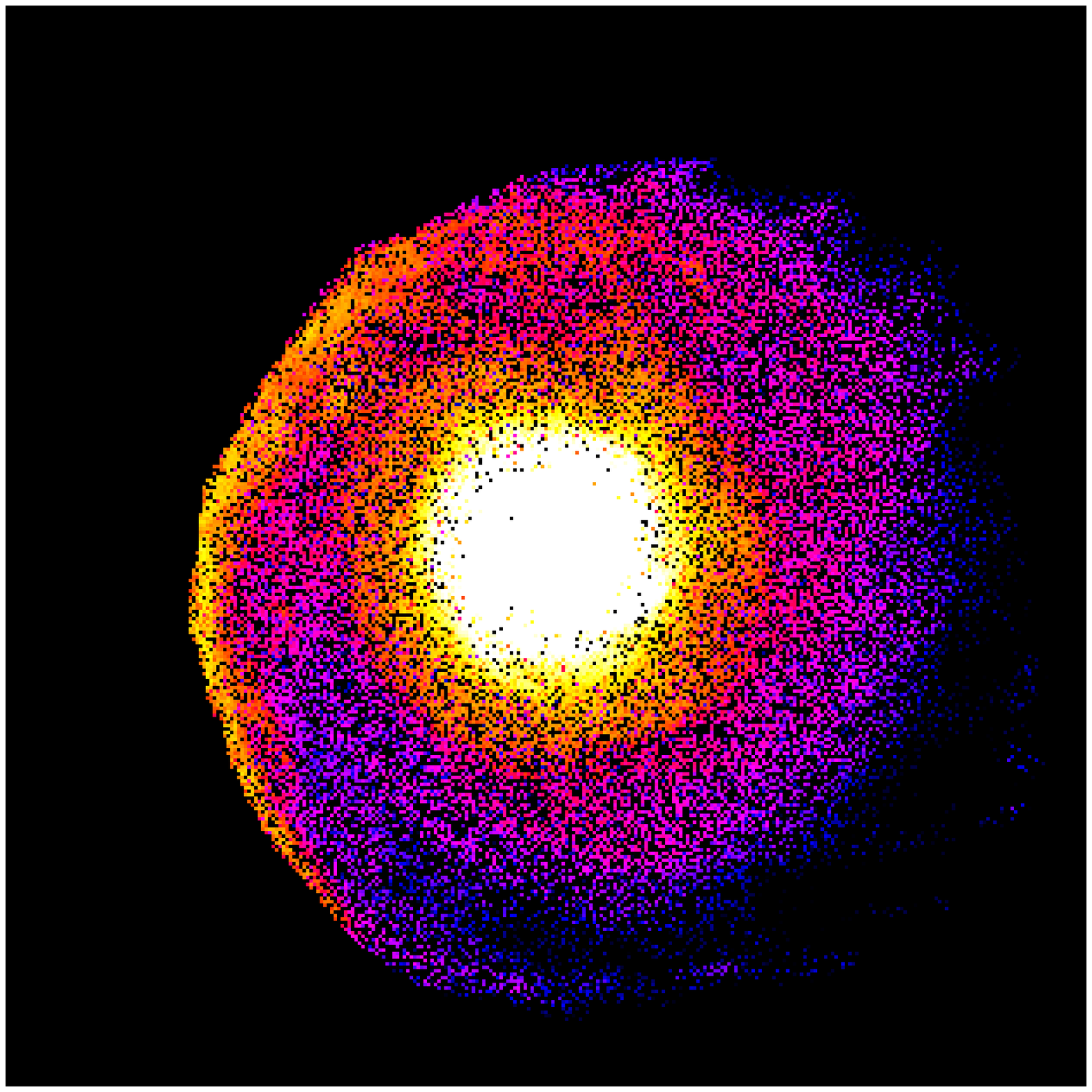}
\includegraphics[%
  scale=0.27]{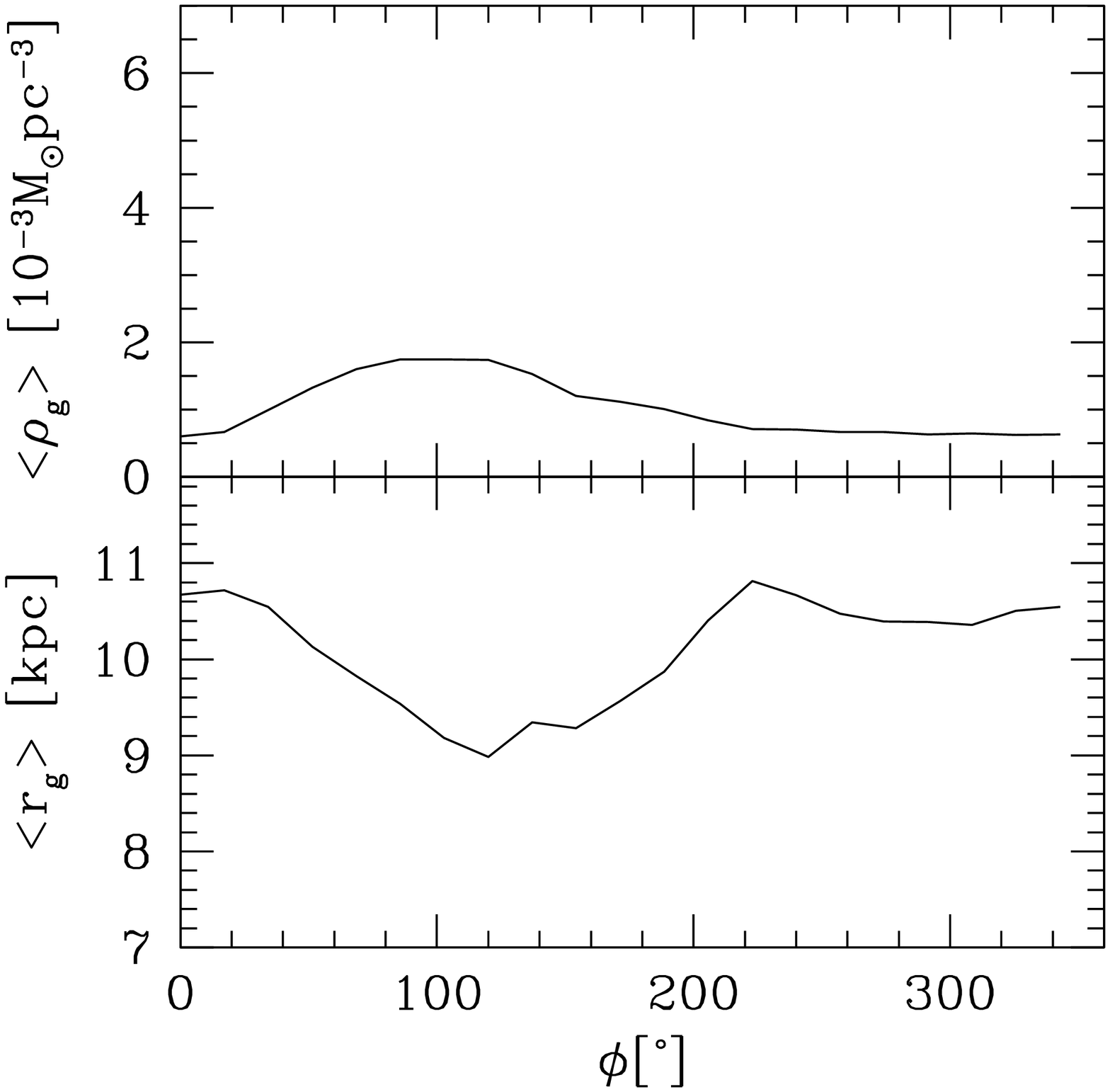}
\hspace {3cm}
\includegraphics[%
  scale=0.32]{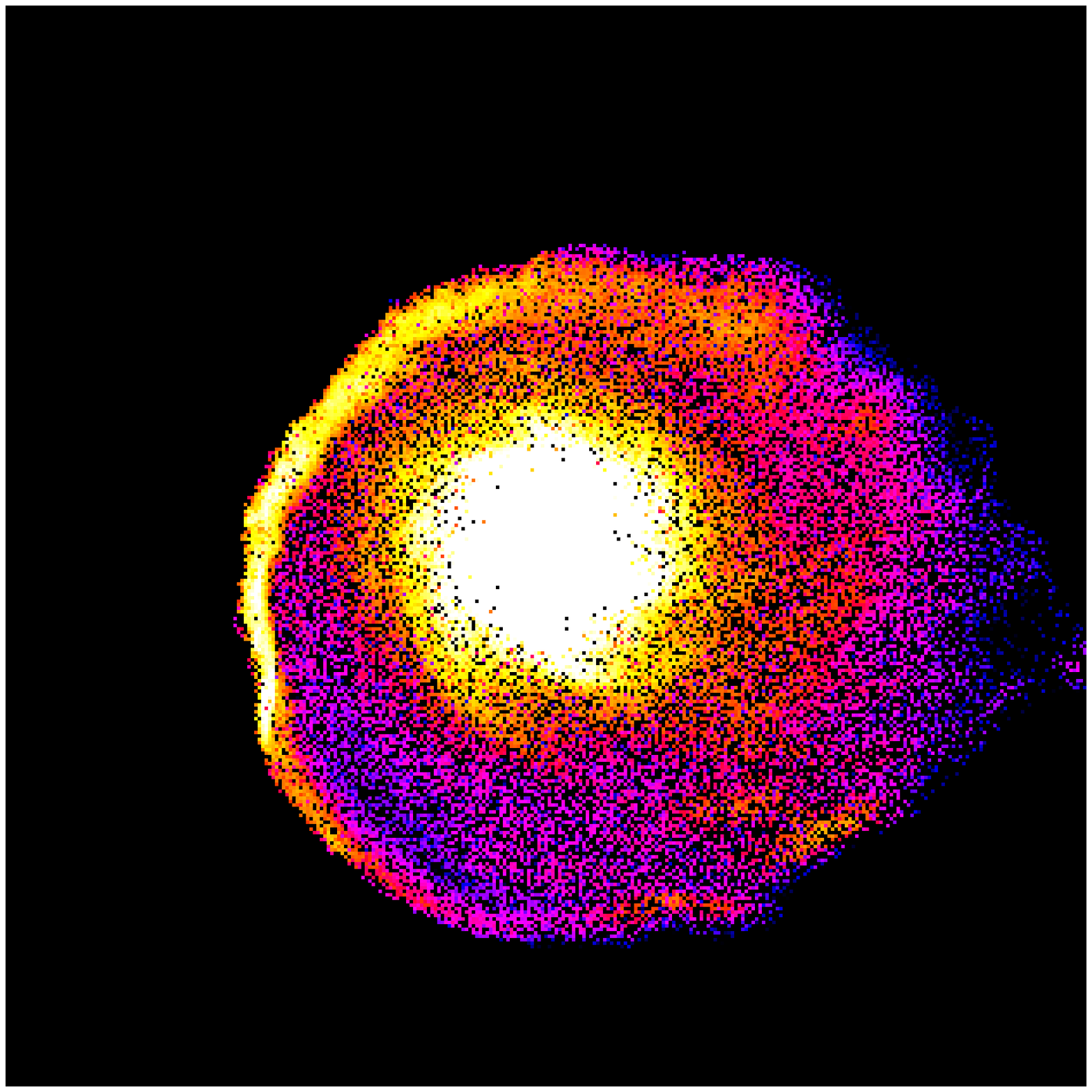}
\includegraphics[%
  scale=0.27]{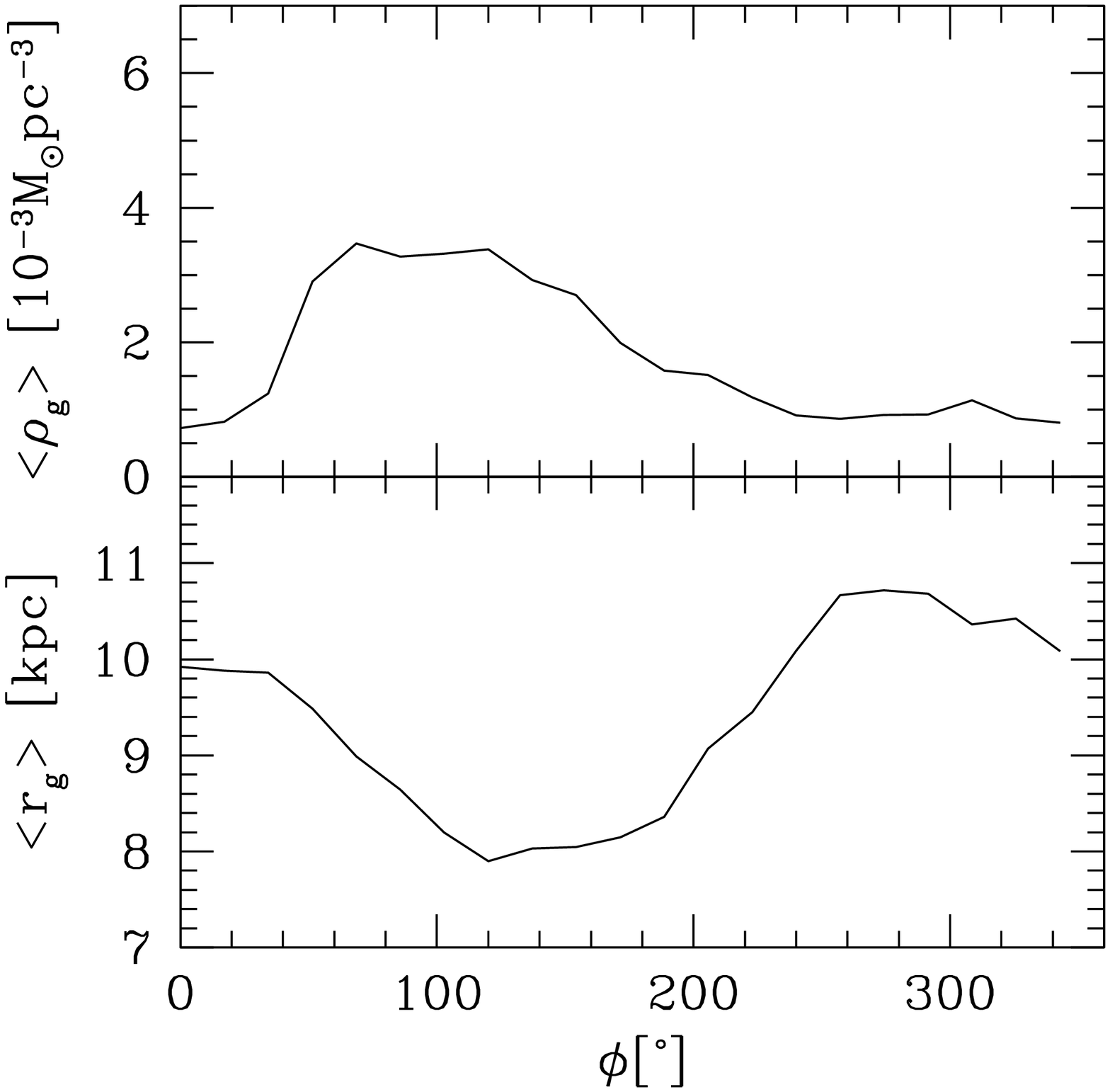}
\hspace {3cm}
\includegraphics[%
  scale=0.32]{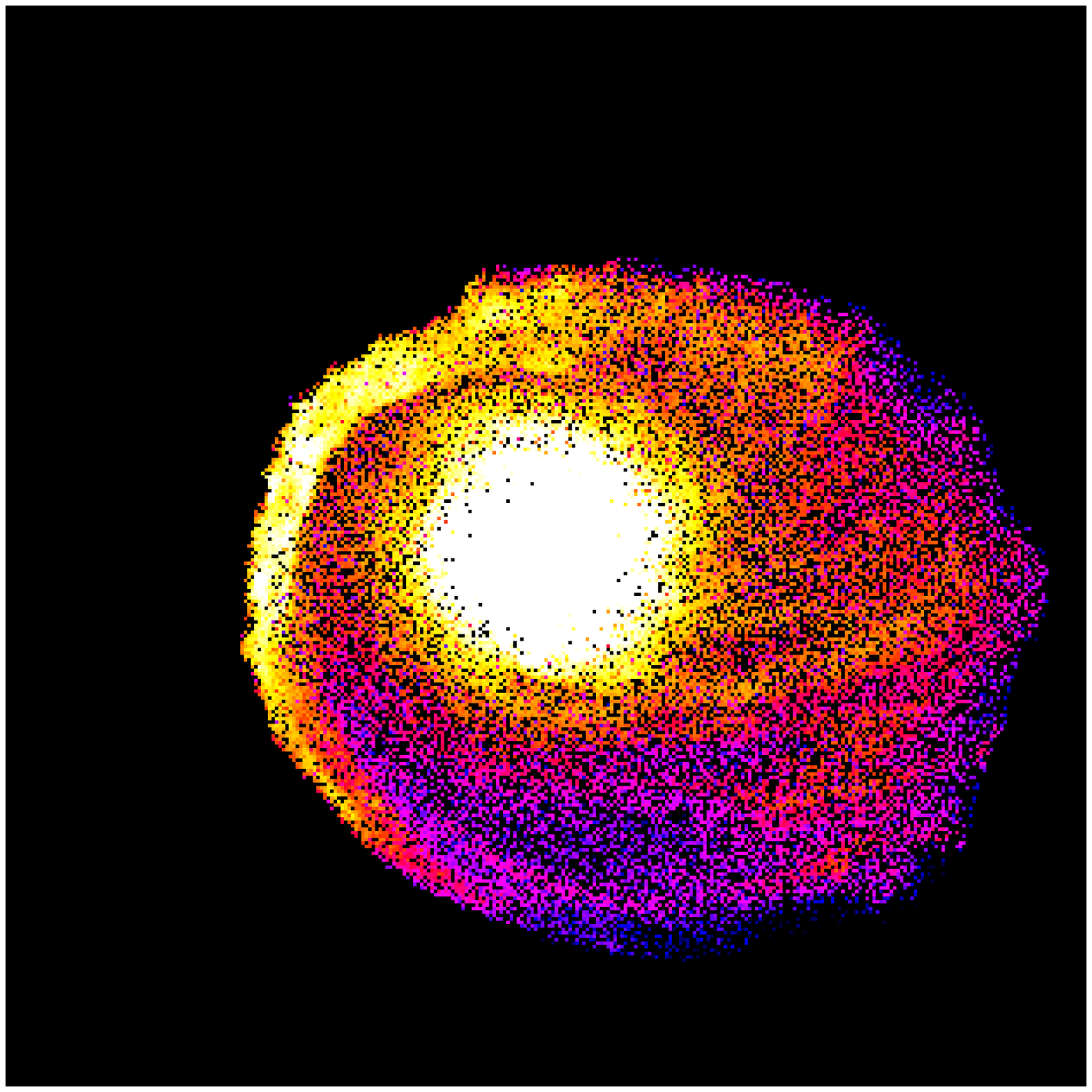}
\includegraphics[%
  scale=0.27]{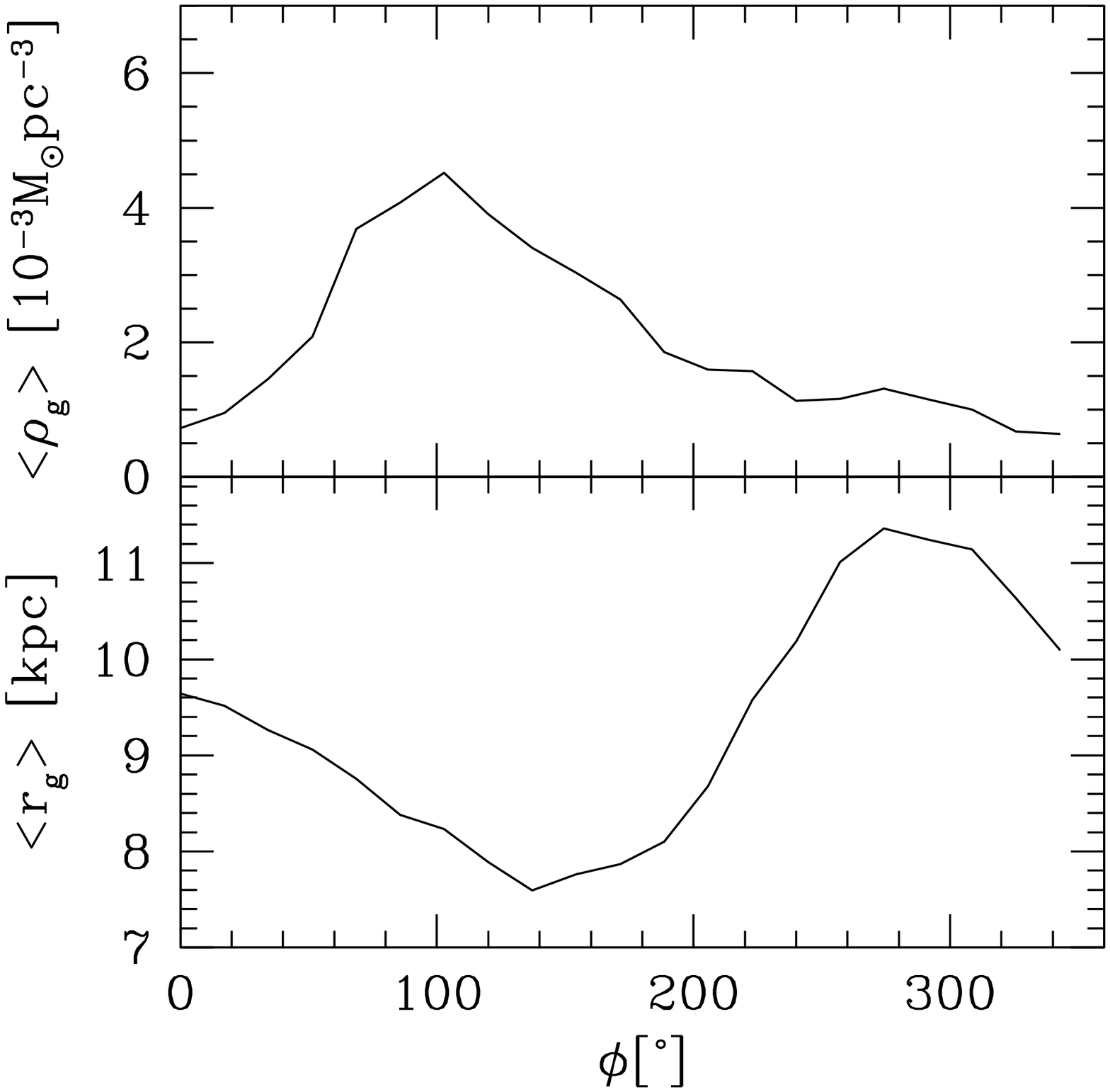}
\hspace {3cm}
\includegraphics[%
  scale=0.32]{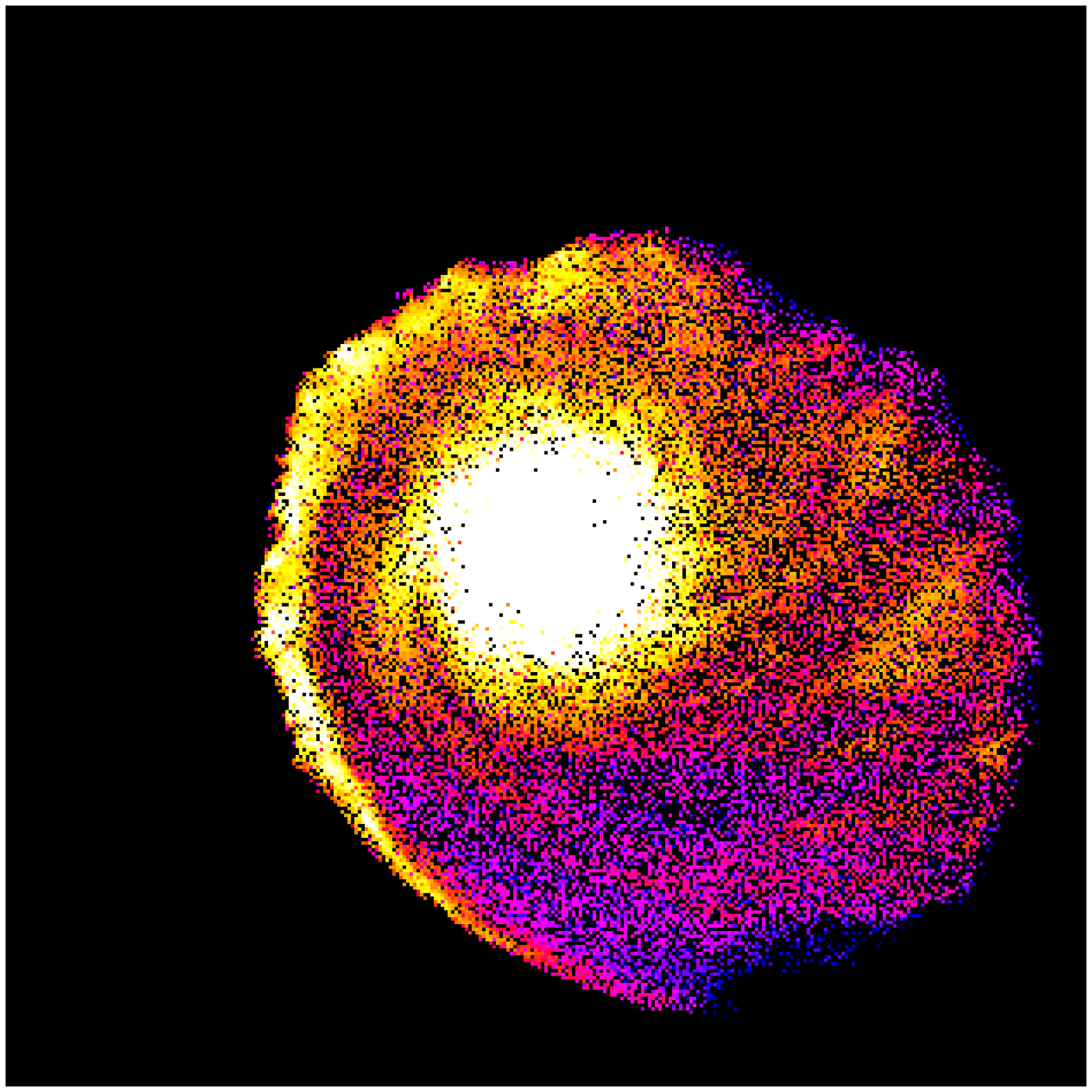}
\includegraphics[%
  scale=0.27]{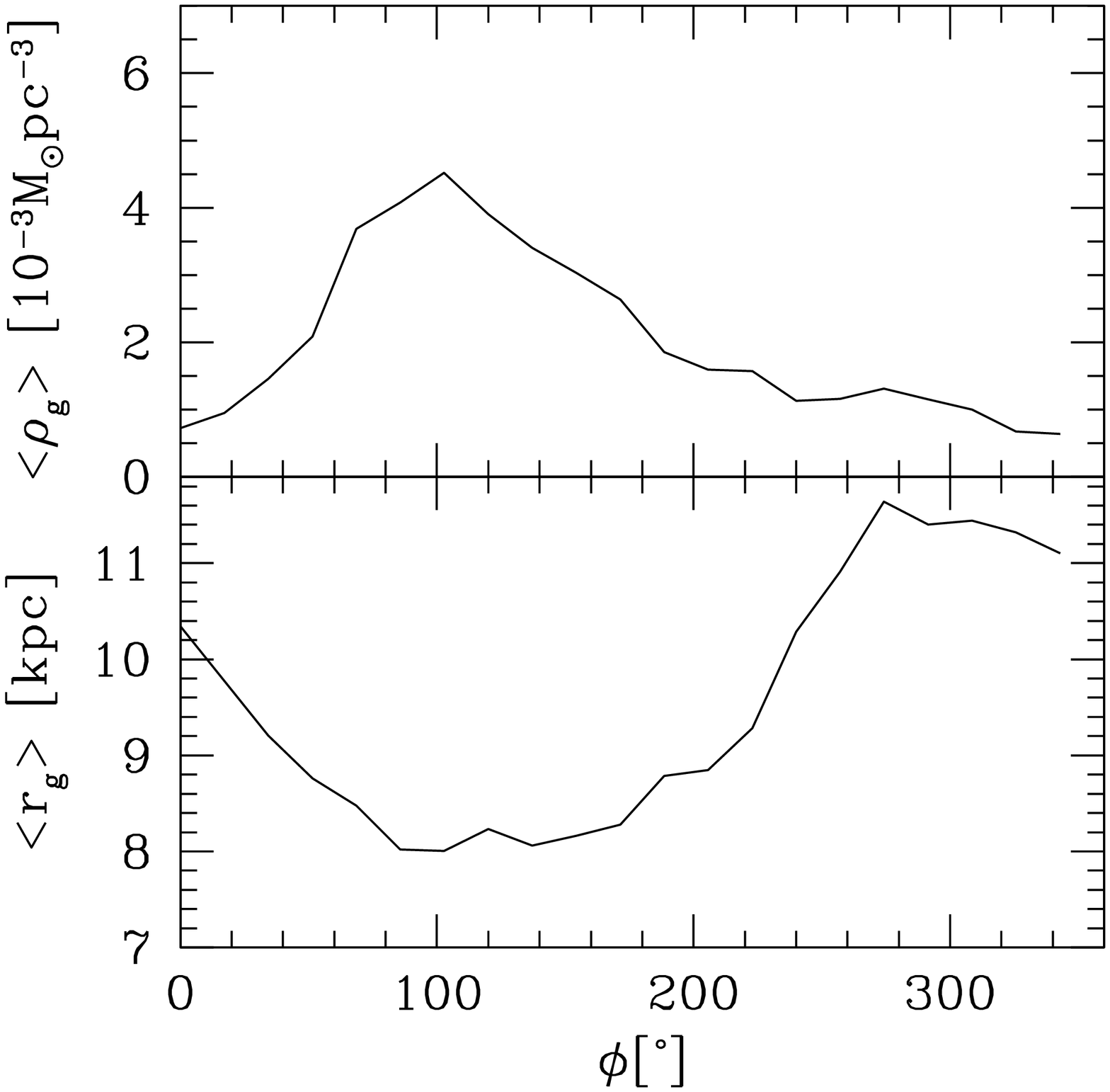}
\hspace {5cm}
 \includegraphics[%
  scale=0.32]{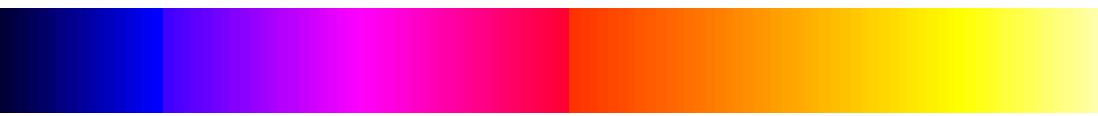}
\hspace {5.5cm} 
\caption{Edge-on model (run cool90). Evolution of the disk gas density distribution. Left panels: HI column density maps (the galaxy is
  moving towards the left of the page). The color-scale is logarithmic with limits $5 \times 10^{19}$ and $1.2 \times10^{21}$ cm$^{-2}$. 
Right panels: mean gas density and radius of the external disk as a function of the azimuthal angle $\phi$. Each pair of plots (from the top to the bottom) represents the state of the disk at increasing times along the orbit: Time$=0.4,0.6, 0.8, 1$Gyr.}
\label{densitymaps}
\end{figure*}

\begin{figure*}
\includegraphics[%
  scale=0.32]{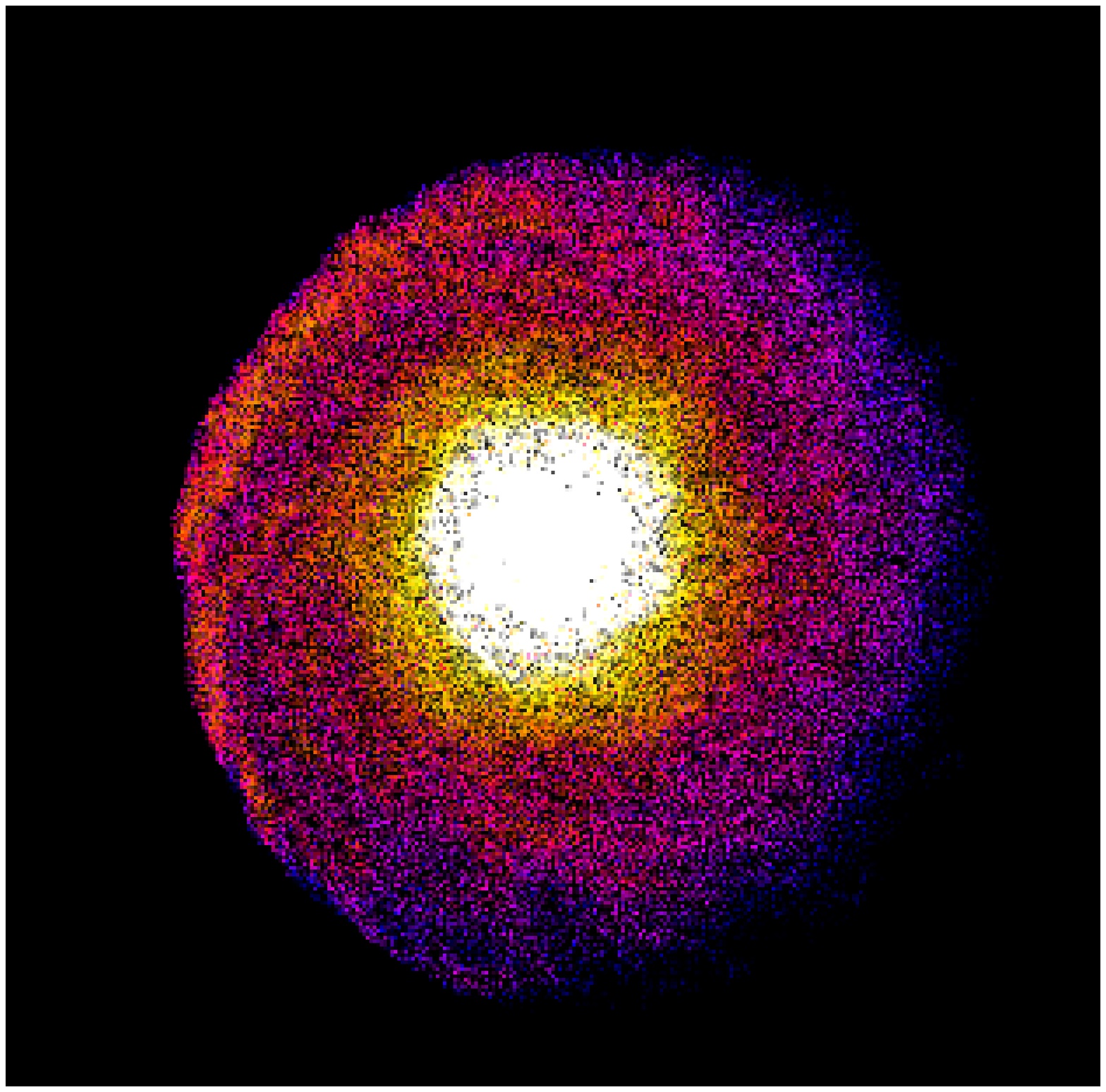}
\includegraphics[%
  scale=0.27]{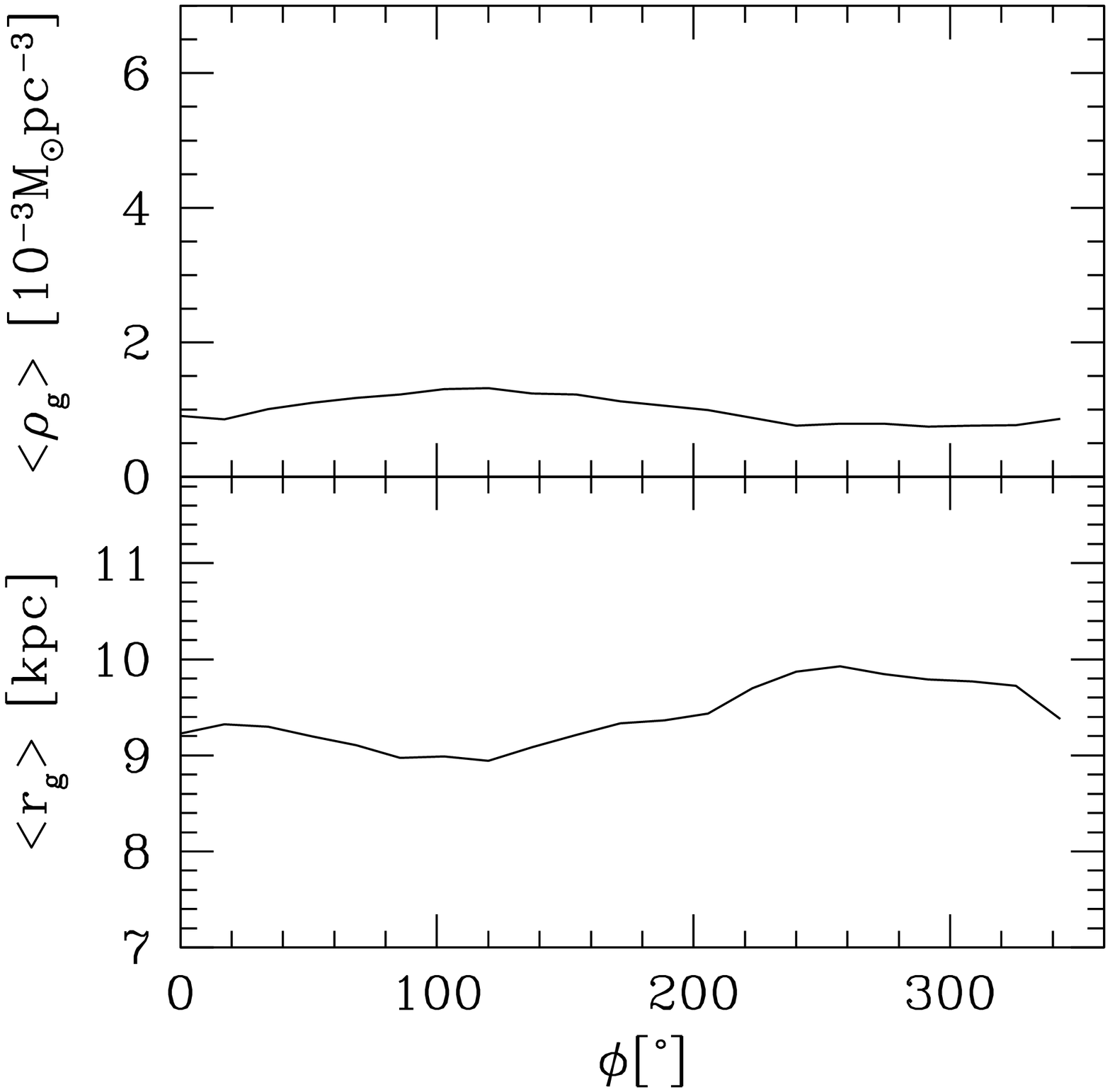}
\hspace {3cm}
\includegraphics[%
  scale=0.32]{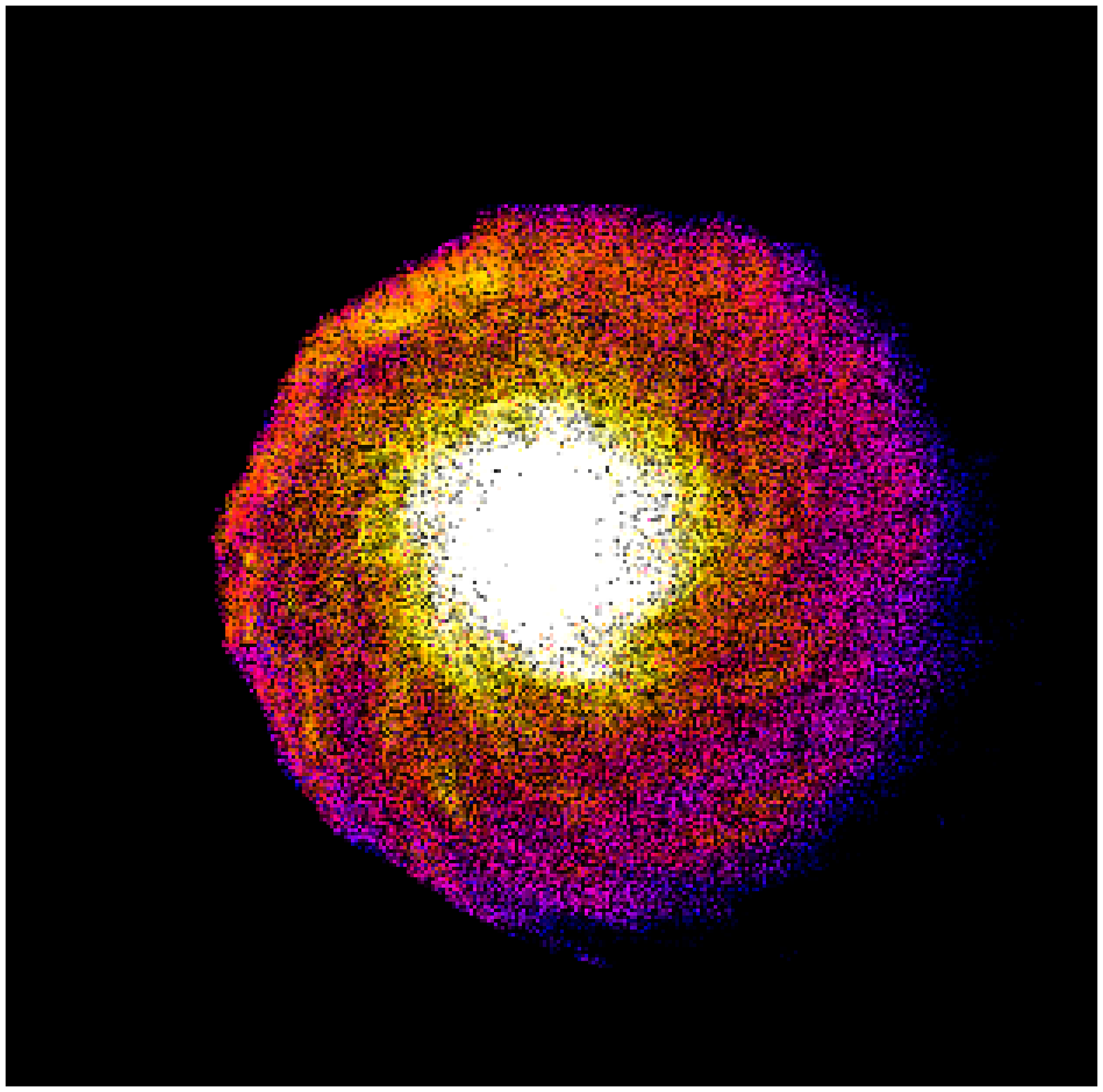}
\includegraphics[%
  scale=0.27]{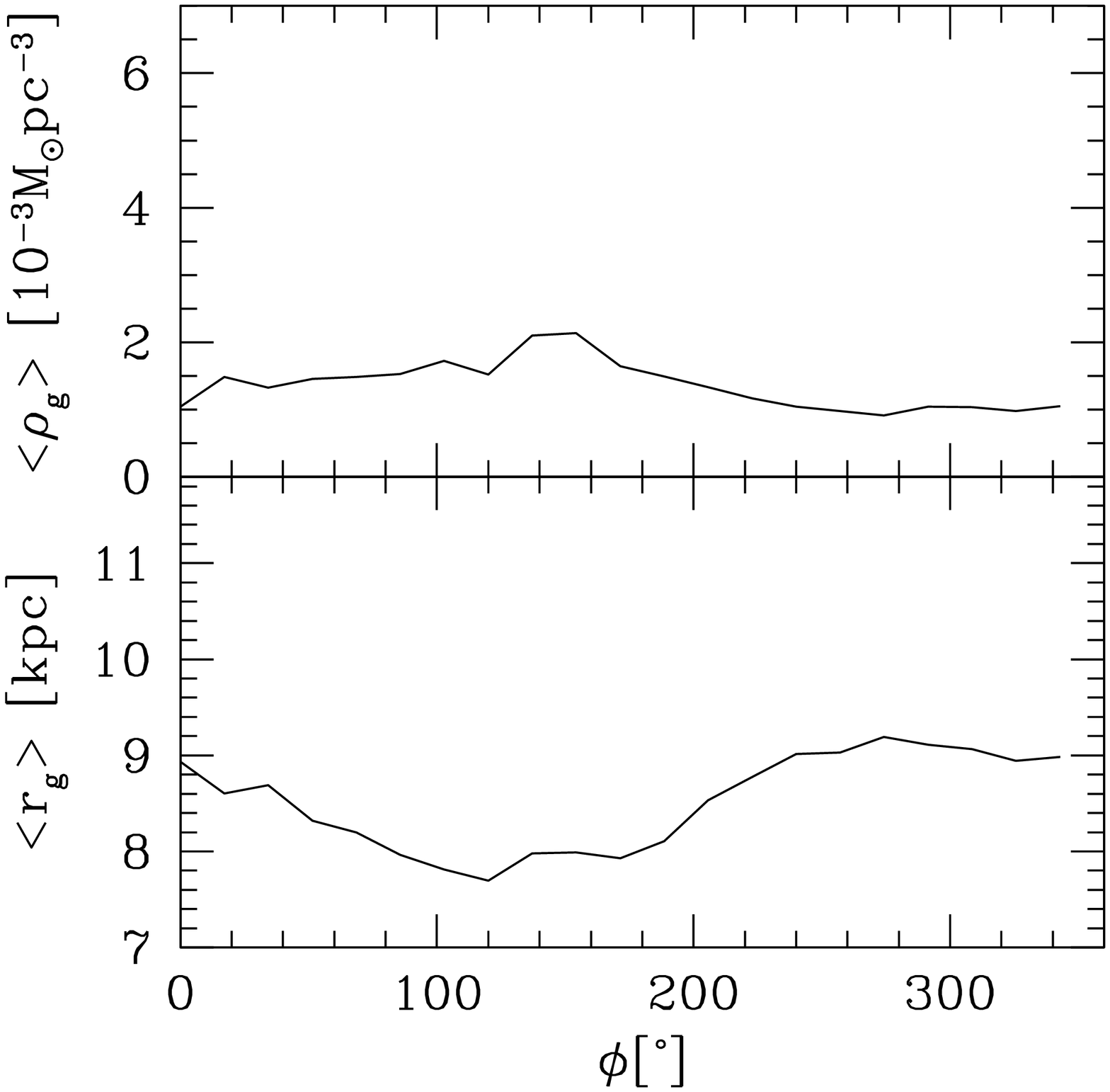}
\hspace {3cm}
\includegraphics[%
  scale=0.32]{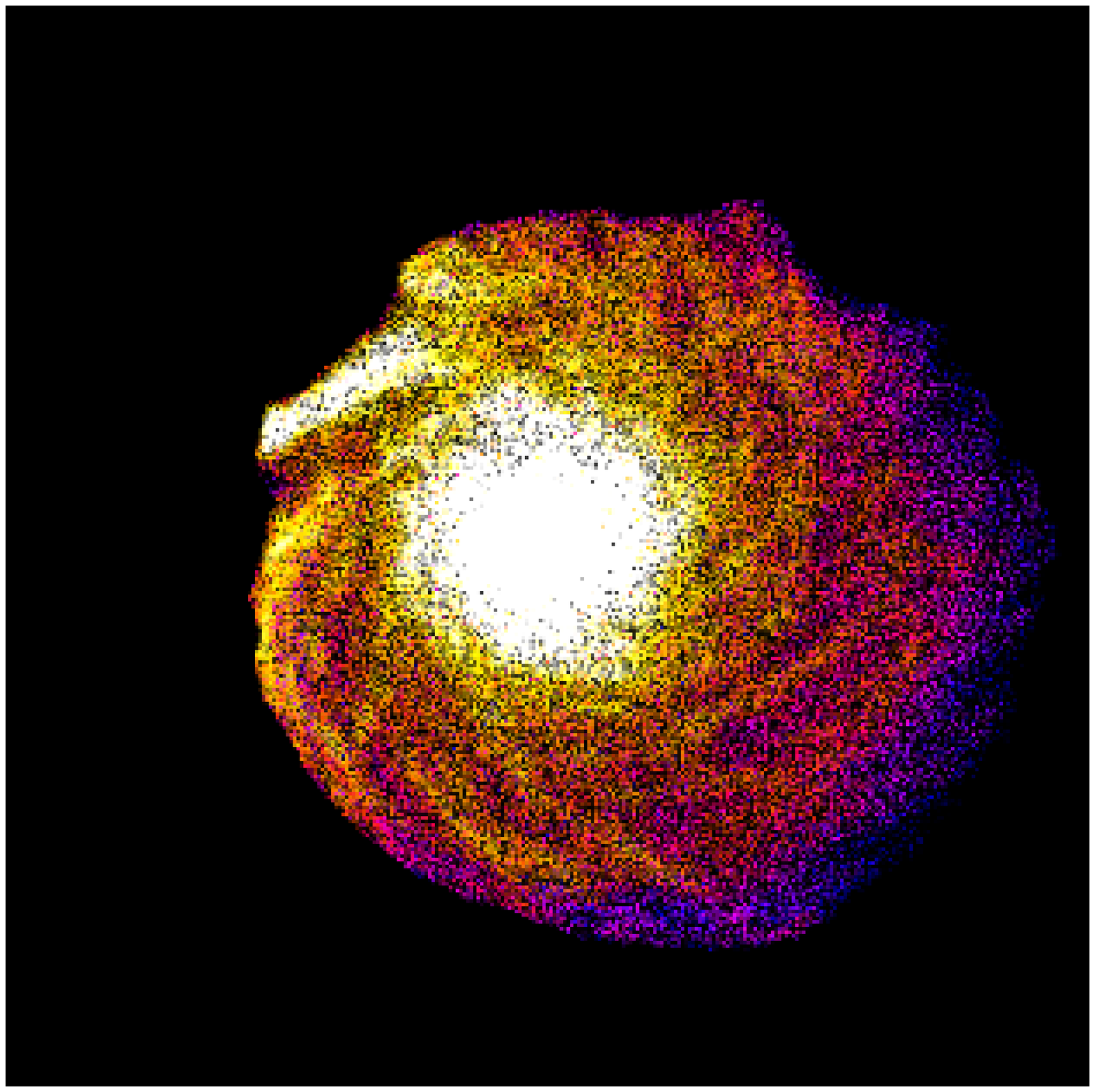}
\includegraphics[%
  scale=0.27]{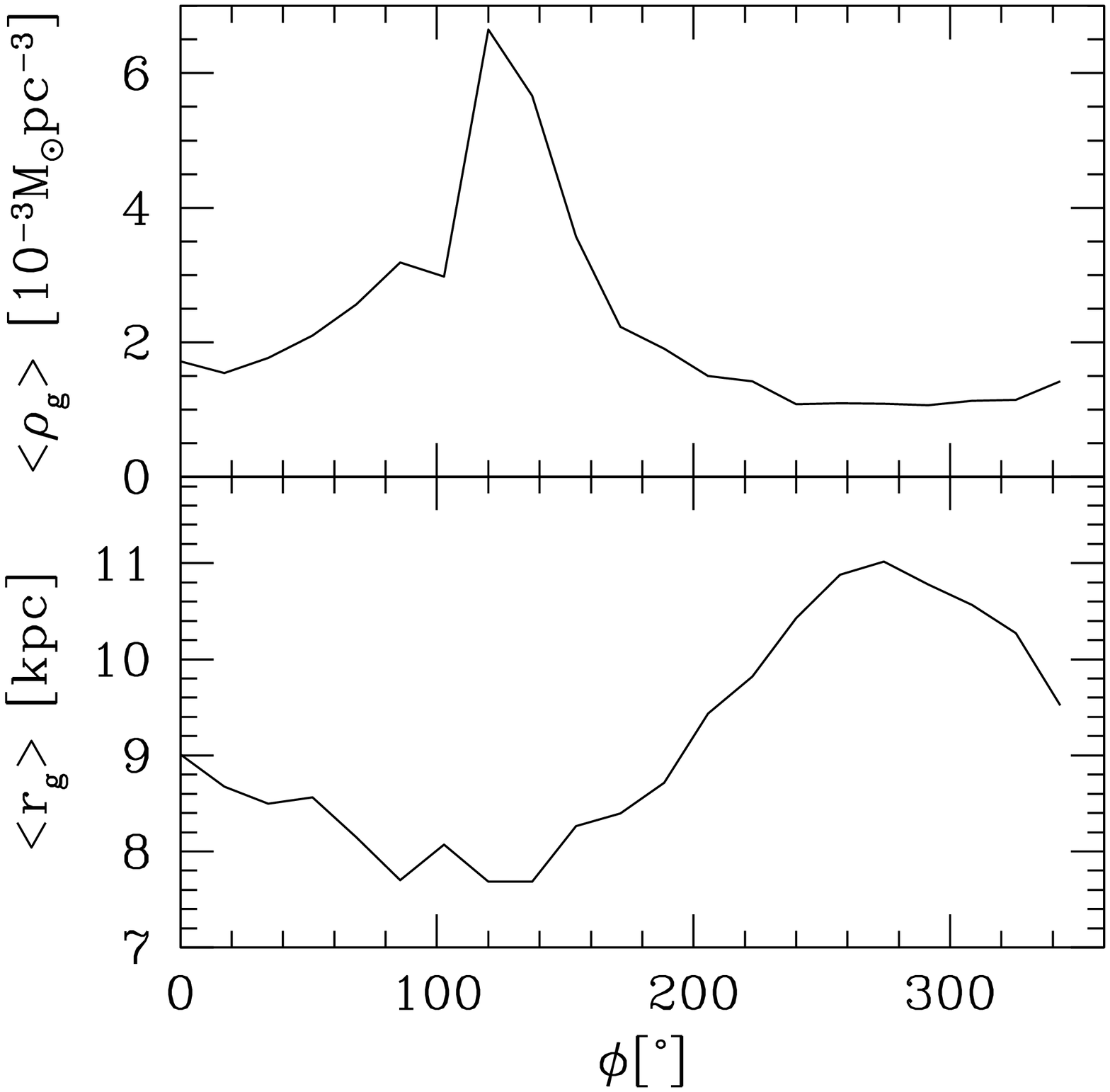}
\hspace {3cm}
\includegraphics[%
  scale=0.32]{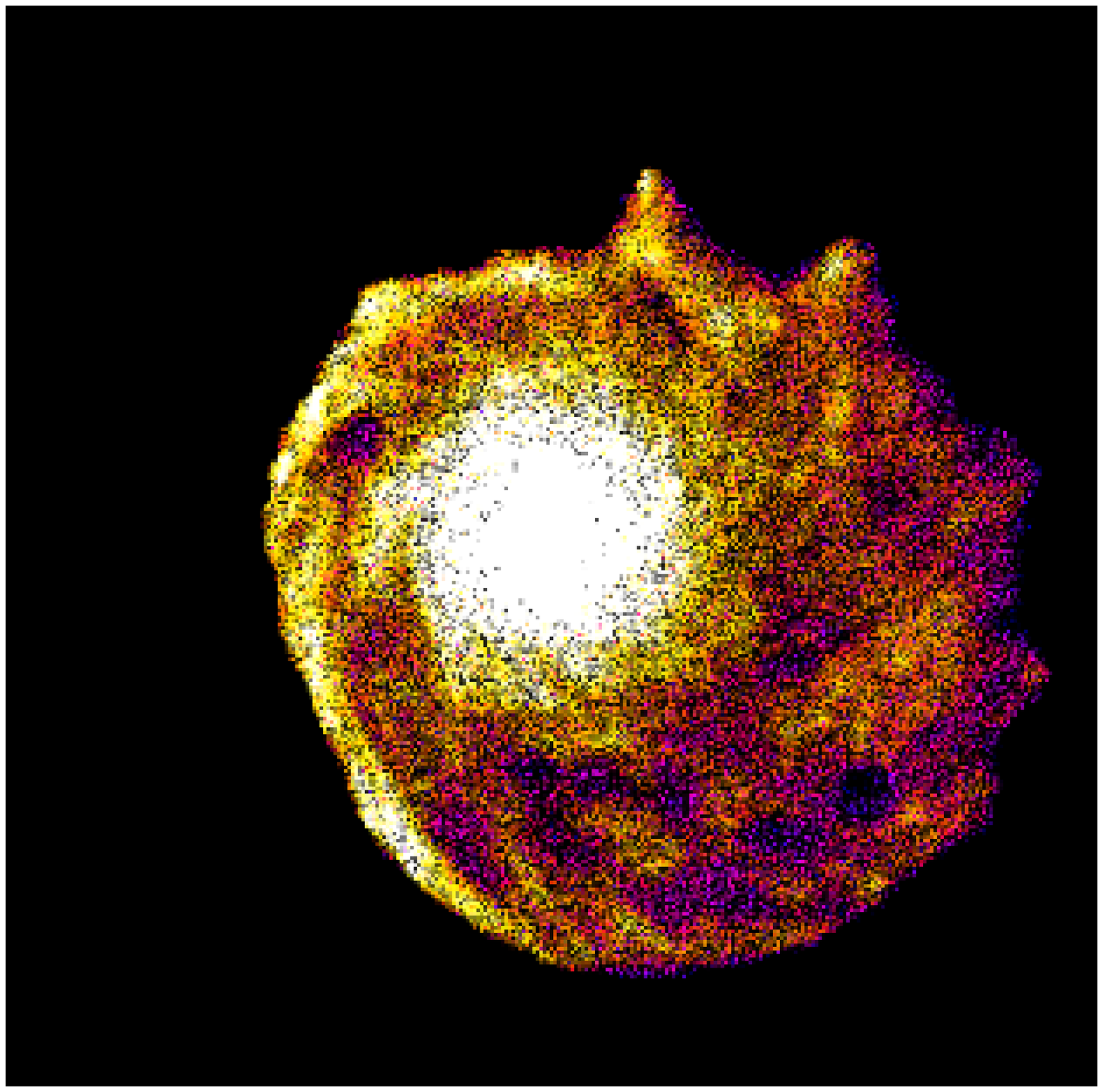}
\includegraphics[%
  scale=0.27]{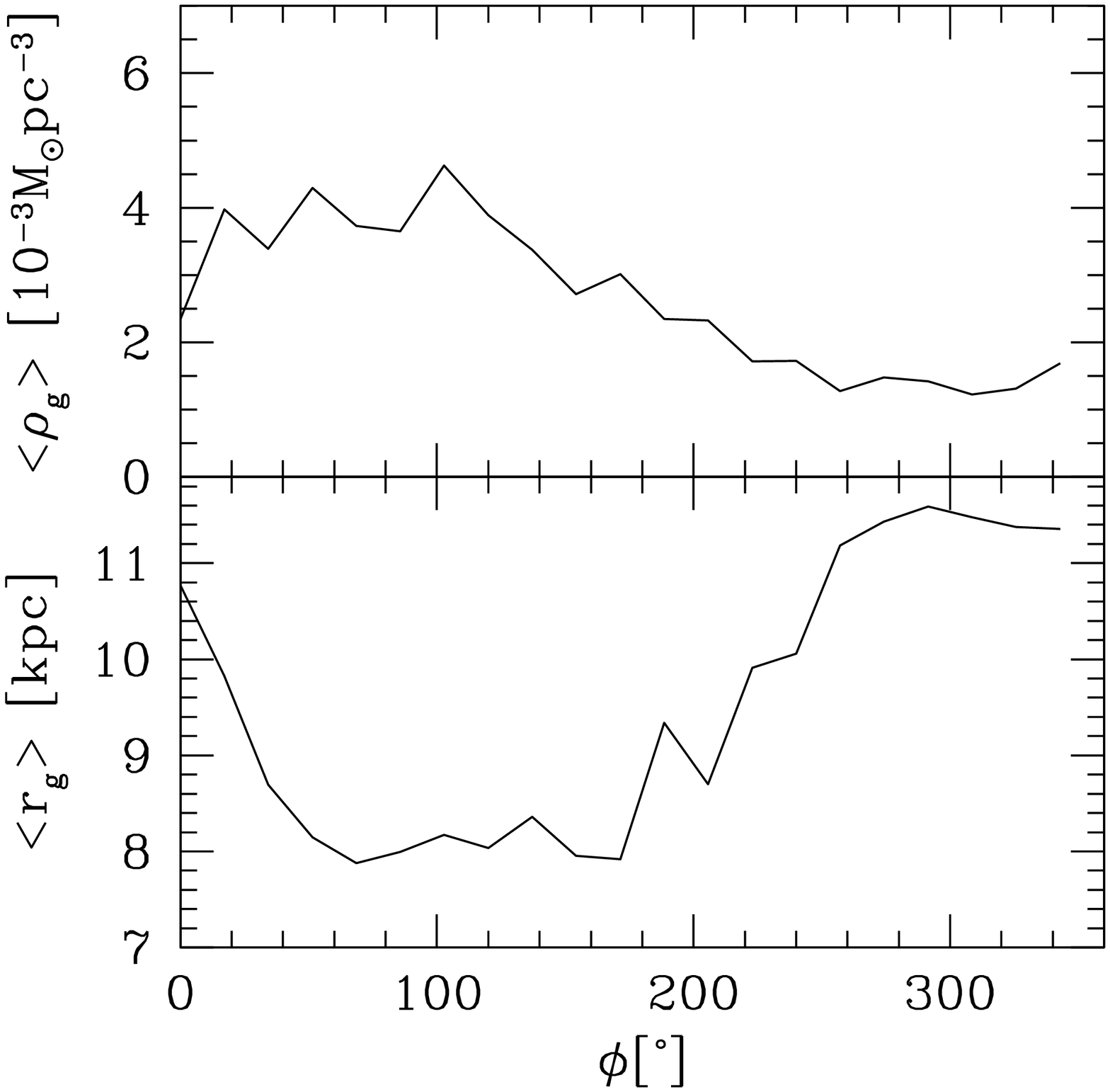}
 \hspace {5cm}
 \includegraphics[%
  scale=0.32]{bar1.eps}
\hspace {5.5cm} 

\caption{The same as in Fig. \ref{densitymaps} for the model with inclination angle $i= 45^{\circ}$ (run cool45)..}
\label{densitymaps45}
\end{figure*}

\begin{figure*}
\includegraphics[%
  scale=0.32]{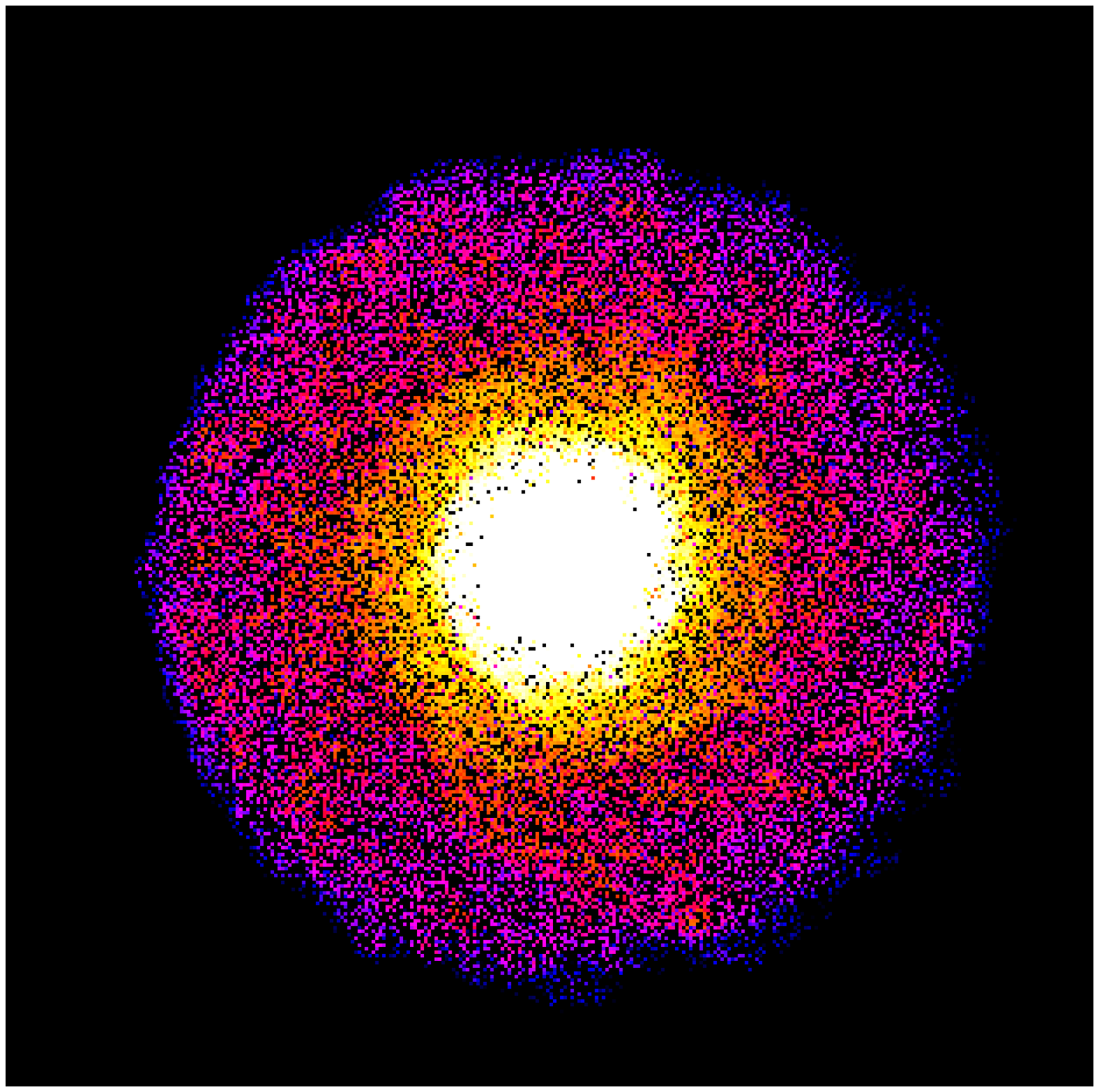}
\includegraphics[%
  scale=0.27]{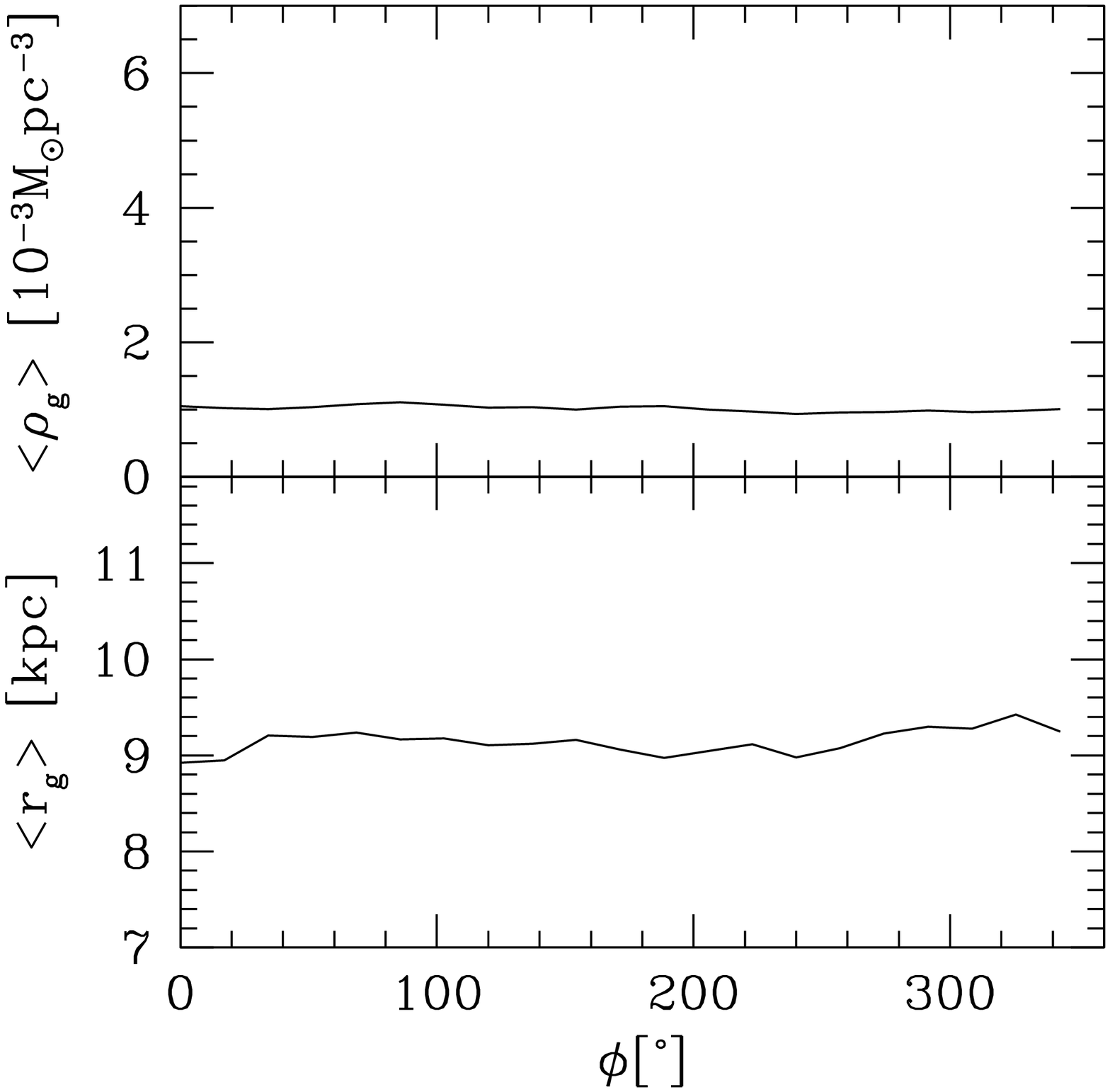}
\hspace {3cm}
\includegraphics[%
  scale=0.32]{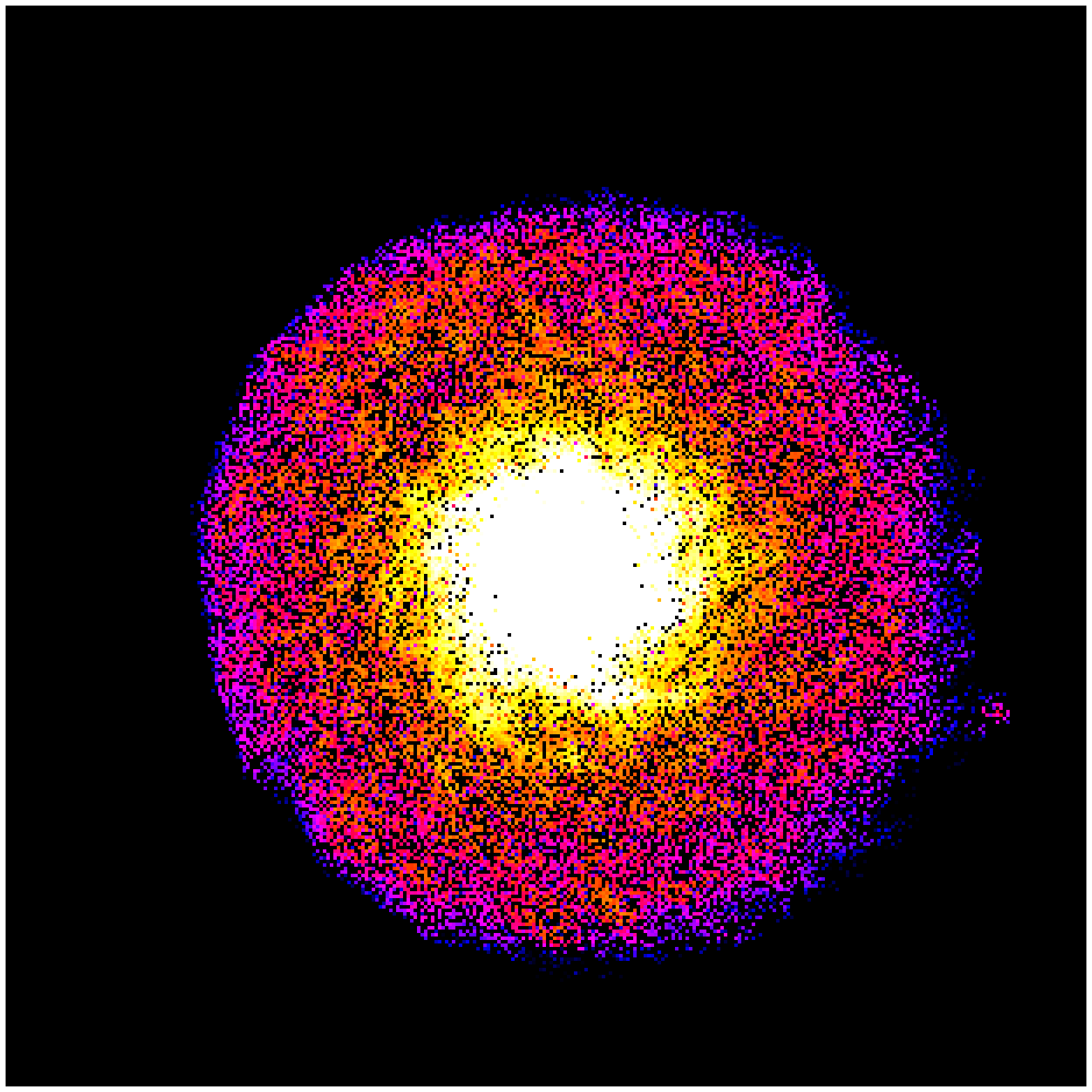}
\includegraphics[%
  scale=0.27]{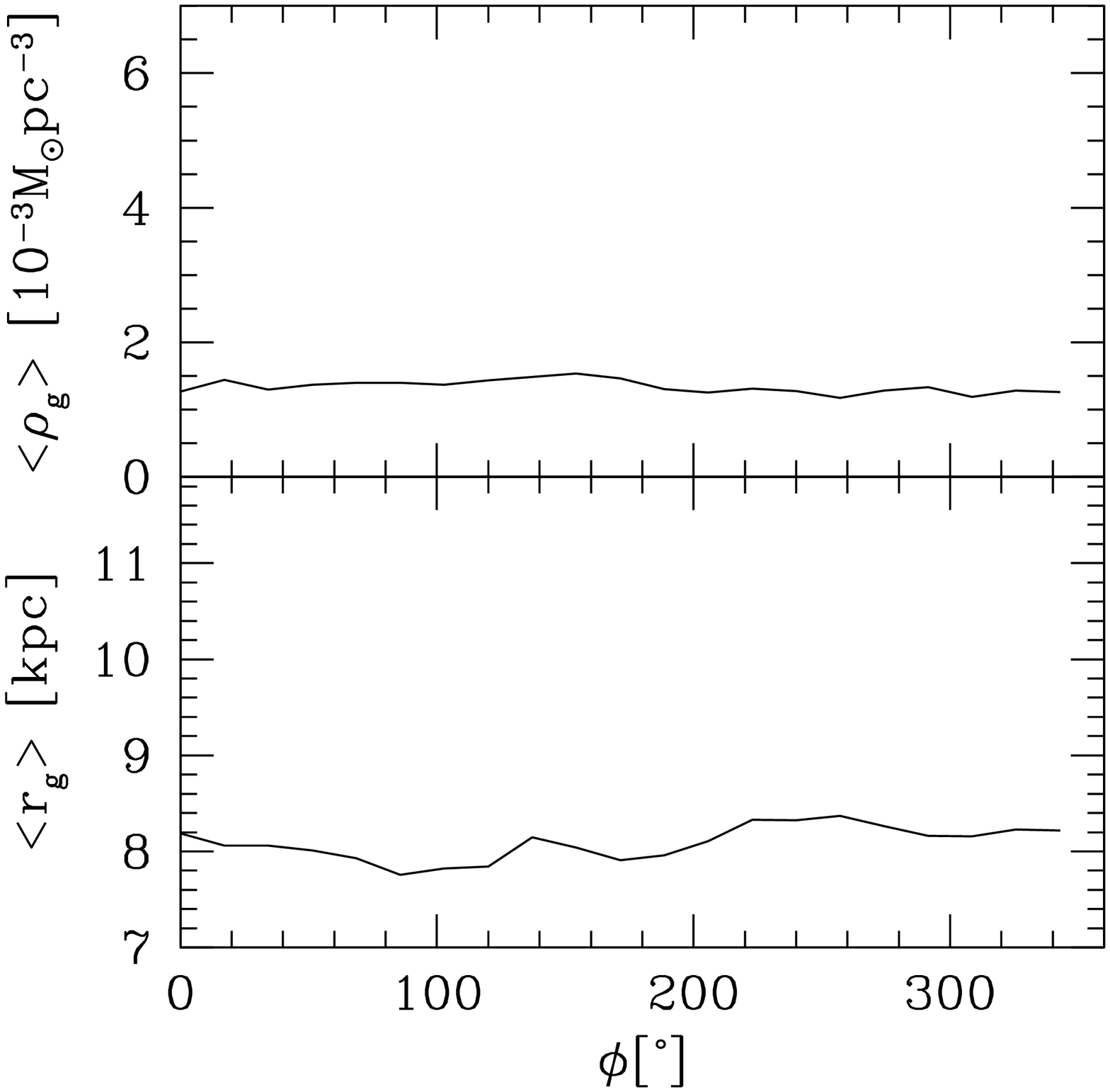}
\hspace {3cm}
\includegraphics[%
  scale=0.32]{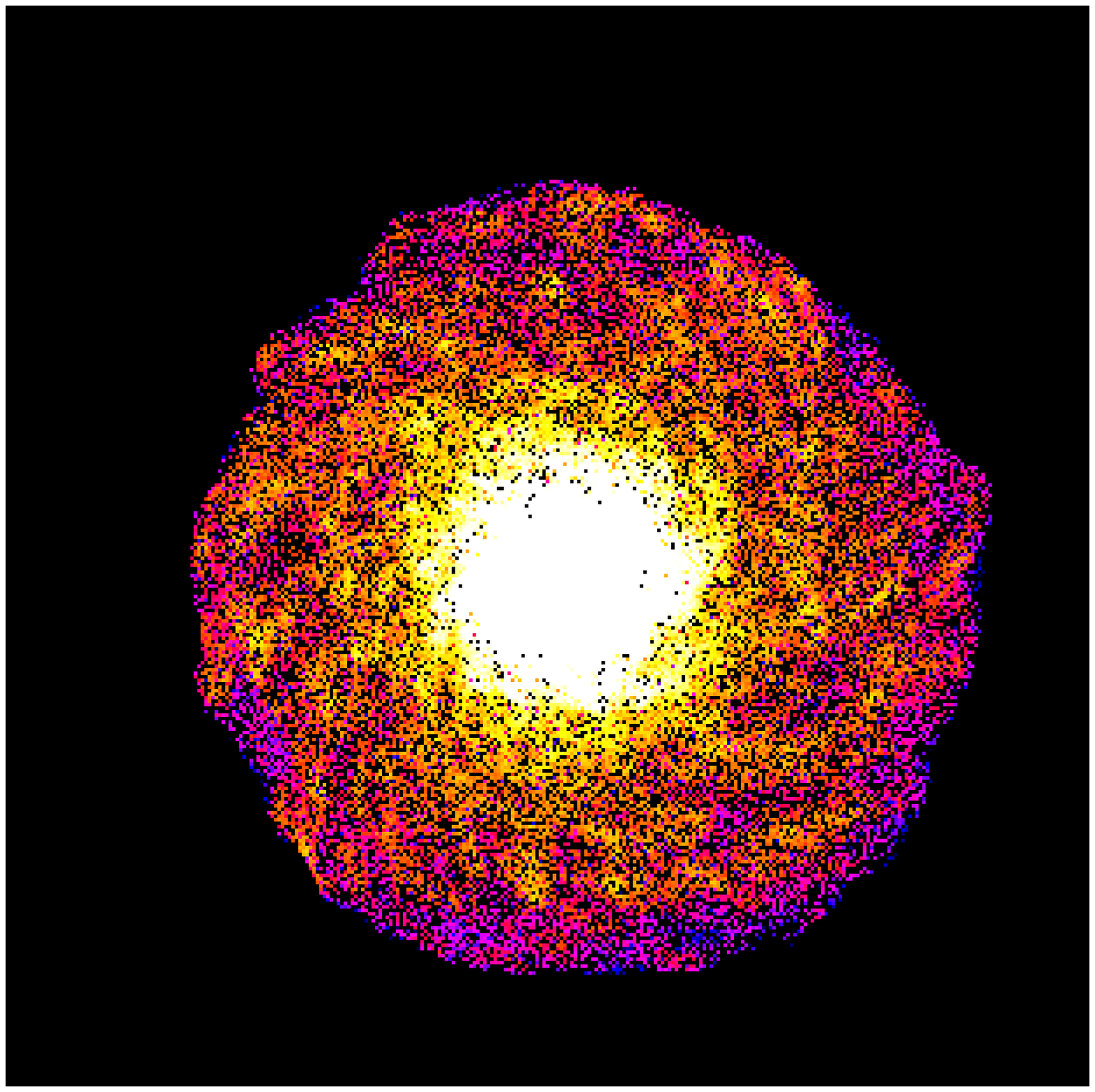}
\includegraphics[%
  scale=0.27]{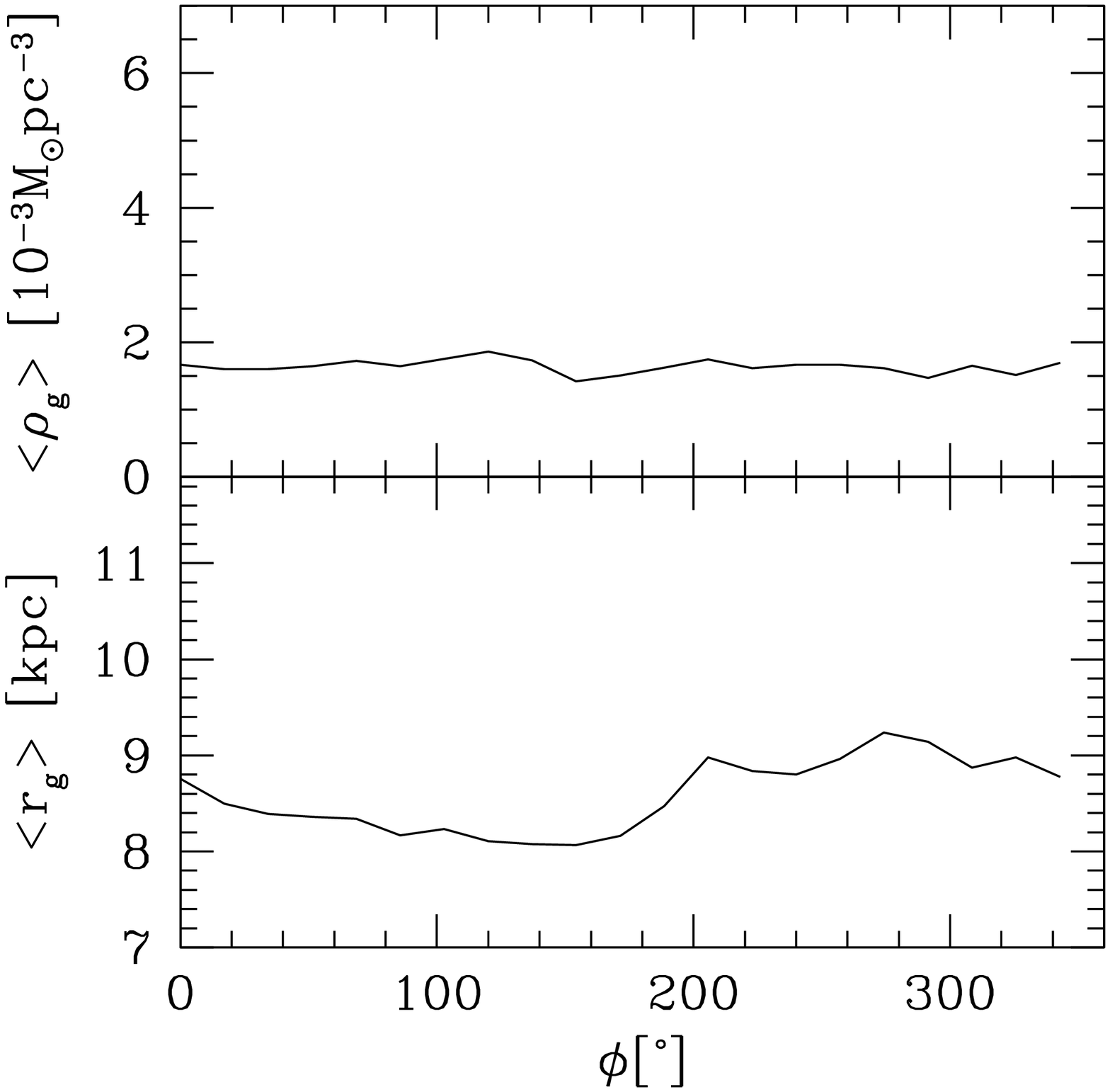}
\hspace {3cm}
\includegraphics[%
  scale=0.32]{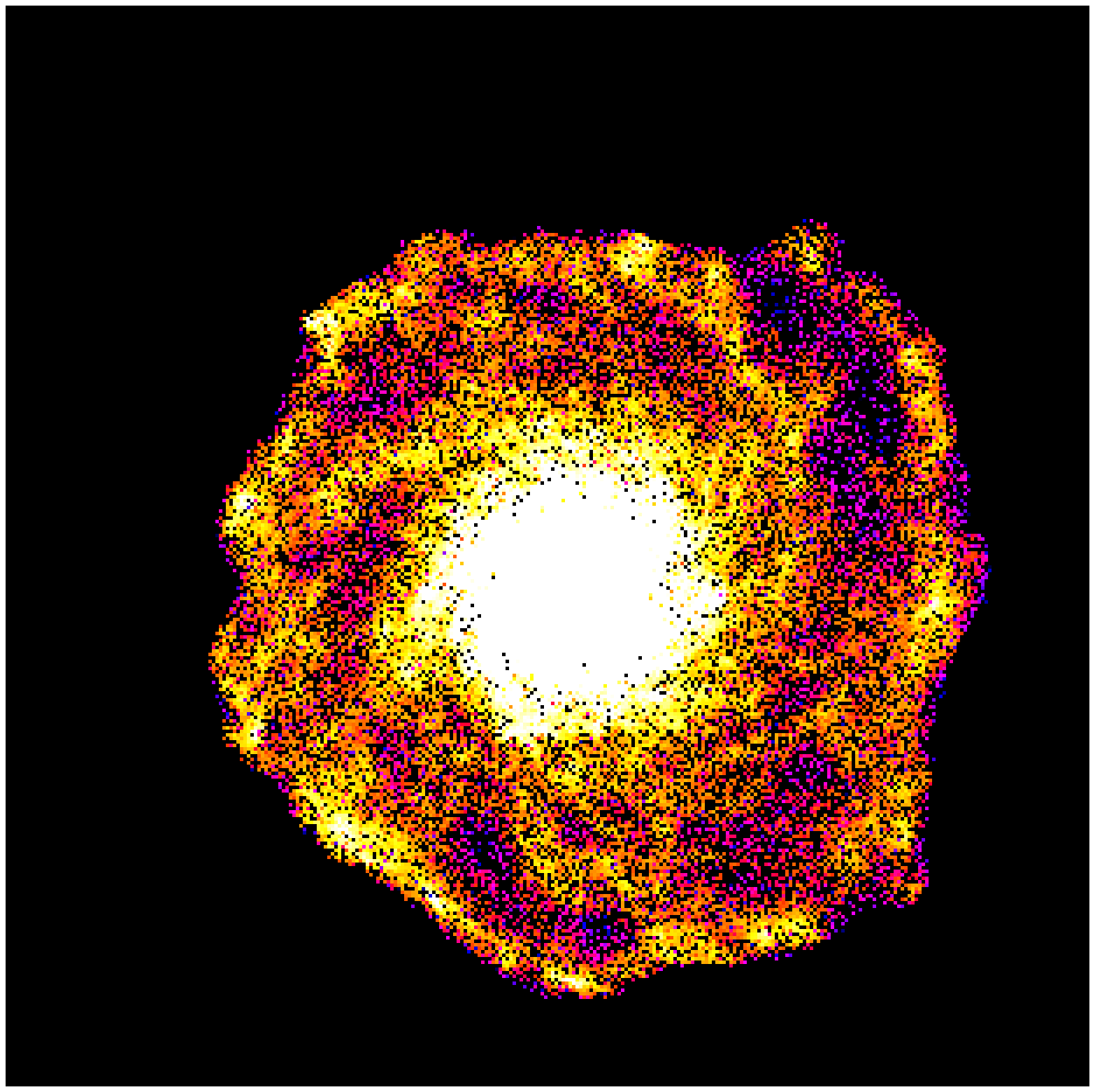}
\includegraphics[%
  scale=0.27]{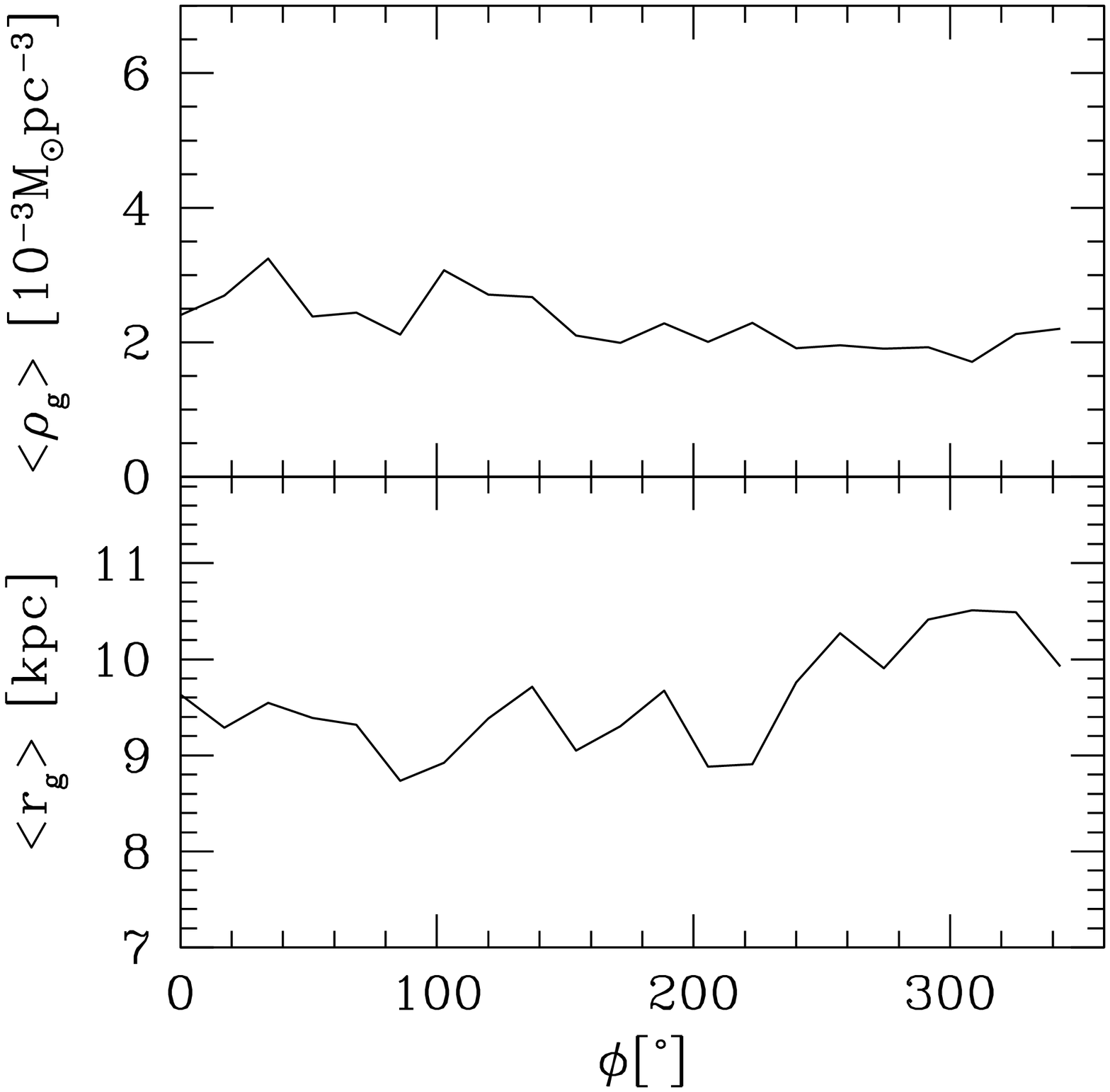}
  \hspace {5cm}
 \includegraphics[%
  scale=0.32]{bar1.eps}
\hspace {5.5cm} 

\caption{The same as in Fig. \ref{densitymaps} for the model with inclination angle $i= 10^{\circ}$ (run cool10).}
\label{densitymaps10}
\end{figure*}

Figs. \ref{densitymaps}--\ref{densitymaps10} illustrate -- for
different values of the inclination angle $i$ --
changes in the disk gas density distribution as the satellite passes
through increasing values of the external pressure, moving towards the perigalacticon. 
Each couple of panels illustrates the state of the disk 
at increasing (from the top to the bottom) times along the orbit. 
Panels on the left represent HI column density maps. 
The density
contrast is chosen in order to highlight the density gradient in the external
disk, since the gas distribution in the central regions of the LMC is
dominated by the presence of the bar and a direct comparison with pure
ram-pressure simulations is not possible. 
The color scale is logarithmic, with white
corresponding to a density larger than $1.2 \times10^{21}$ cm$^{-2}$ and blue
to values lower than $5 \times 10^{19}$ cm$^{-2}$. 
Hot gas
particles flow on to the disk from the left to the right of each plot, 
with increasing ram-pressure values from the first to the
fourth image. 
The disk is seen face on and rotating clockwise.
In the case of the edge-on run (Fig. \ref{densitymaps})
cold gas particles lying in the left-bottom quadrant of the
disk feel the largest ram-pressure, due to the fact that their relative
velocity with respect to the external medium is maximum. 
The rotational velocity of the external disk is 
$\sim 55$ km s$^{-1}$, which implies a relative velocity at the
pericenter of $\sim 350 $ km s$^{-1}$ (450 km s$^{-1}$ in the case of
a high velocity orbit).
Panels on the right represent the change in mean density
and radius of the gaseous external disk as a function of the azimuthal angle
$\phi$. Referring to the geometry of the HI density images on the left,
$\phi=0 ^{\circ}$ corresponds to the bottom of the disk and  increases
clockwise in such a way that the disk moves in the direction of $\phi=90
^{\circ}$. The gas density and the mean radius are both calculated within
sections of a three dimensional annulus with internal and external radius
equal to 7 and 15 kpc, respectively.  
The initial azimuthal profiles (not represented in the plots) are flat since
both these quantities have only radial dependence. As soon as the satellite
starts moving through the surrounding medium, the external gas density develops a
peak centered on $\sim \phi=90^{\circ}$: disk particles localized in regions
of maximum ram-pressure get compressed and move on inner orbits, while their
circular velocity increases consequently. After about a quarter of the orbital
period the gas has reached its minimum radius and maximum local density. 
The gaseous disk becomes strongly asymmetric: compression at the 
front edge produces a density increase along the left border of the disk,
evident in the HI maps even at early times. 
The high density region forms a thin ($\sim1.5$ kpc) 
but continuous and well defined arc which has not an equivalent in the stellar distribution. 
At the perigalacticon this feature extends for almost
$160^{\circ}$ with a density more than one order of magnitude higher than gas located at smaller radii. 
Its average thickness ($\sim $ 1.5 kpc) and velocity dispersion along the line of sight are larger than the average values in the rest of the disk.

In the case of a satellite moving through the hot medium with an 
inclination angle different from $90^{\circ}$ the external pressure 
directed perpendicularly to the plane of the disk increases as cos$i$
\citep{Roediger&Brueggen06} while compression at the leading edge is 
much less pronounced. 
Figs. \ref{densitymaps45} and \ref{densitymaps10} refer to runs with
inclination angles $i=45^{\circ}$ and $i=10^{\circ}$. Disks are
shown face-on.
The increment in density along the leading edge is smaller (cool45) than in the  
edge-on model and almost absent for $i=10^{\circ}$ (cool10), while compression perpendicular to the plane of the disk produces local 
gravitational instabilities in the external gaseous disk (also
Mayer, Mastropietro \& Tran in preparation). 
This effect is more evident in the nearly face-on run cool10 where high density
filaments delimitate regions where the local density is almost one order of magnitude lower.
Despite the absence of a peak in the azimuthal  mean density profile, 
the integrated final density of cool10 is comparable to the other runs.

\begin{figure}
\epsfxsize=8truecm \epsfbox{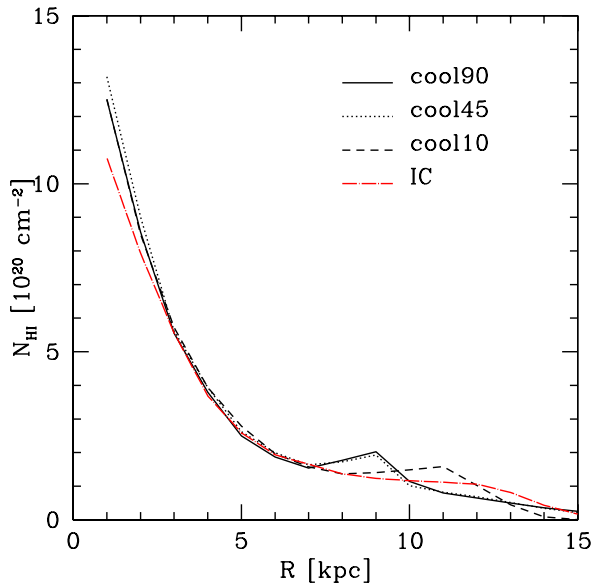} 
\caption{HI column density profile of the final LMC disk
  in units of $10^{20}\textrm{cm}^{-2}$. The curves refer to
  different values of the inclination angle $i$.
  The red long-dashed line represents the column density
  profile of the initial disk.}
\label{dennosf}
\end{figure}

Fig. \ref{dennosf} represents the azimuthally averaged HI column
density profile of the final disk configuration for the three simulations. As a result
of the increase in density along the edge of the disk, the mean column density
shows a secondary peak at large radius. 
A limb-brightened density profile has actually been observed by
\citet{StaveleySmithetal03} using the Parkes multibeam HI survey of the LMC. 
The outer profiles of cool90 and cool45  are very similar, while in run
cool10 the increment in density with respect to the original profile of the disk
(long-dashed red curve in the plot) is located at larger radii ($\sim 8.5$ kpc). 
Indeed, due to compression of the leading
edge the final gas distribution of cool90 and
cool45 is asymmetric, the dense front edge being much closer to the center
than the opposite border of the disk. Therefore the HI peak in the external
disk is located at relatively small radii,
while the azimuthally averaged gas distribution is more extended than in the
case of the run cool10 since ram pressure elongates the back side of
the disk.
Again we stress the fact that our gaseous disk is more extended than it would be in a fully self consistent simulation including both gravitational and hydrodynamical forces.
As seen in M05 and \citet{Mastropietro08} the combined effect of ram-pressure and tidal stripping is quite efficient in stripping gas from the outer satellite's disk, creating the tip of the Magellanic Stream already at large distances from the MW.
The high density feature in cool90 would then form along the border of the disk at smaller radii and would not be easily subjected to further stripping due to the relatively ram-pressure values.

\section{Simulations with star formation}

\begin{table*}
\caption{Star formation simulations. The second column indicates the use of
  the converging flow criterium in the star formation recipe, the third and
  the fourth columns represent the star formation efficiency $\epsilon$ and
  the star formation rate parameter $c_{\star}$ (Eq. \ref{starformeq}),
  respectively, while $i$ is the inclination angle of the disk, i.e. the
  angle between the angular momentum of the disk and the flux of hot
  particles. A model with $i=90$ is indeed a disk moving edge-on in the
  external medium. 
  The sixth and seventh columns 
  represent the minimum and maximum ram pressure suffered by the satellite during 
  its approach to perigalacticon. The last column indicates the presence of a
  temperature floor for the disk gas.  }
\begin{tabular}{l|c|c|c|c|c|c|c|}
\hline
Run& conv & $\epsilon$ & $c_{\star}$ & $i$ & [$^{\circ}]$ $P_{min}$ [$10^{-13}$
  dyn cm$^{-2}$] & $P_{max}$ [$10^{-13}$ dyn cm$^{-2}]$ & $T_{min}$[kelvin] \\
\hline
SF90 & no & 0.33 & 0.02 & 90 & $4.83\times 10^{-2}$ & 1.50& 0 \\ 
SF45 &no& 0.33 & 0.02 & 45 & $4.83\times 10^{-2}$ & 1.50& 0\\
SF10 & no& 0.33 & 0.02 & 10 & $4.83\times 10^{-2}$ & 1.50& 0\\
SFconv &yes&  0.33 & 0.02 & 90 & $4.83\times 10^{-2}$ & 1.50& 0\\
SF$\epsilon$1 & no& 1 &  0.02 & 90& $4.83\times 10^{-2}$ & 1.50& 0 \\
SFconv$\epsilon$1 & yes & 1 &  0.02 & 90& $4.83\times 10^{-2}$ & 1.50& 0 \\
SF90c0.01 & no & 0.33 & 0.01 &  90& $4.83\times 10^{-2}$ & 1.50& 0 \\
SF10c0.01 & no & 0.33 & 0.01 &  90& $4.83\times 10^{-2}$ & 1.50& 0 \\
SF90c0.05 & no & 0.33 & 0.05 & 90 & $4.83\times 10^{-2}$ & 1.50& 0\\
SF10c0.05 & no & 0.33 & 0.05 & 90 & $4.83\times 10^{-2}$ & 1.50& 0\\ 
SF90v400 & no& 0.33& 0.02& 90& $4.83\times 10^{-2}$ & 2.67 & 0  \\ 
SF10v400 & no& 0.33 & 0.02& 10& $4.83\times 10^{-2}$ & 2.67 & 0  \\
SF10v400t12000 & no& 0.33& 0.02& 10& $4.83\times 10^{-2}$ & 2.67 & $1.2 \times 10^{4}$  \\
SF10v400t15000 & no& 0.33& 0.02& 10& $4.83\times 10^{-2}$ & 2.67 & $1.5 \times 10^{4}$  \\
SFld90 & no& 0.33& 0.02& 90& $4.83\times 10^{-3}$ & $1.5\times 10^{-1}$ & 0  \\
SFld10& no& 0.33& 0.02& 10& $4.83\times 10^{-3}$ & $1.5\times 10^{-1}$ & 0 \\

\hline
\end{tabular}
\label{sfruns}
\end{table*}

The compressive increase in HI density is naturally associated with excess star formation.
SPH simulations cannot follow the formation of molecular clouds  but in first approximation the molecular 
gas fraction can be related to the density of atomic gas \citep{Vollmeretal08}. 

The main parameters of star formation simulations are summarized
in Table \ref{sfruns}. 

As we already mentioned in Section 3, our standard star formation model 
(SF) does not include the converging flow criterium and is characterized by an
efficiency $\epsilon= 0.33$ \citep{Katz92}. The star formation rate parameter 
$c_{\star}$ is initially set equal to $0.02$.  
This model was adopted to run wind tube simulations with inclination angles 
$i=90,45,10^{\circ}$ (SF90, SF45, SF10) (Table \ref{sfruns}).
We also investigated different star formation recipes requiring converging 
flows and assuming different values of $c_{\star}$ and $\epsilon$. 
We explored star formation rate parameter values in the range  from 0.01 to 
0.05, that produce a star formation rate integrated over the entire disk 
comparable with the $0.1 M_{\odot}$ yr$^{-1}$ provided by \citet{Sandage86}.
An efficiency $\epsilon=1$ implies that whenever a gas particle
satisfies the star formation requirements, it is immediately turned into a
single star particle of the same mass \citep{Kaufmannetal06}.  

In the last six simulations listed in Table \ref{sfruns} we used our standard
star  formation model SF to investigate the effects of different orbital
parameters and gas halo densities. 

Runs SFv400 are characterized by a maximum ram pressure value 
corresponding to a perigalactic velocity of 400 km s$^{-1}$
\citep{Kallivayaliletal06}. 
Such high velocity disks -- when moving face-on through the external hot
medium -- are strongly affected by local instabilities and star
formation is consequently enhanced.
The introduction of an artificial lower limit for the satellite
gas temperature (in runs SF10v400t12000 and SF10v400t15000), higher than
the cut-off in the cooling function, has the effect of reducing   gravitational 
instabilities and fragmentation in the disk.
This temperature threshold can be justified in order to crudely model
the effect of the UV background and stellar feedback \citep{Barnes02}.

Finally, with simulations SF90ld and SF10ld in Table \ref{sfruns} we also
consider the possibility of a Galactic hot halo ten times less dense
than our standard model.

\begin{figure*}
\includegraphics[%
  scale=0.18]{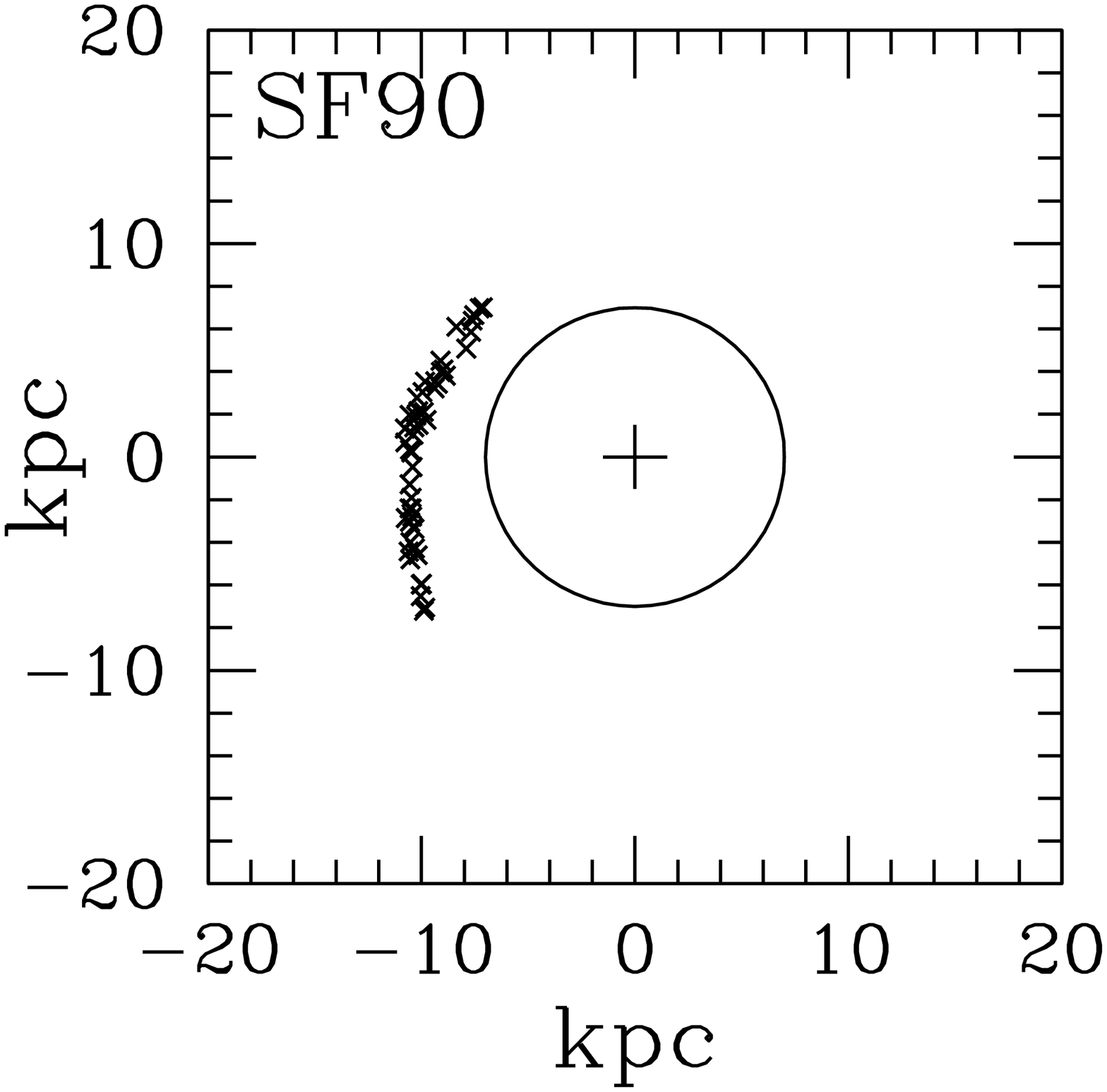}
\includegraphics[%
  scale=0.18]{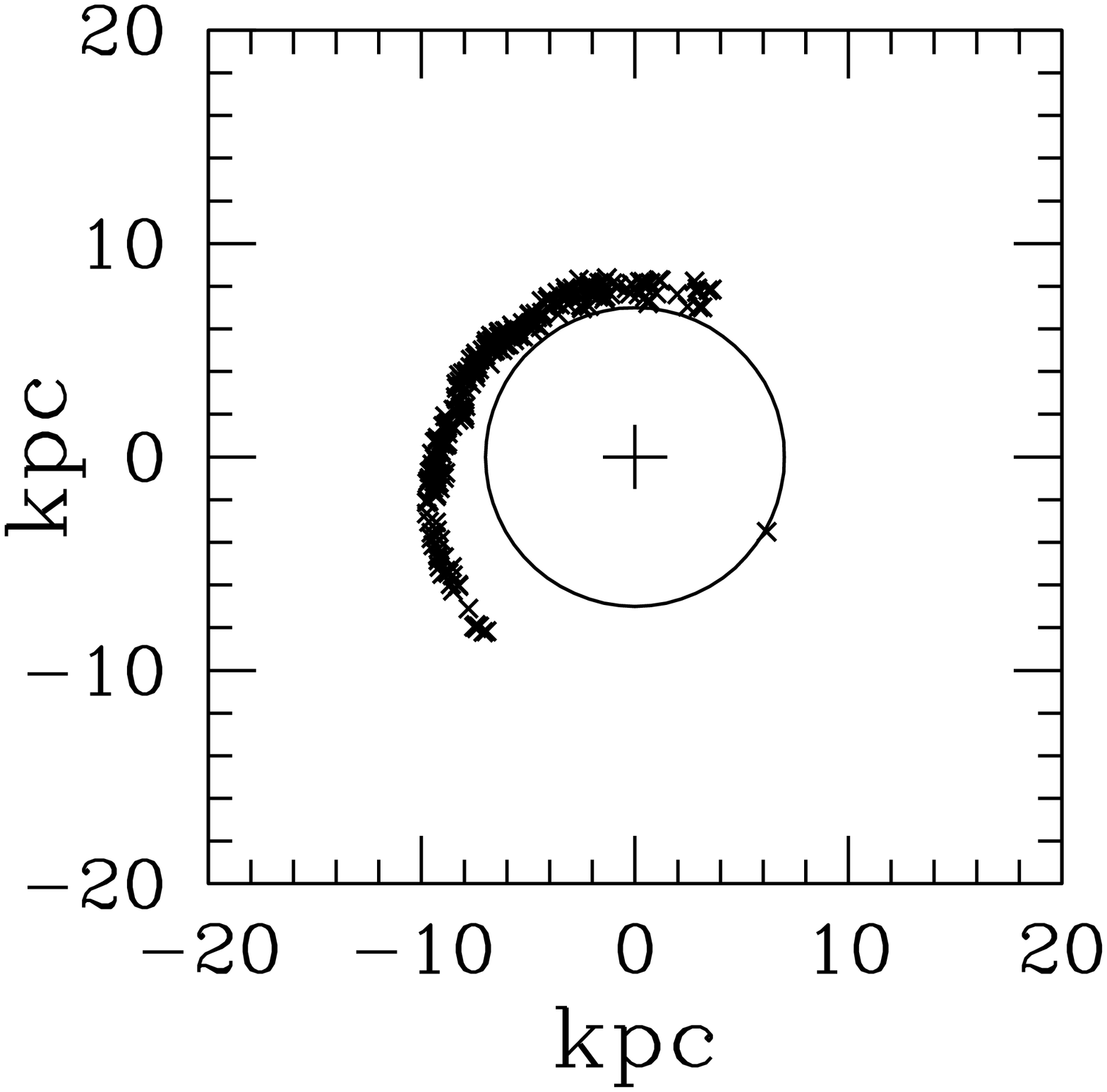}
\includegraphics[%
  scale=0.18]{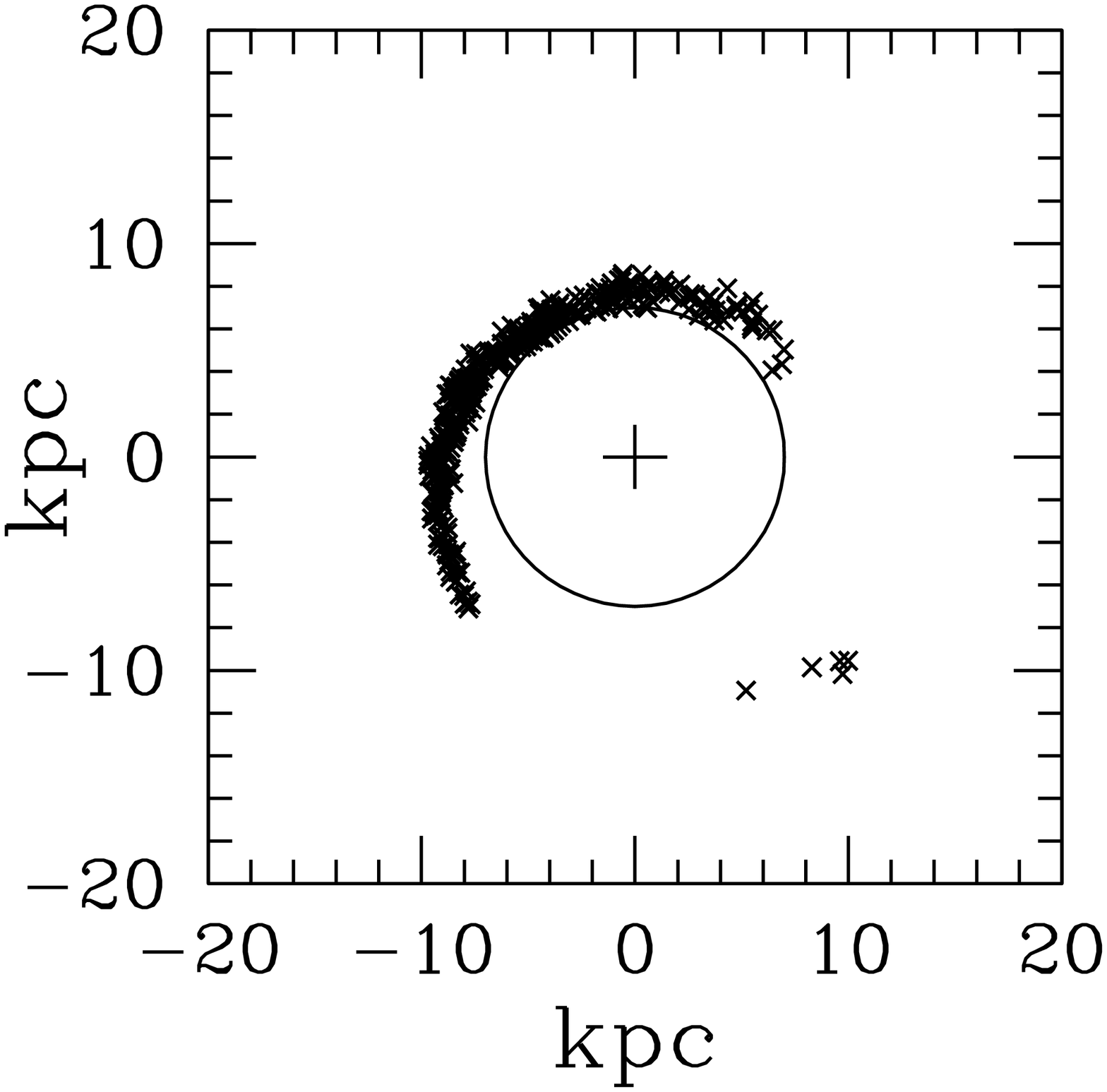}
\includegraphics[%
  scale=0.18]{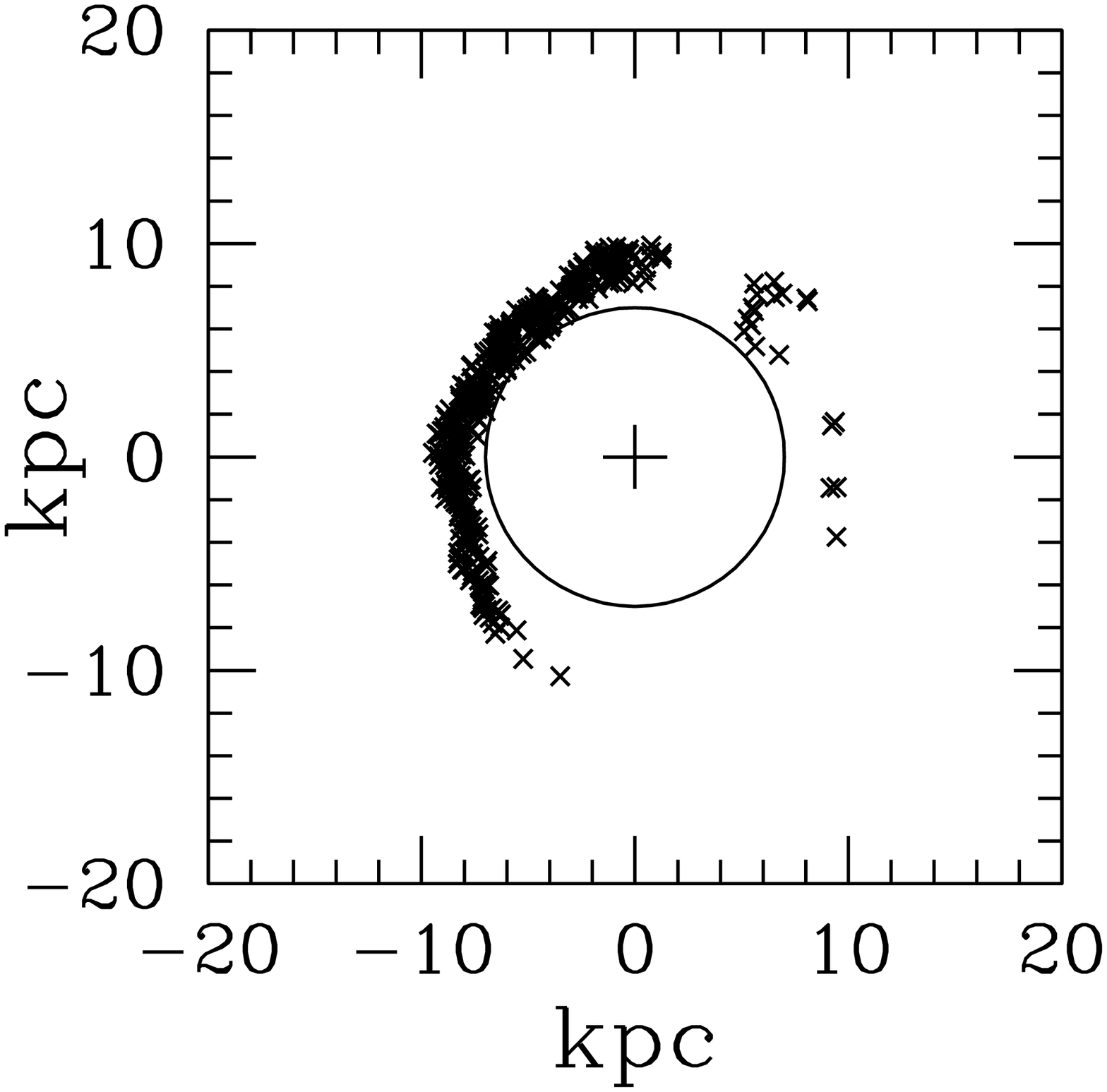}
\includegraphics[%
  scale=0.18]{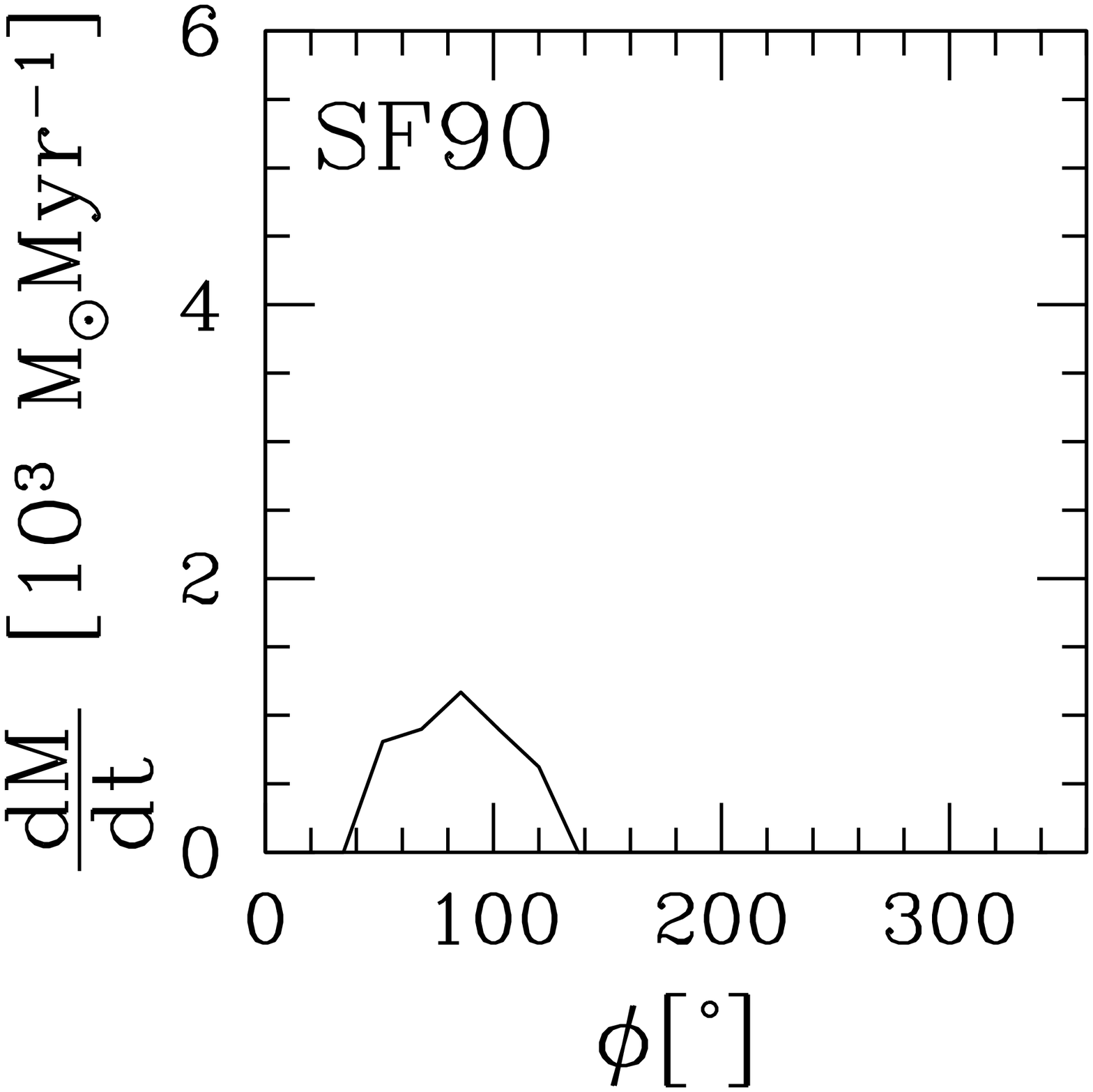}
\includegraphics[%
  scale=0.18]{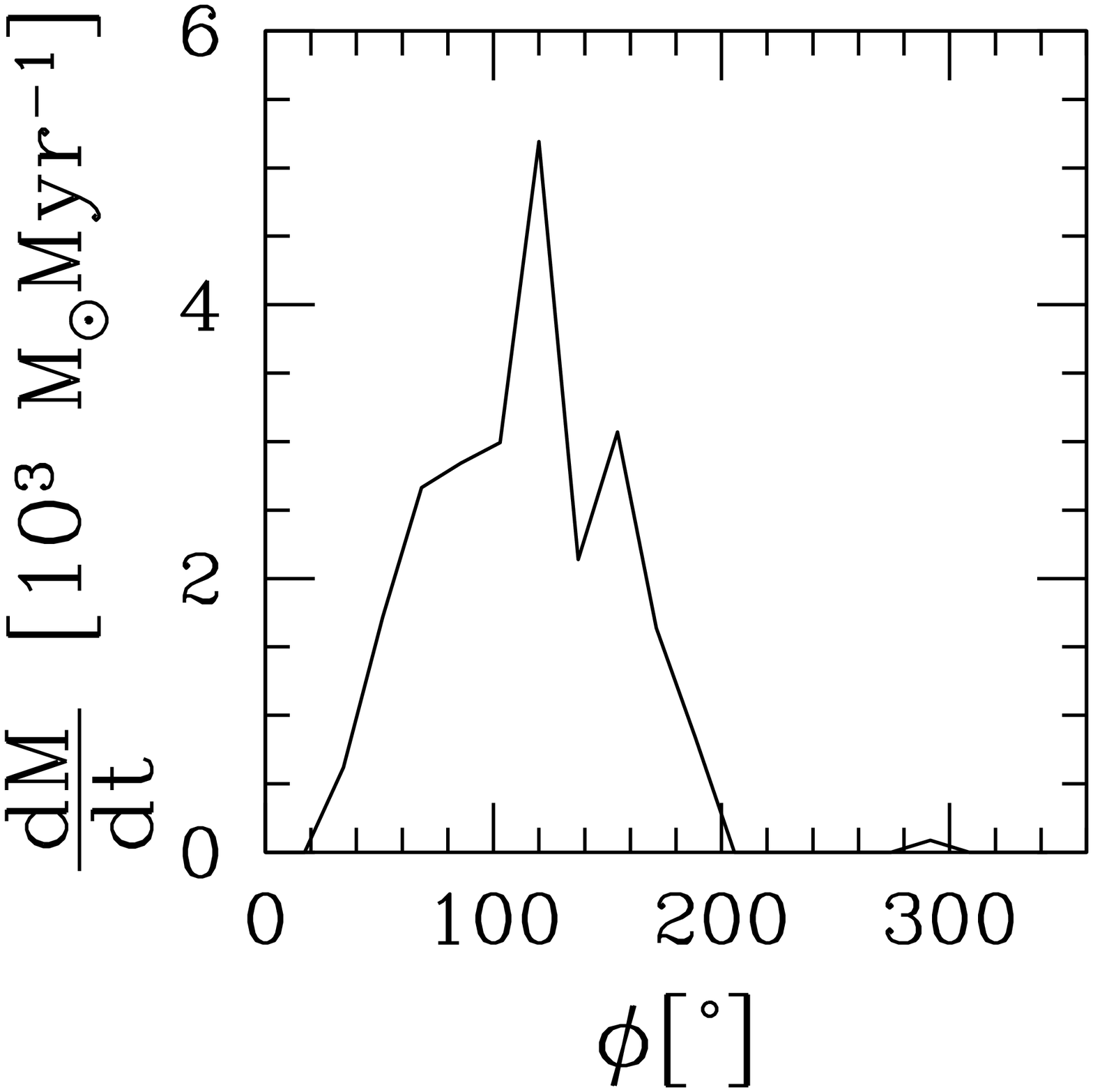}
\includegraphics[%
  scale=0.18]{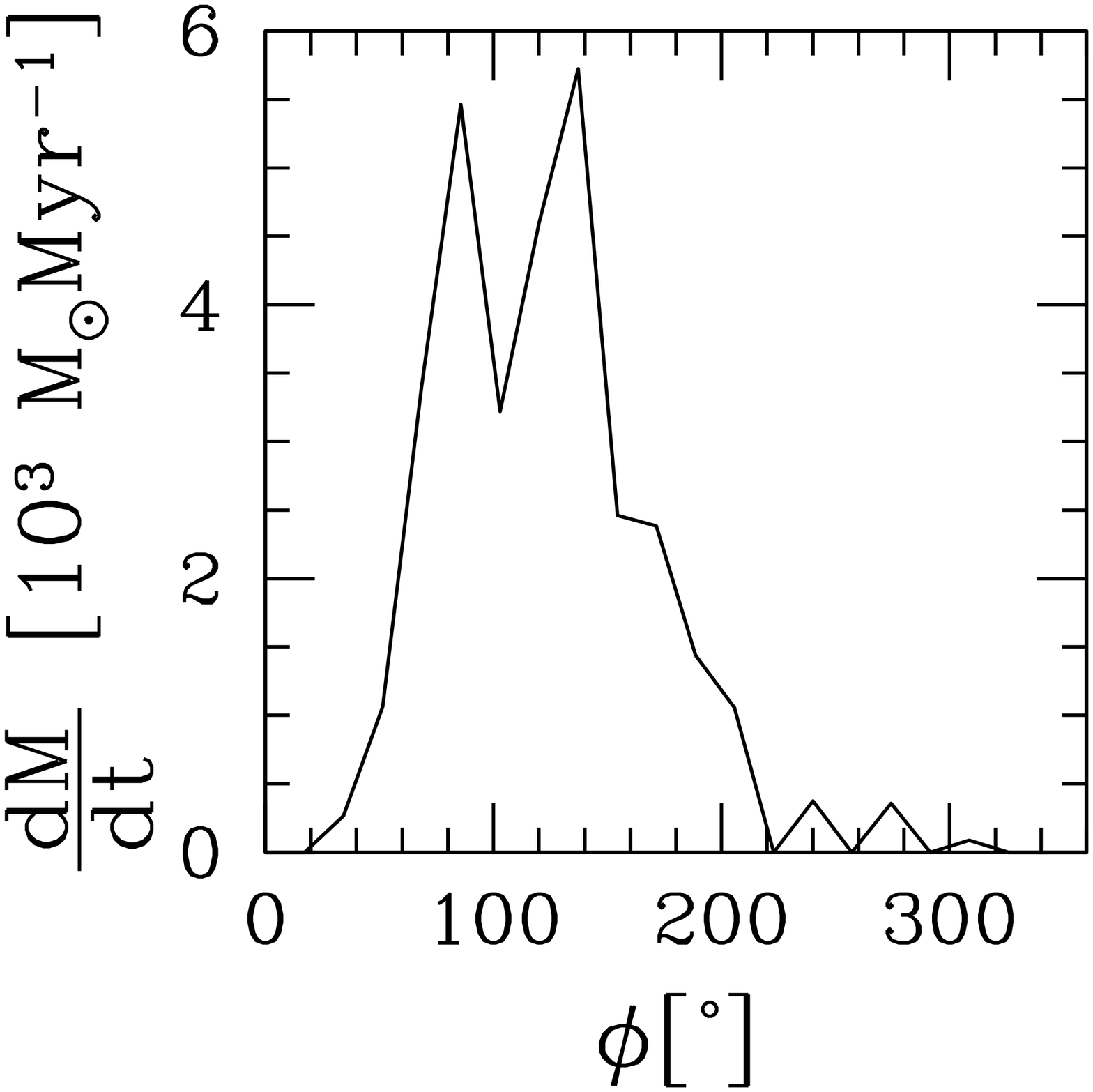}
\includegraphics[%
  scale=0.18]{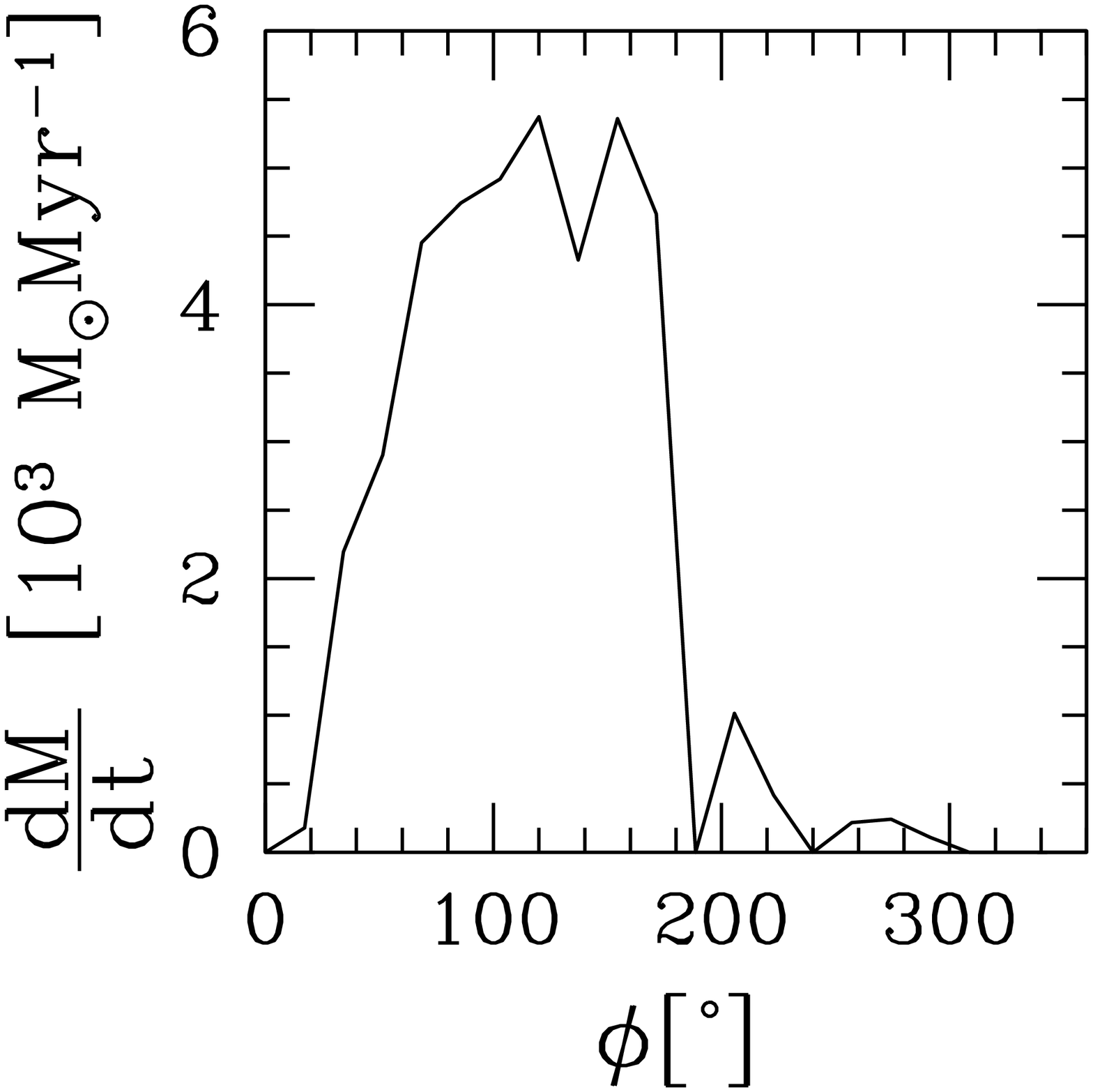}
\includegraphics[%
  scale=0.18]{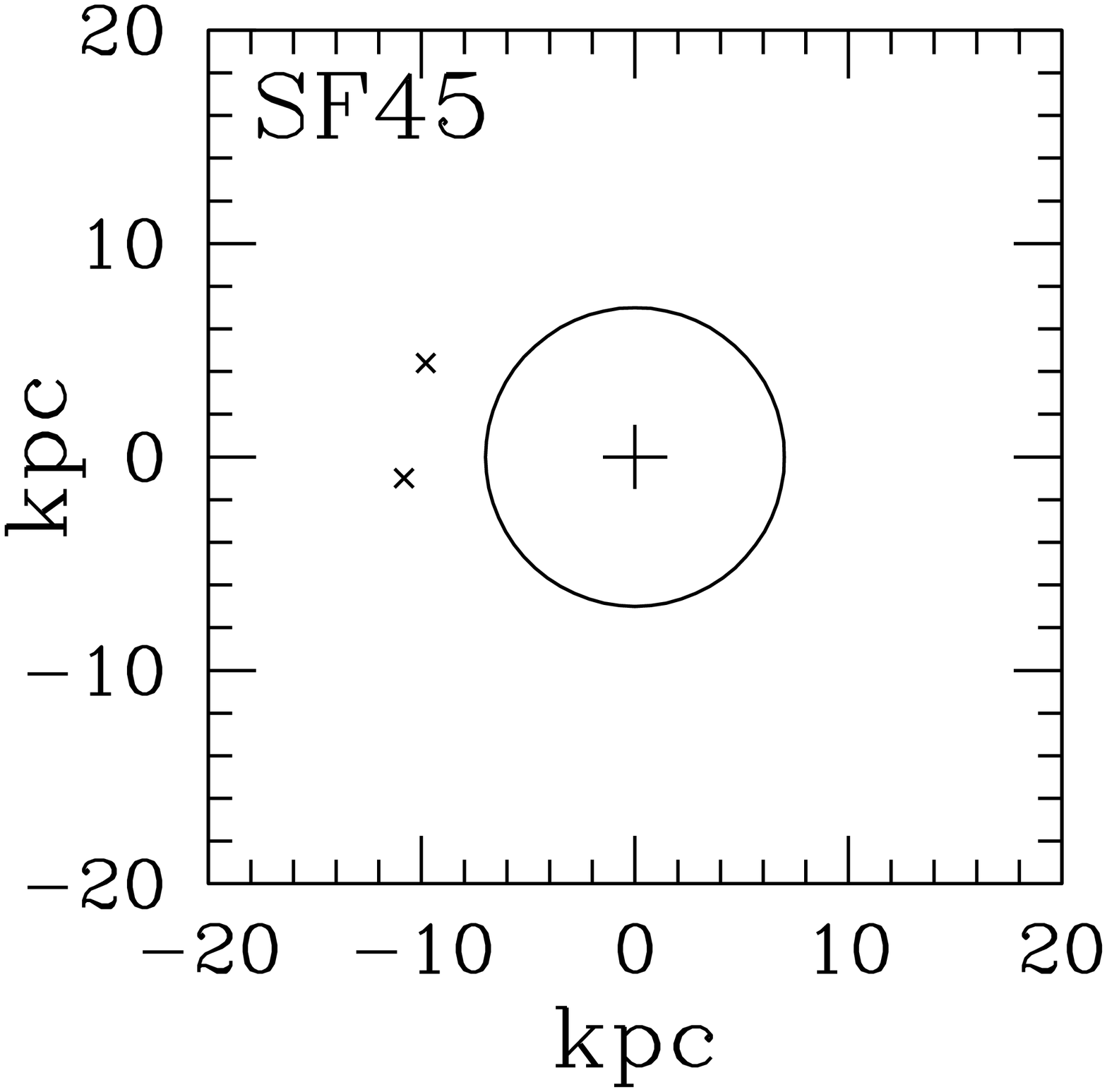}
\includegraphics[%
  scale=0.18]{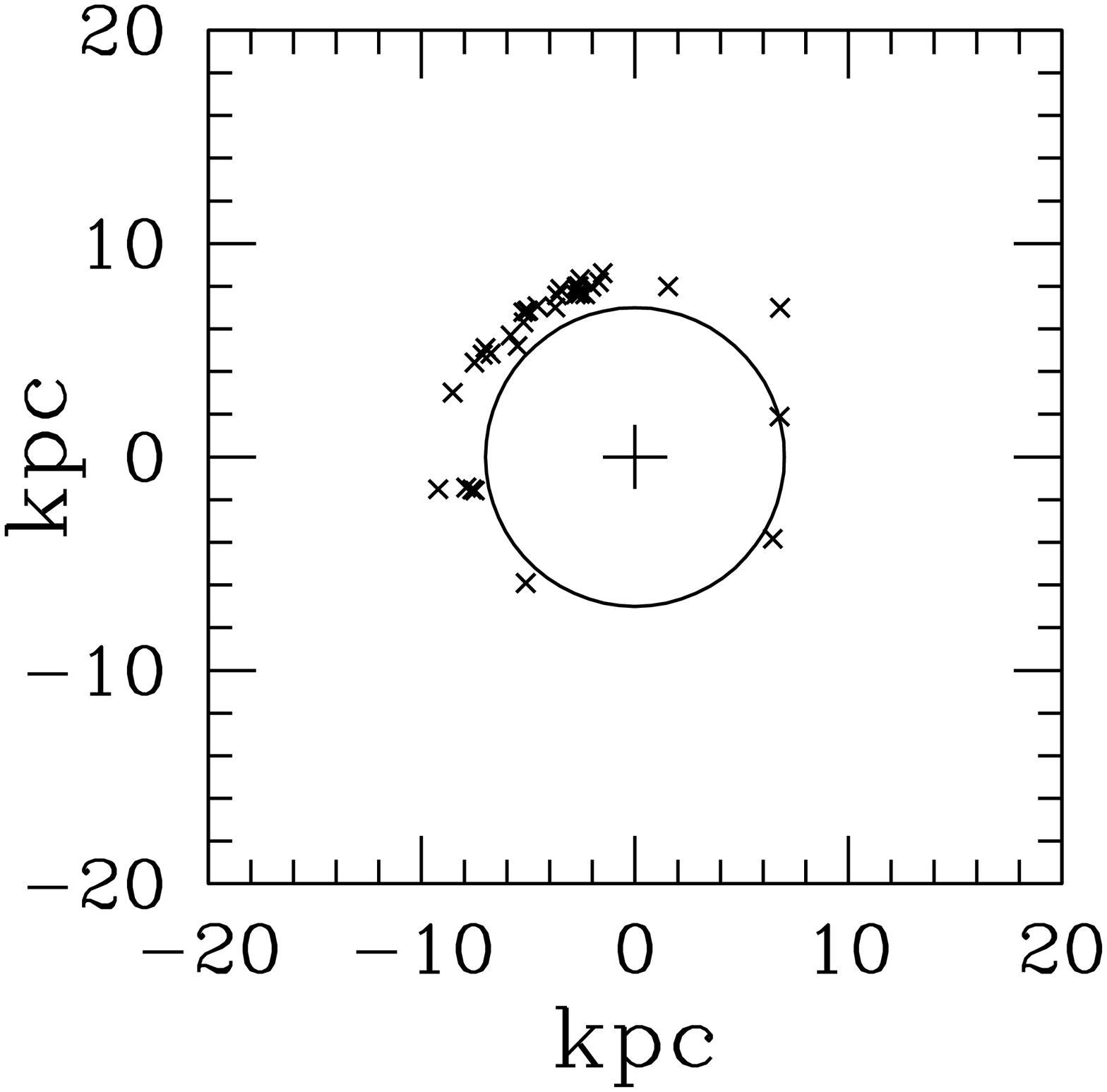}
\includegraphics[%
  scale=0.18]{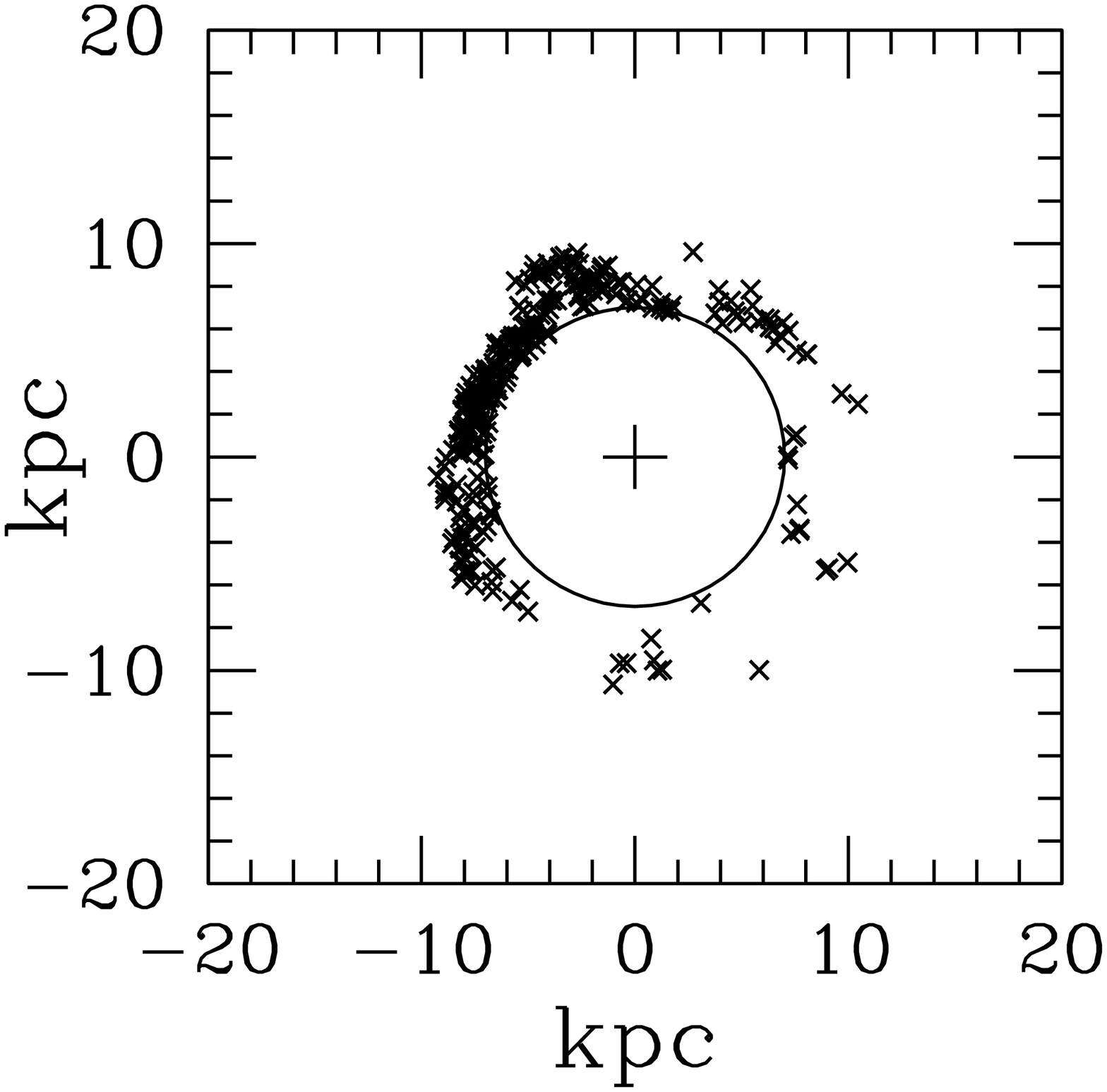}
\includegraphics[%
  scale=0.18]{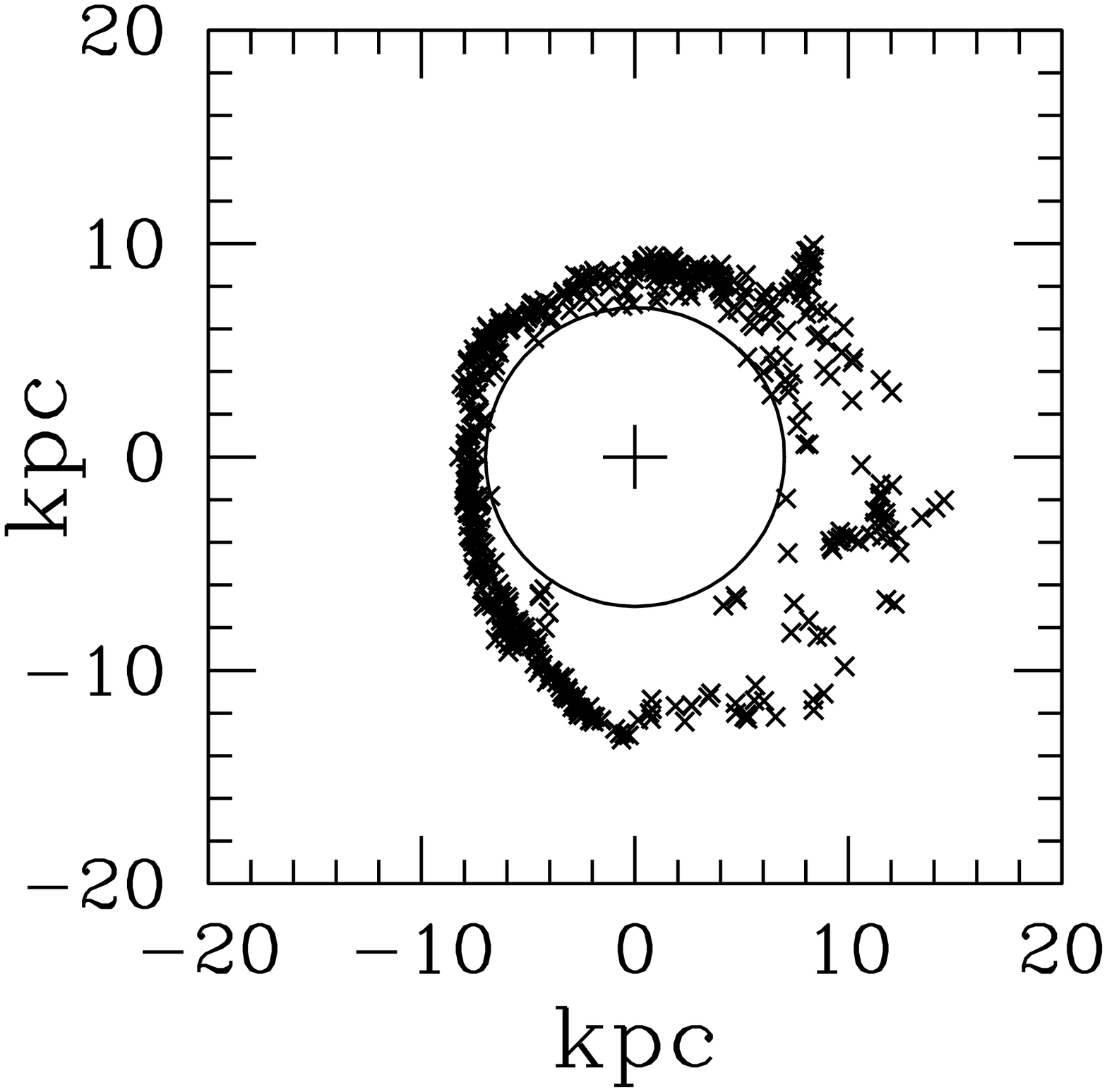}
\includegraphics[%
  scale=0.18]{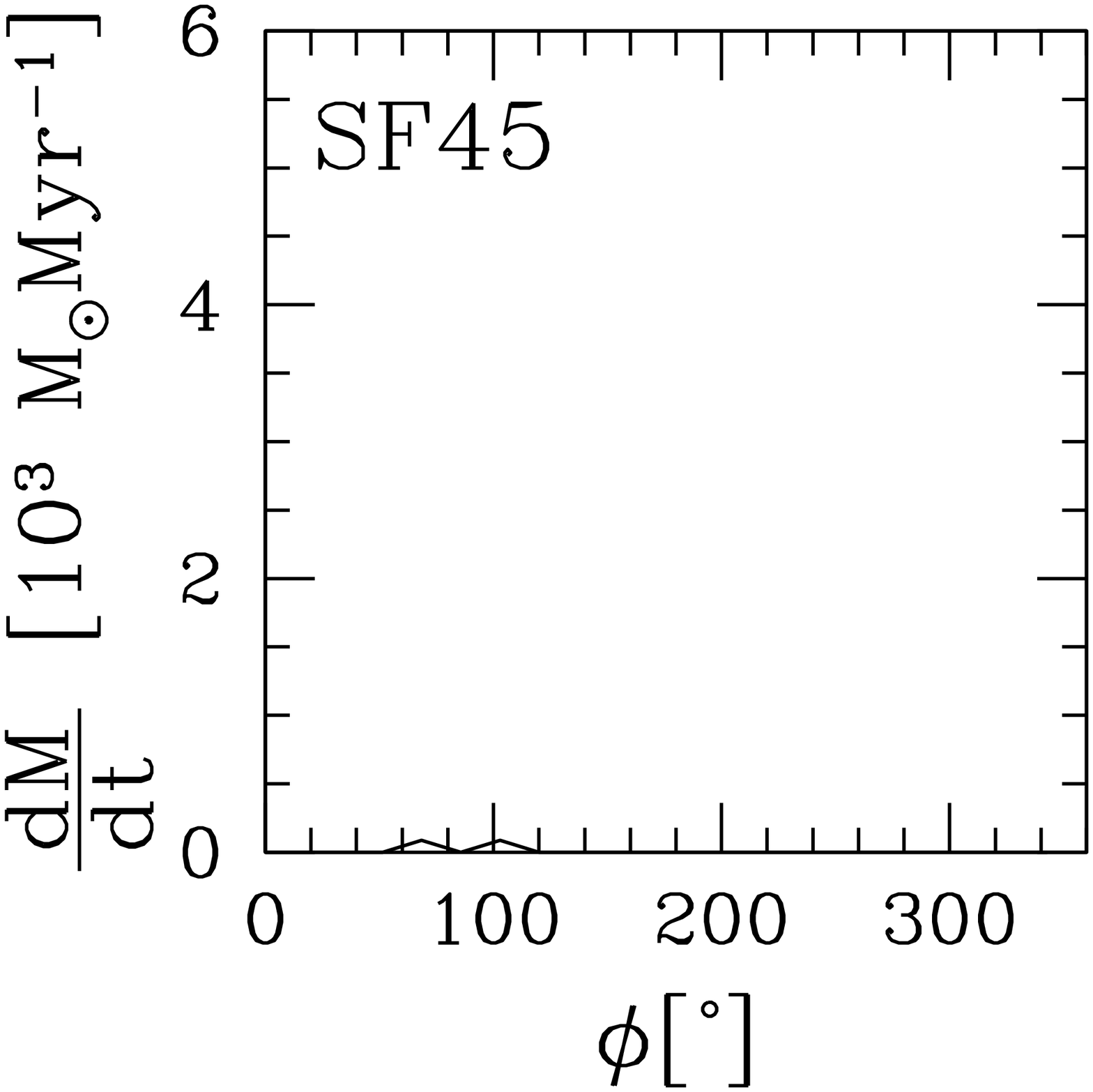}
\includegraphics[%
  scale=0.18]{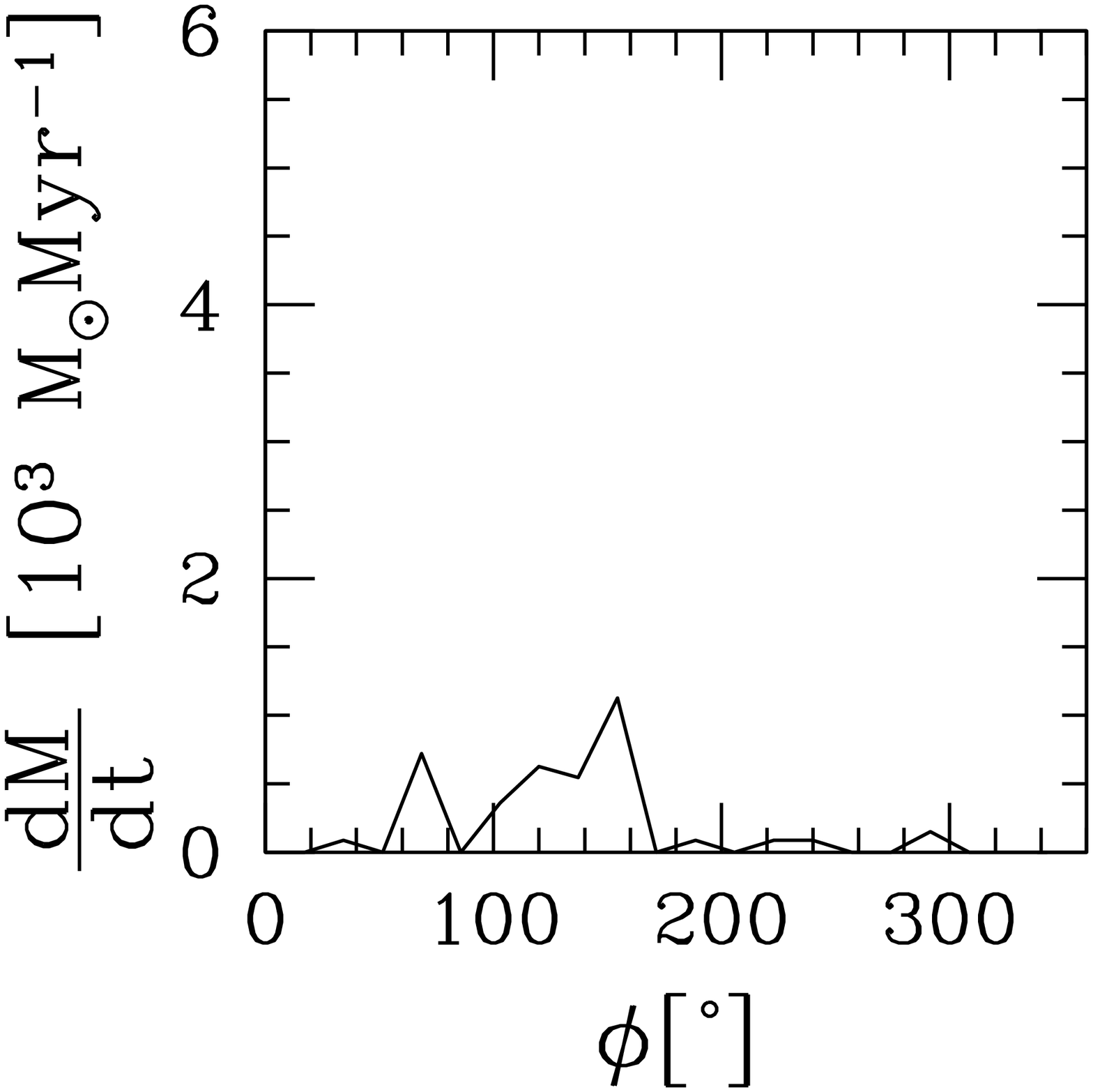}
\includegraphics[%
  scale=0.18]{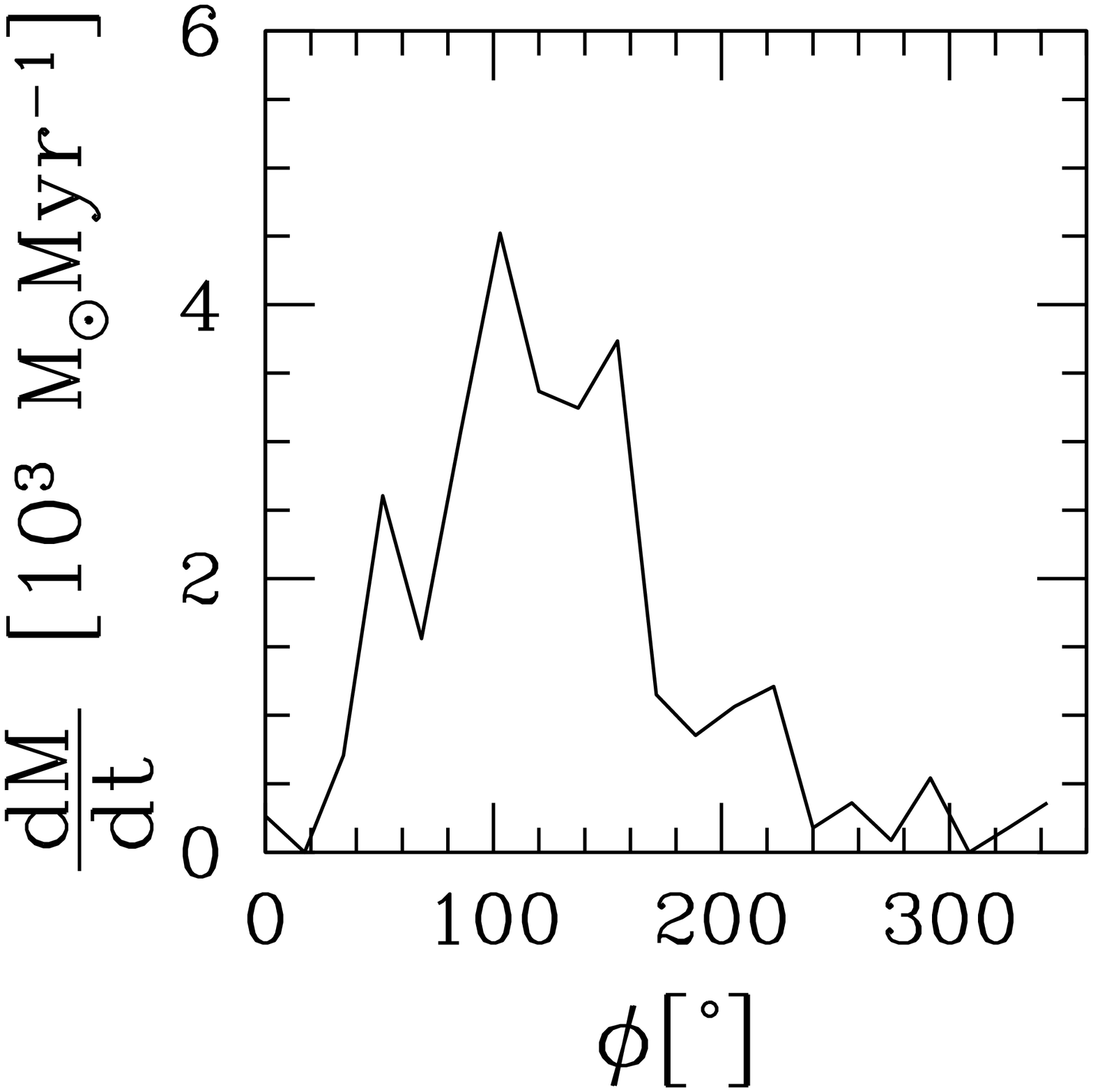}
\includegraphics[%
  scale=0.18]{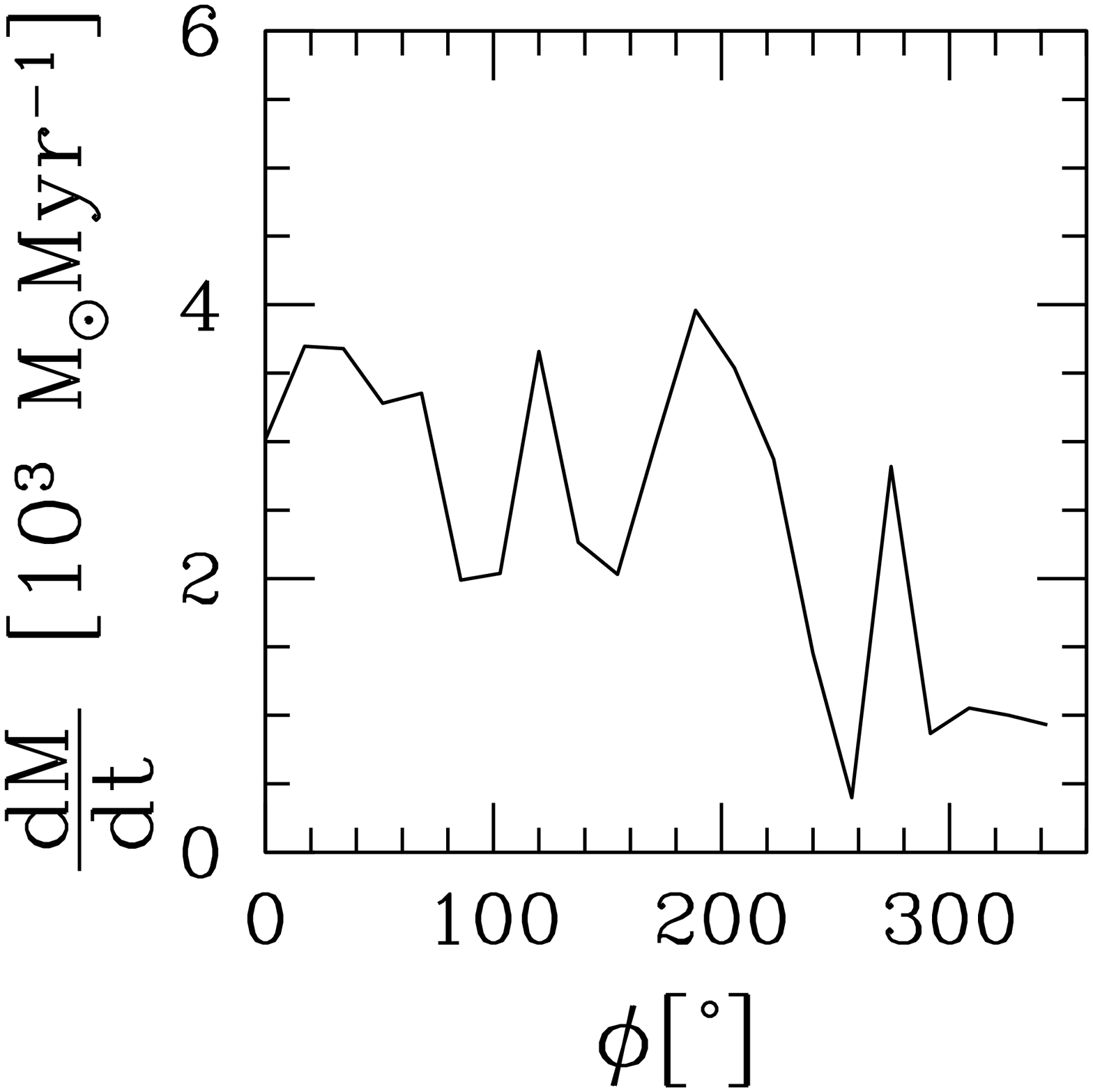}
\includegraphics[%
  scale=0.18]{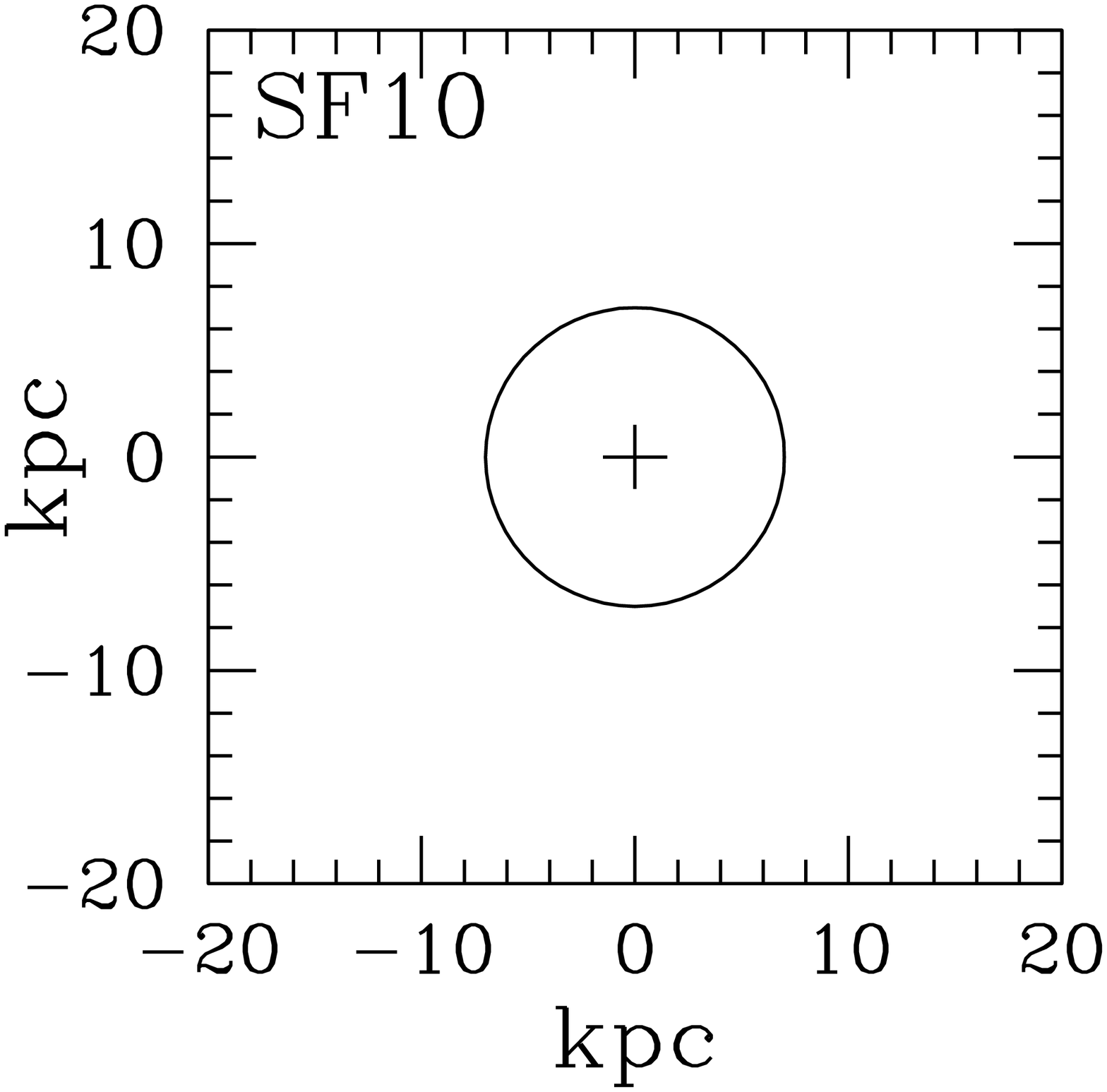}
\includegraphics[%
  scale=0.18]{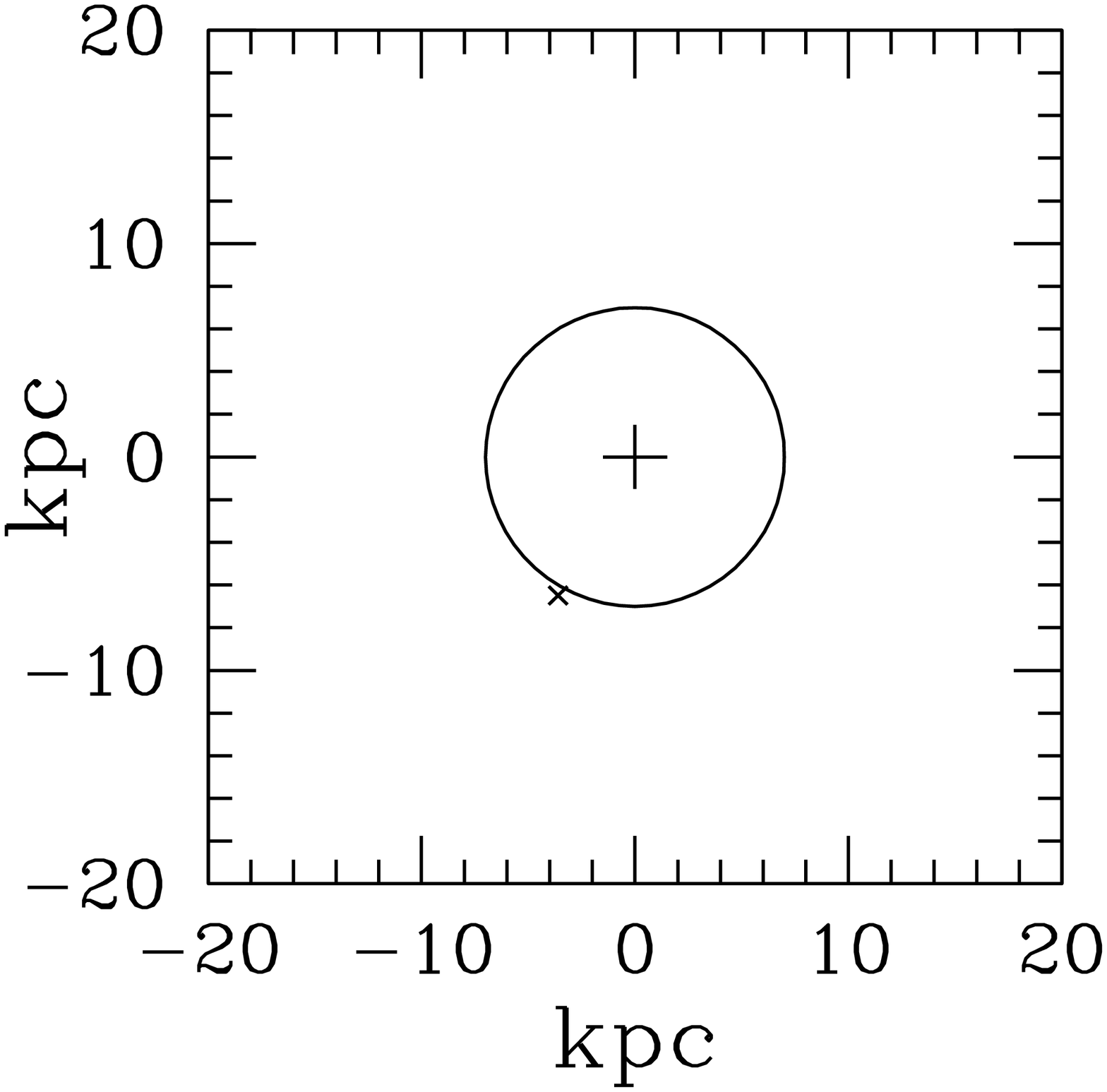}
\includegraphics[%
  scale=0.18]{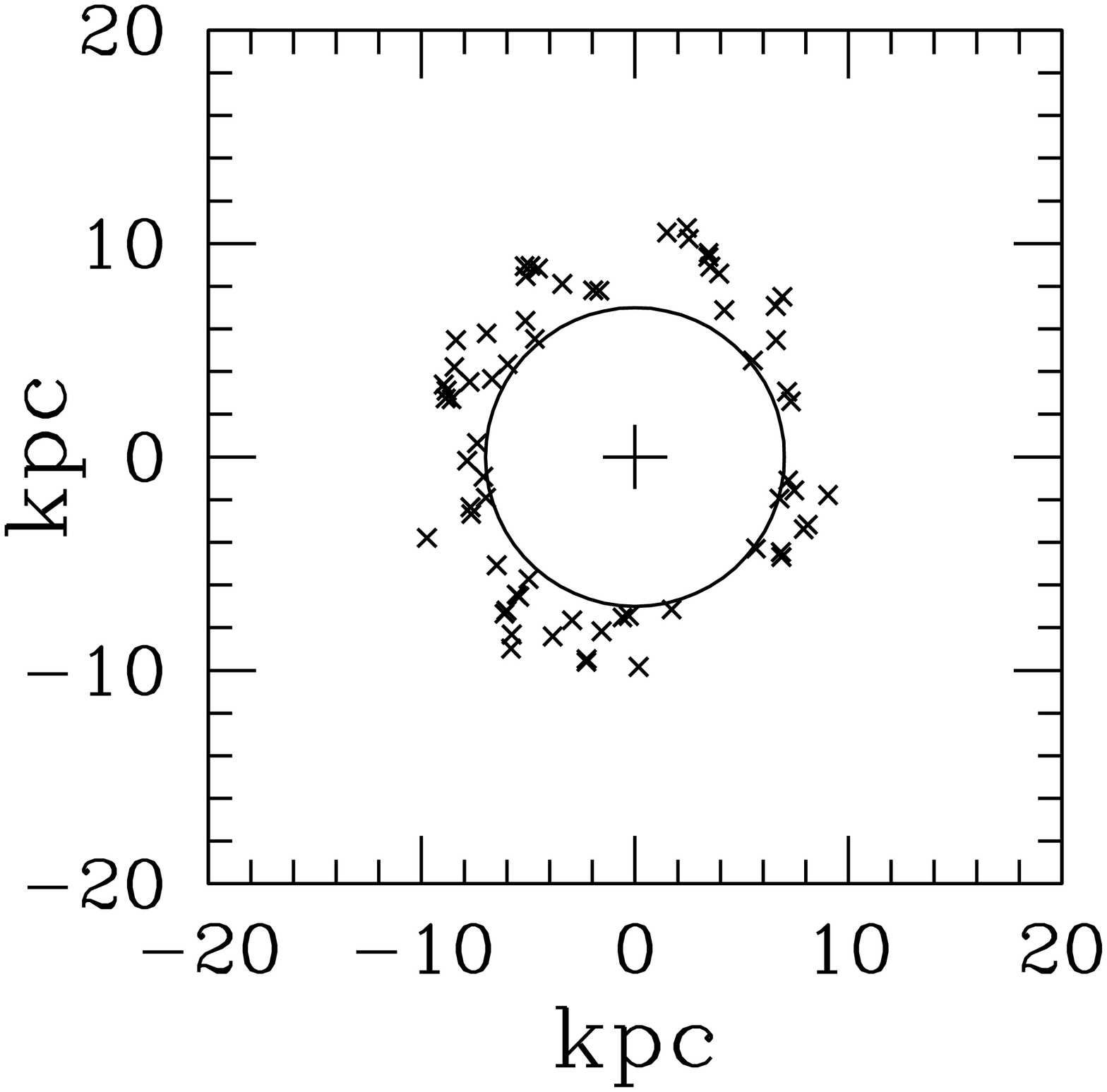}
\includegraphics[%
  scale=0.18]{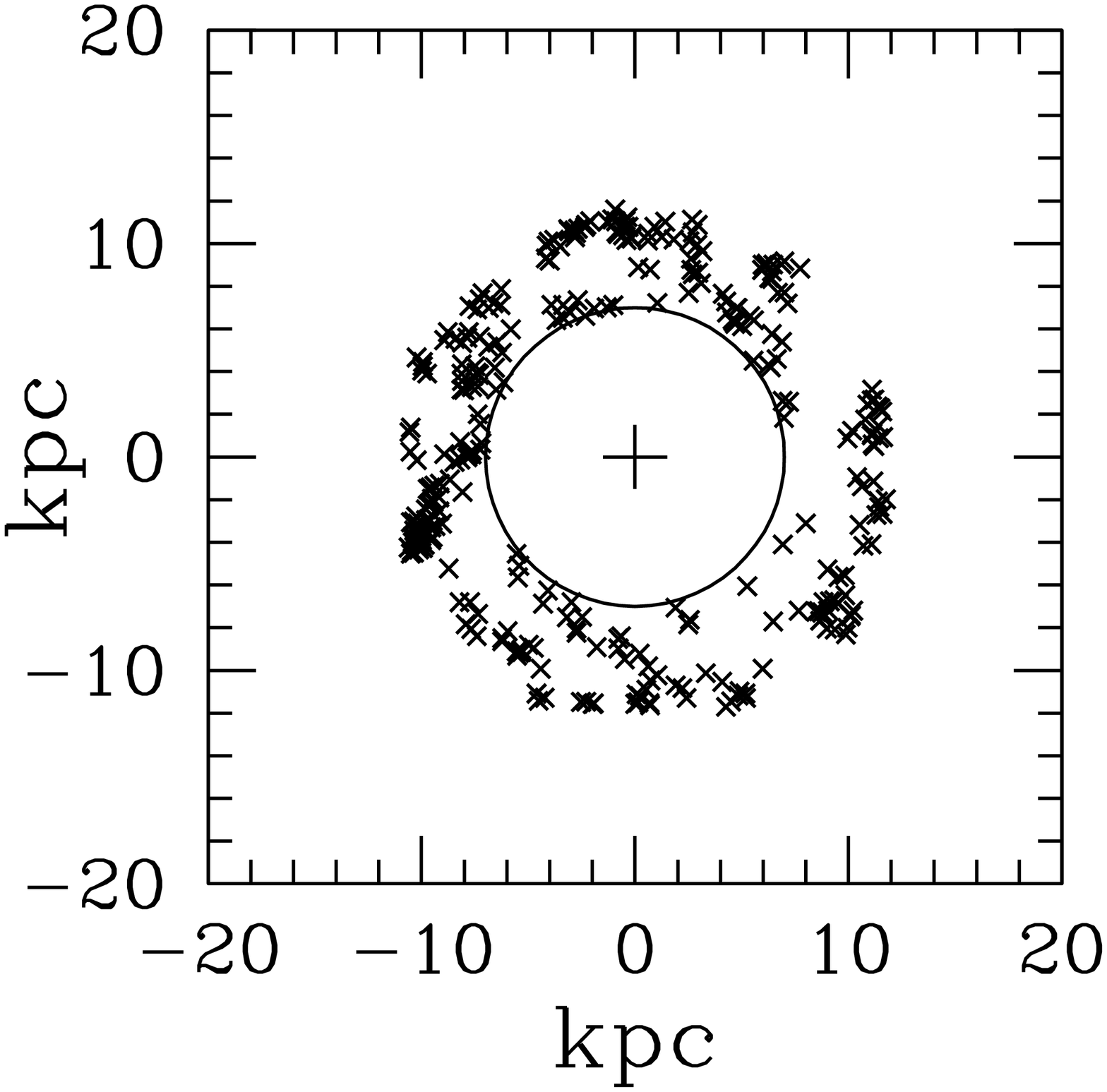}
\includegraphics[%
  scale=0.18]{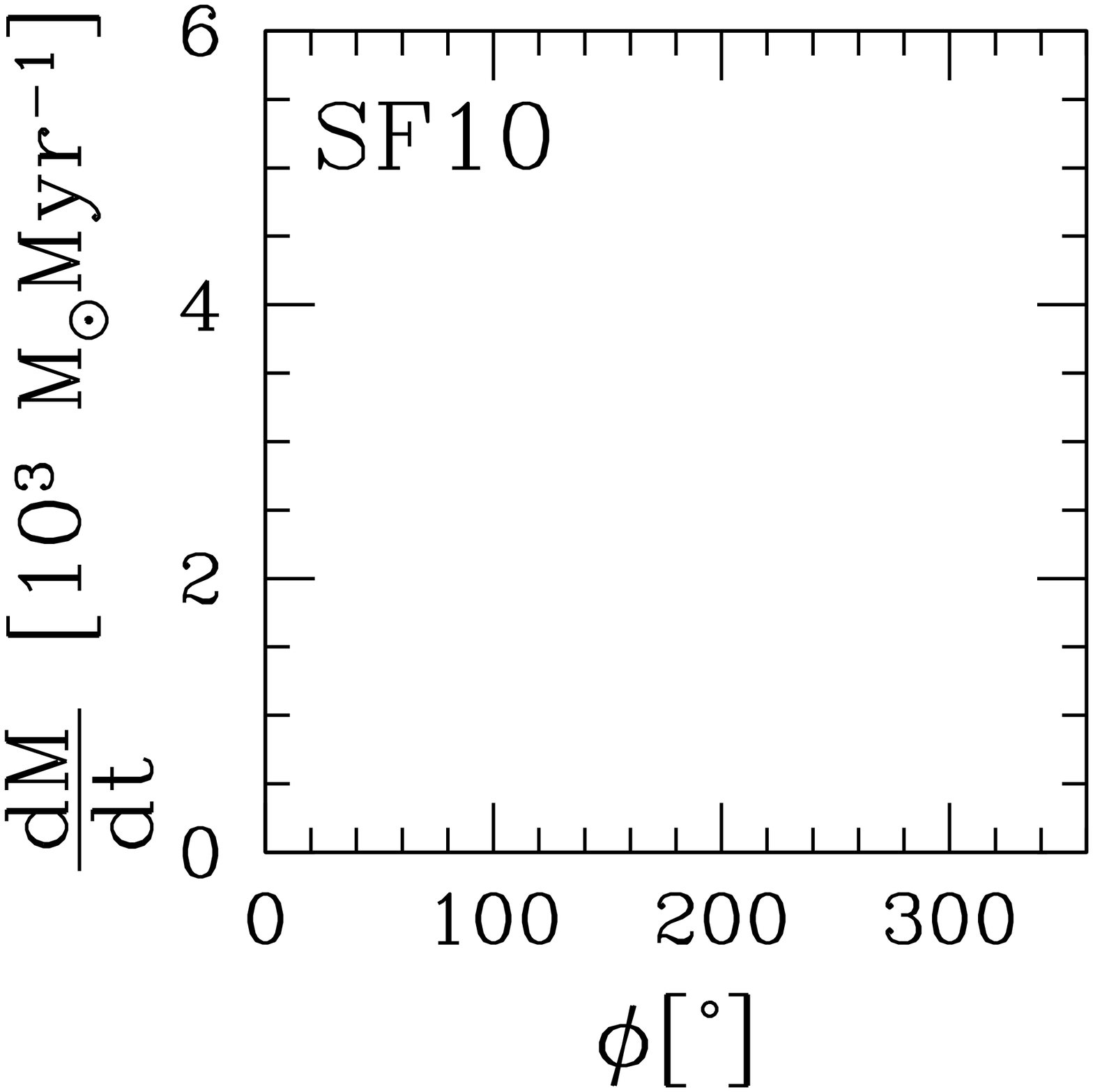}
\includegraphics[%
  scale=0.18]{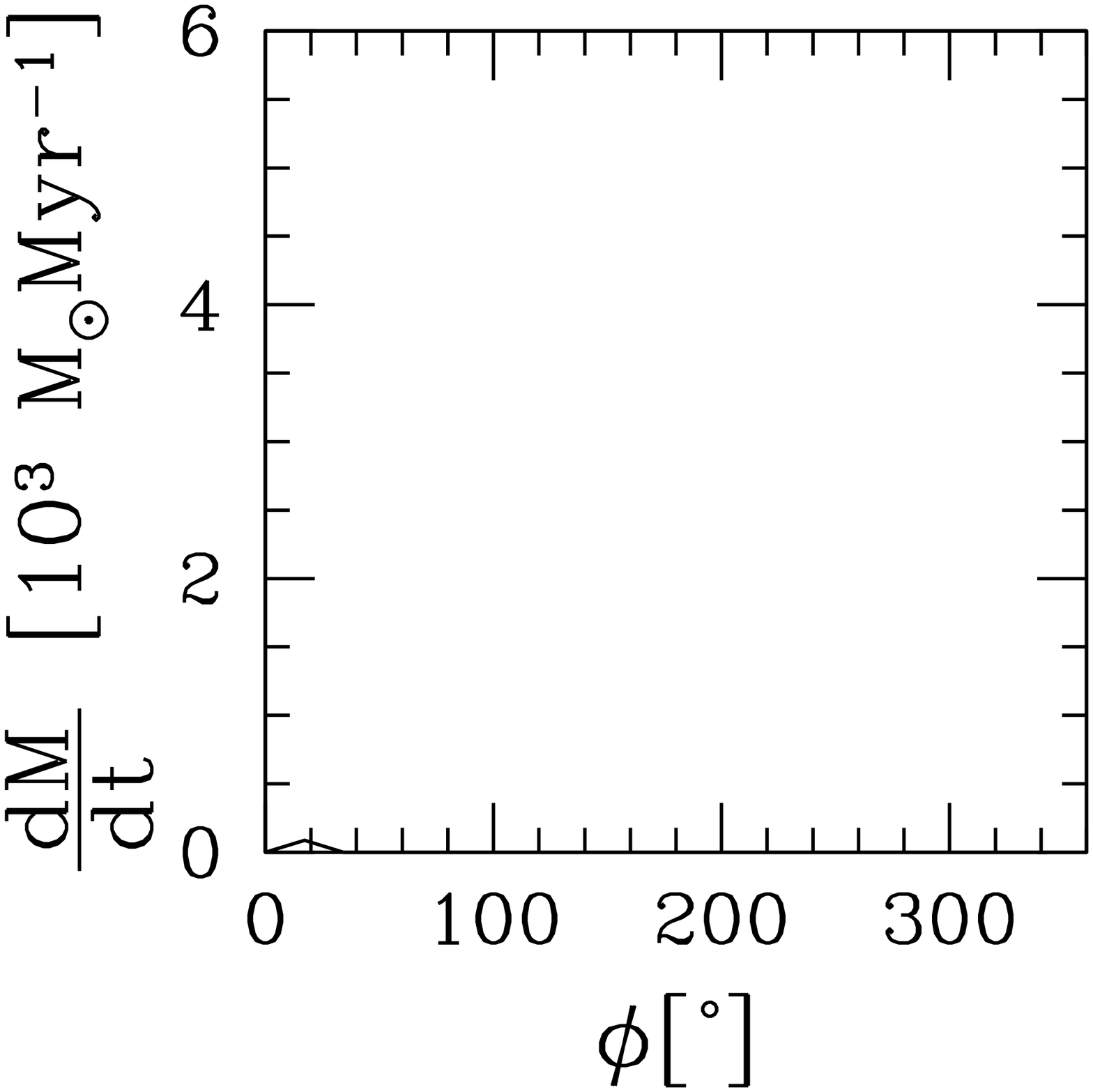}
\includegraphics[%
  scale=0.18]{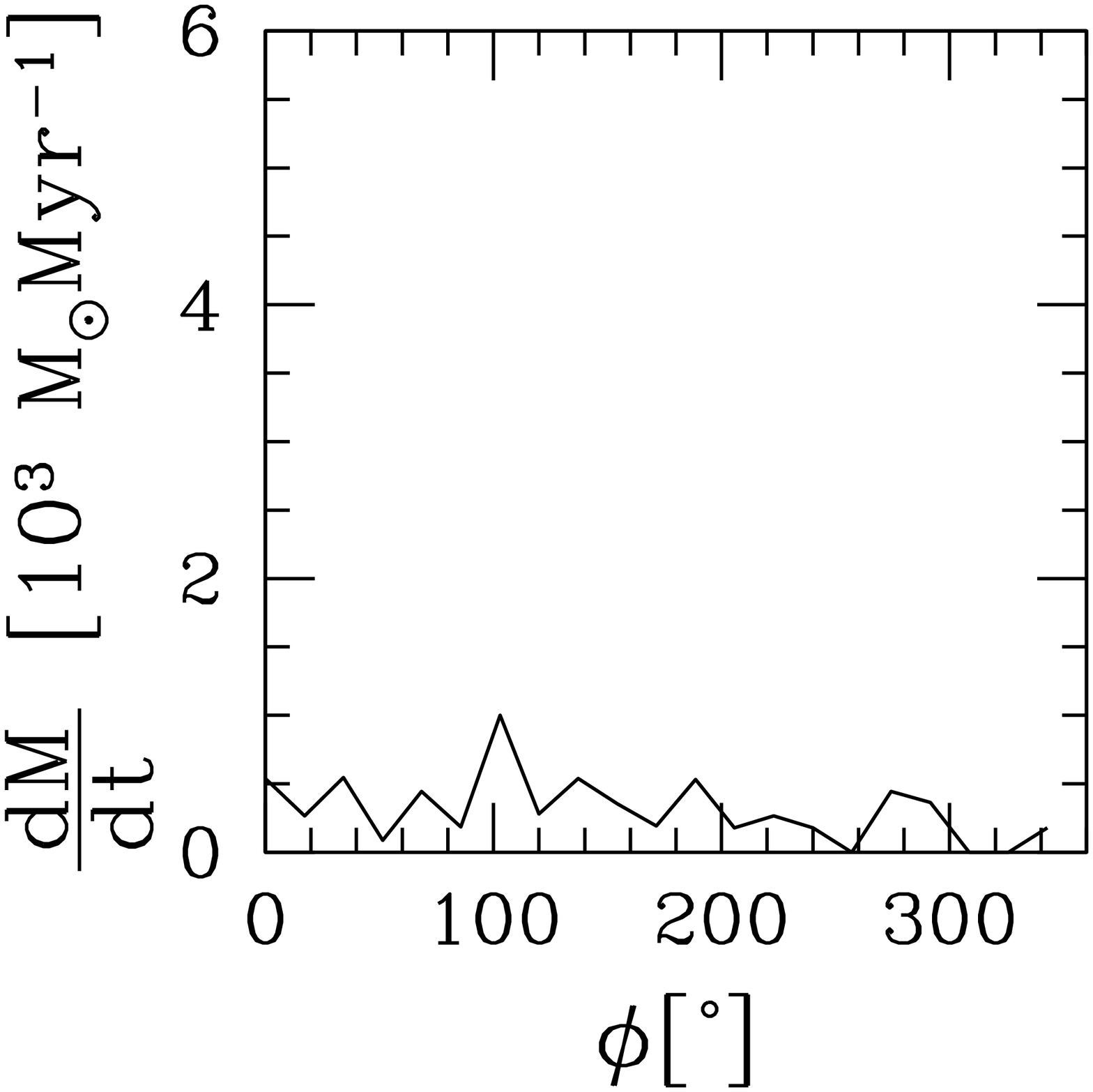}
\includegraphics[%
  scale=0.18]{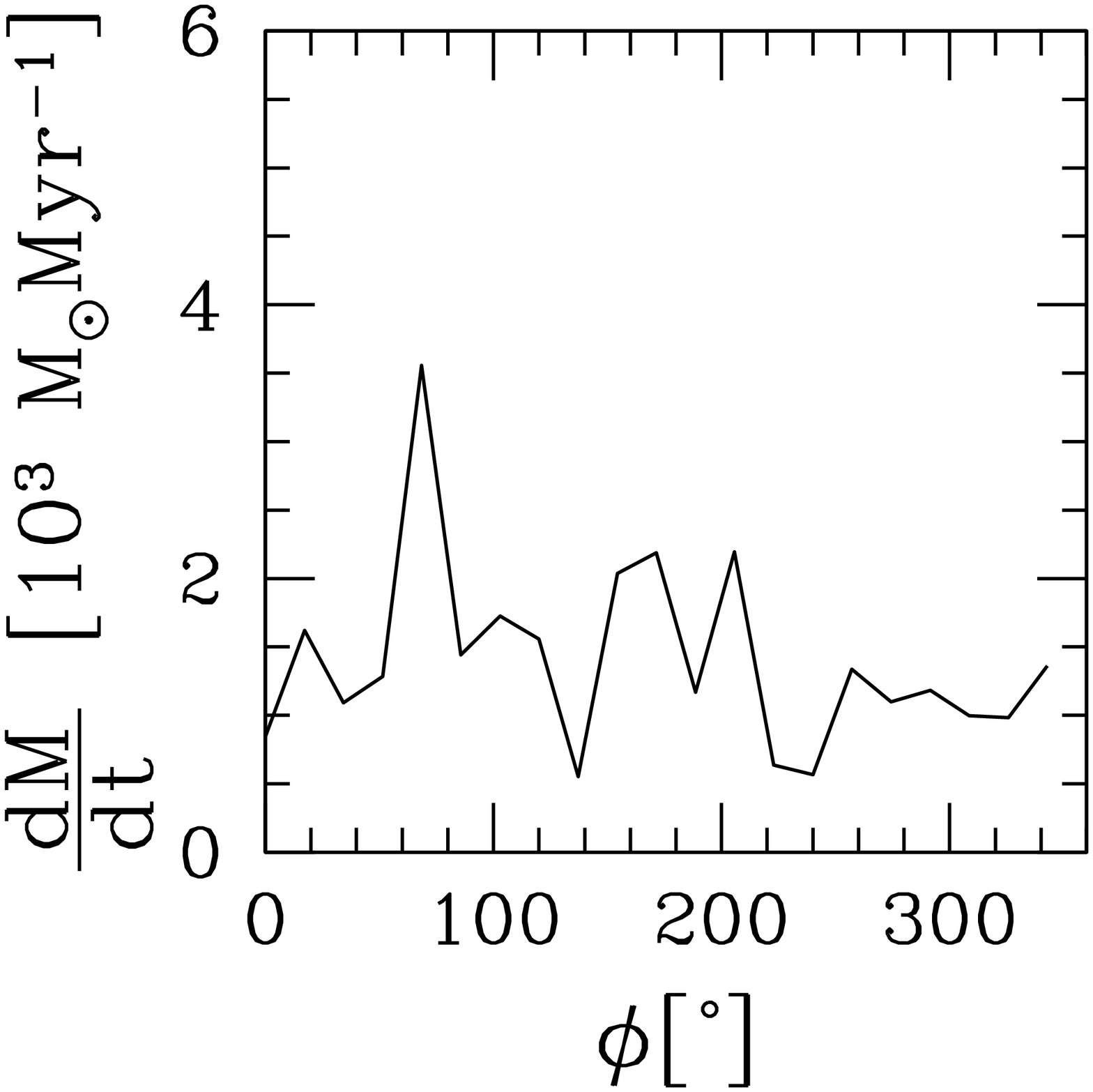}

\caption{Location of the instantaneous star formation events in the external 
disk. Each couple of rows corresponds to a different inclination angle
of the disk. Time increases from the left to the right: Time$=0.4,0.6,0.8,1$ Gyr. 
The first row of the couple represents the disk face on projection,
the second one the star formation rate versus the azimuthal angle $\phi$. }
\label{inclstarform}

\end{figure*}

Fig. \ref{inclstarform} illustrates the state of the newly formed stellar disk
at increasing times along the satellite orbit. 
Each couple of rows corresponds to one of the first three runs of
Table \ref{sfruns} and is associated with a different inclination
angle $i$.
The first row of the pair represents the face-on projection of the disk 
with the galaxy moving towards the left of the page and the same
geometry as in Fig. \ref{densitymaps}. 
Time increases from the left to the right.
Each small cross indicates a new star formation event at the time of the
snapshot (within a time interval of 40 Myr) while the circle
delimitates the external disk ($r>7$ kpc).
Stars form in the central regions as soon as the star formation algorithm is
activated, but here the star formation activity of the inner disk is
not represented. 
The second row represents the total mass $M_{\star}$ of the newly
formed stars in the external disk as a function 
of the disk azimuthal angle $\phi$.

In the case of a galaxy moving edge-on through the external medium (SF90), 
stars form at the leading edge of the disk when the ram-pressure 
becomes larger than $\sim 5 \times 10^{-14}$ dyn cm$^{-2}$, 
at Time $\sim 0.3$ Gyr. 
The location of the star formation events initially corresponds to the HI 
column density peak observed in cool90 around $100^{\circ}$ 
(Fig. \ref{densitymaps}). 
Later on it expands along the 
entire front edge, creating a thin stellar arc well distinct 
from the star formation events that characterize the central disk. 
As soon as the satellite encounters ram pressure values comparable to those
experienced by the LMC at the  perigalacticon (Time $\gsim 0.6$ Gyr) some 
episodes of star formation occur
even on the back side of the disk (last plot on the top right of
Fig. \ref{inclstarform}), although they are not relevant in terms of new
stellar mass formed. Indeed $M_{\star}$ shows a drastic drop at $\phi = 200^{\circ}$.  

Runs with inclination $i<90^{\circ}$ are characterized by significant star
formation only for values of the external pressure larger than $ 10^{-13}$ dyn cm$^{-2}$ .
In the case of the nearly face-on run SF10, 
at $t >0.7$ Gyr star formation occurs in the
entire external disk. Ram-pressure affects the plane of the disk
almost perpendicularly and stars form along the delocalized and filamentary high 
density structures visible in Fig. \ref{densitymaps10} 
(Mayer et al. in prep.).
Contrary to what has been found  by  \citet{Kronberger08} who focused on
higher ram-pressure values ($n \sim 10^{-4}$ cm$^{-3}$ and $v=1000$ km
s$^{-1}$) typical of the outskirts of galaxy clusters,  the newly
  formed stars are all located in the plane of the satellite's disk (with the exception of the high velocity face-on run SF10v400 where about 10$\%$ of the stars forms behind the disk). 
The star
formation events appear to be distributed nearly homogeneously along the azimuthal
profile of the external disk, although a small peak in $M_{\star}$ is
observable near  $i=90^{\circ}$. In fact, the orientation of the
disk with respect to its orbital motion is not exactly face-on.
The case of the intermediate run SF45 is more complex. In a first phase,
for low ram pressure values, star formation is produced by 
compression at the leading edge and a thin star formation front -- although not so well defined as in the case of a pure
edge-on model -- appears on the
east side of the disk. 
As soon as the external pressure reaches a critical
level
compression directed perpendicularly to the disk becomes the dominant
mechanism driving star formation.\\

\begin{figure*}
\includegraphics[%
  scale=0.35]{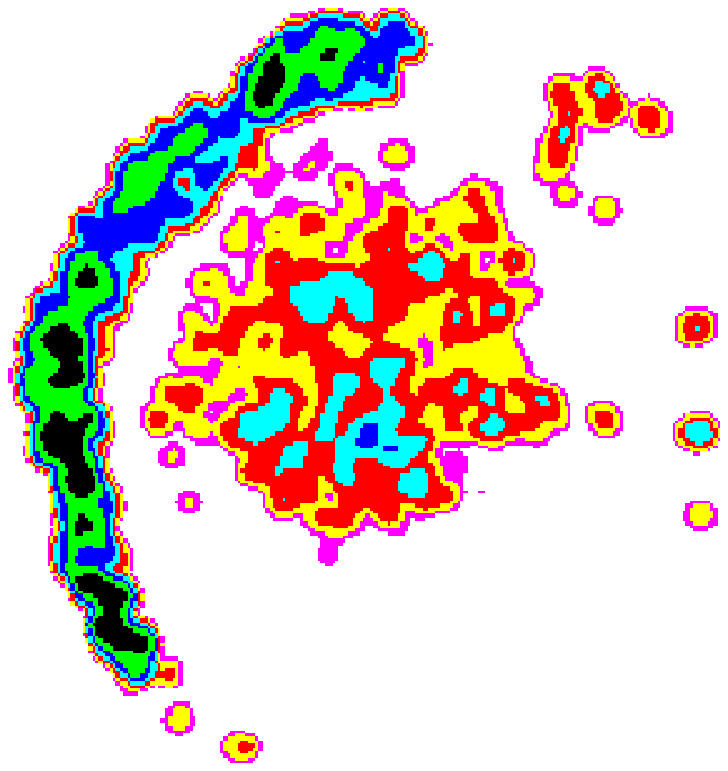}
\includegraphics[%
  scale=0.35]{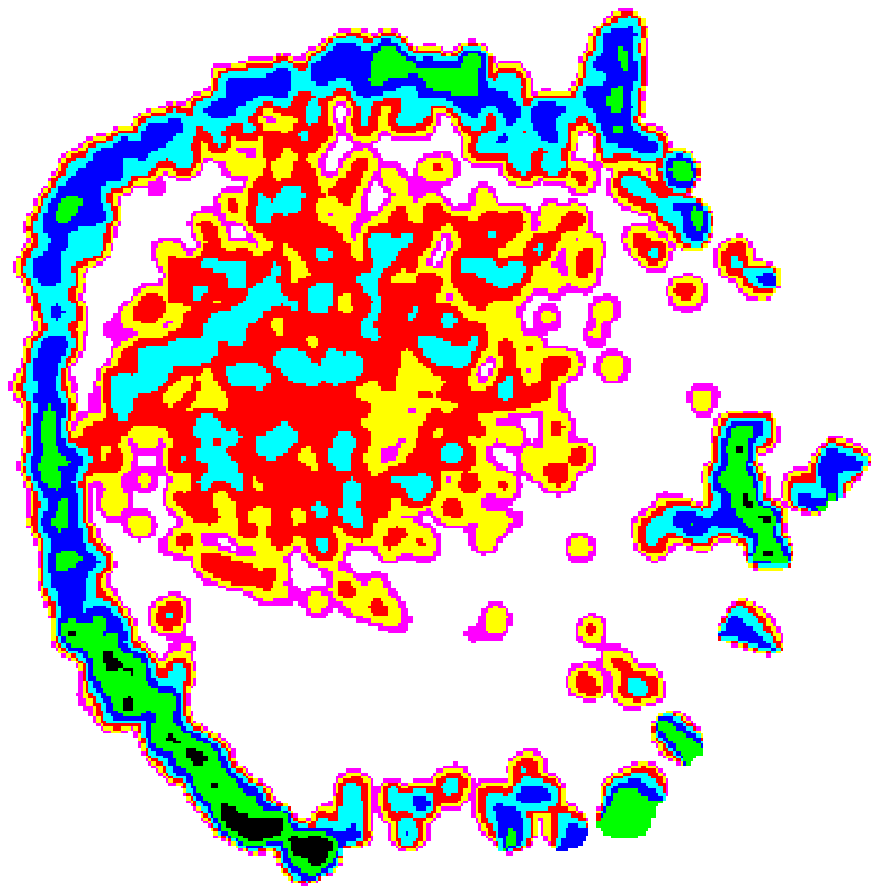}
\includegraphics[%
  scale=0.35]{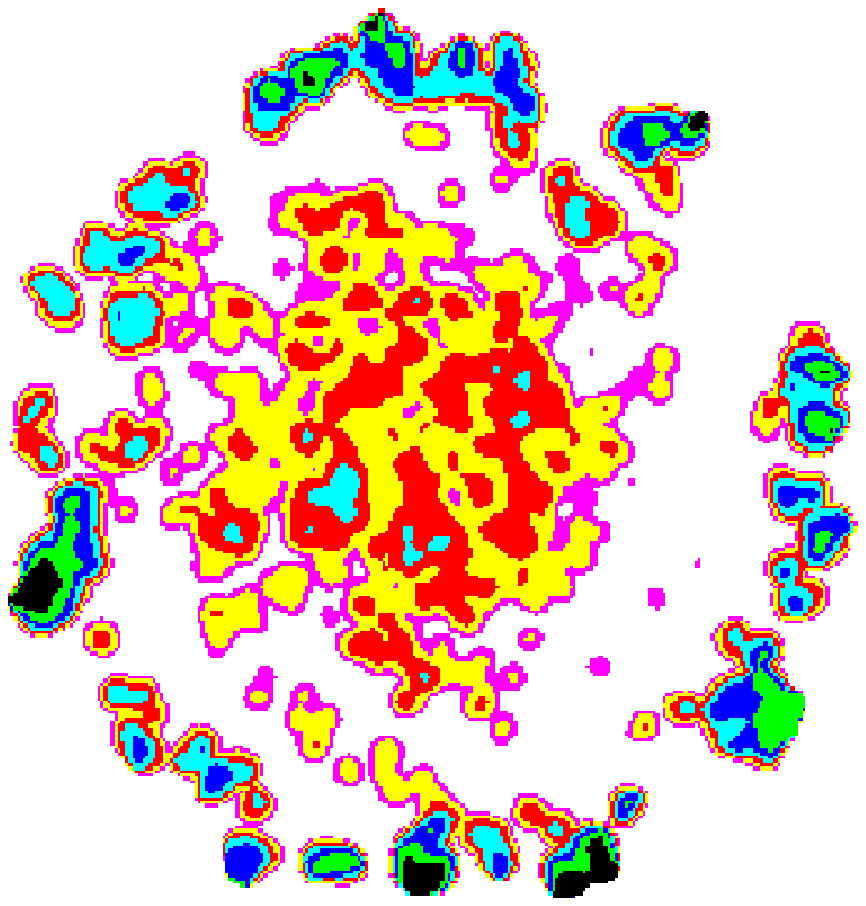}
\includegraphics[%
  scale=0.35]{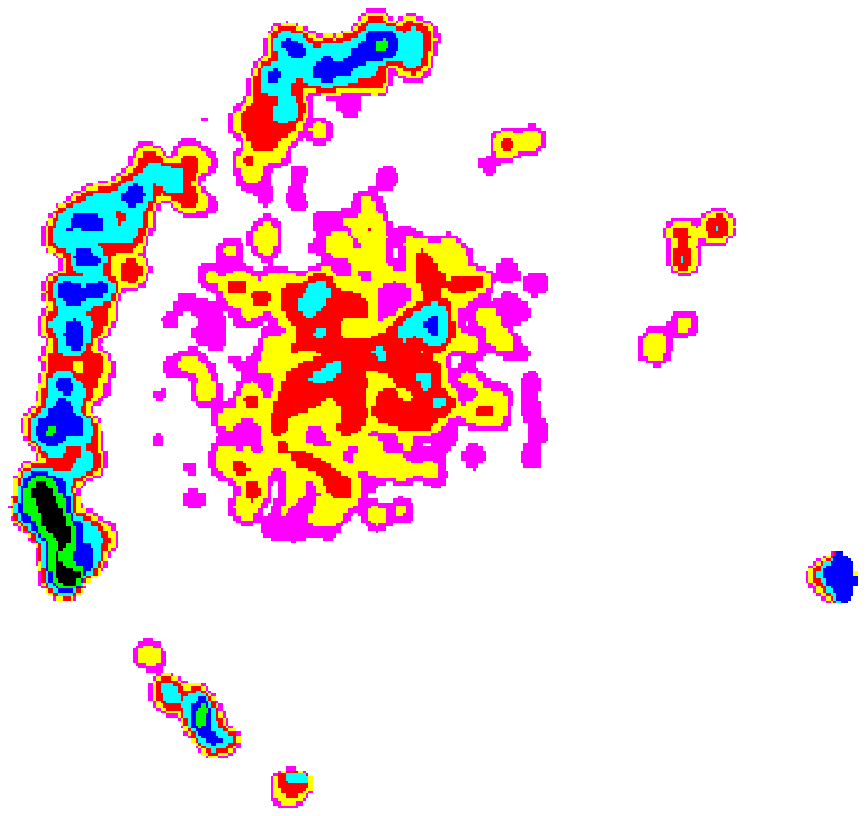}
\includegraphics[%
  scale=0.35]{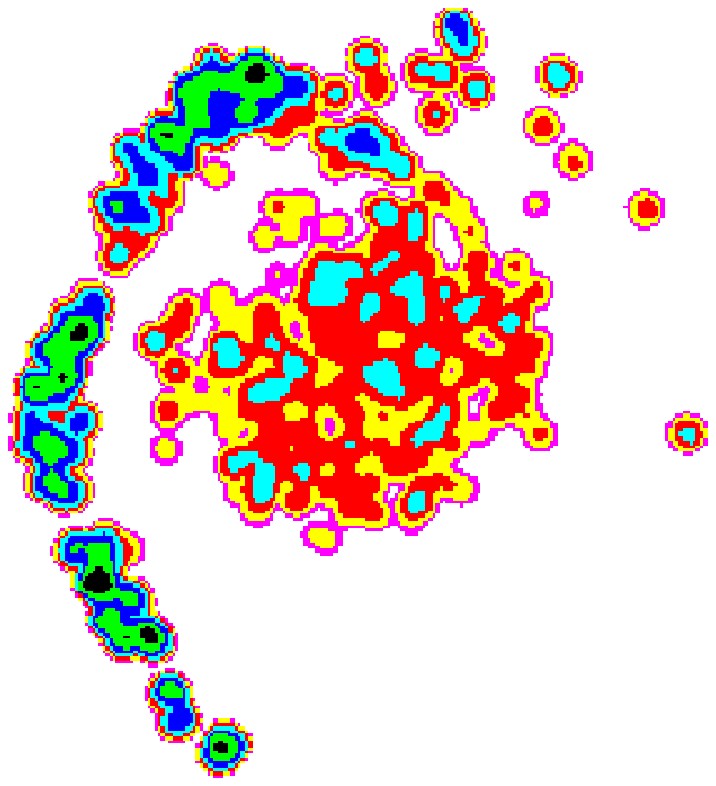}
\includegraphics[%
  scale=0.35]{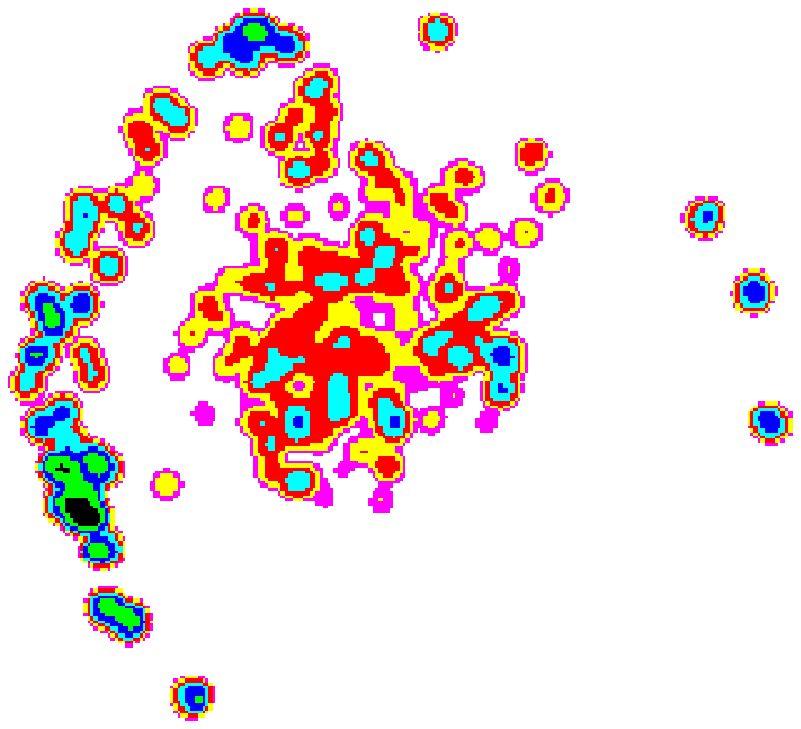}
\includegraphics[%
  scale=0.35]{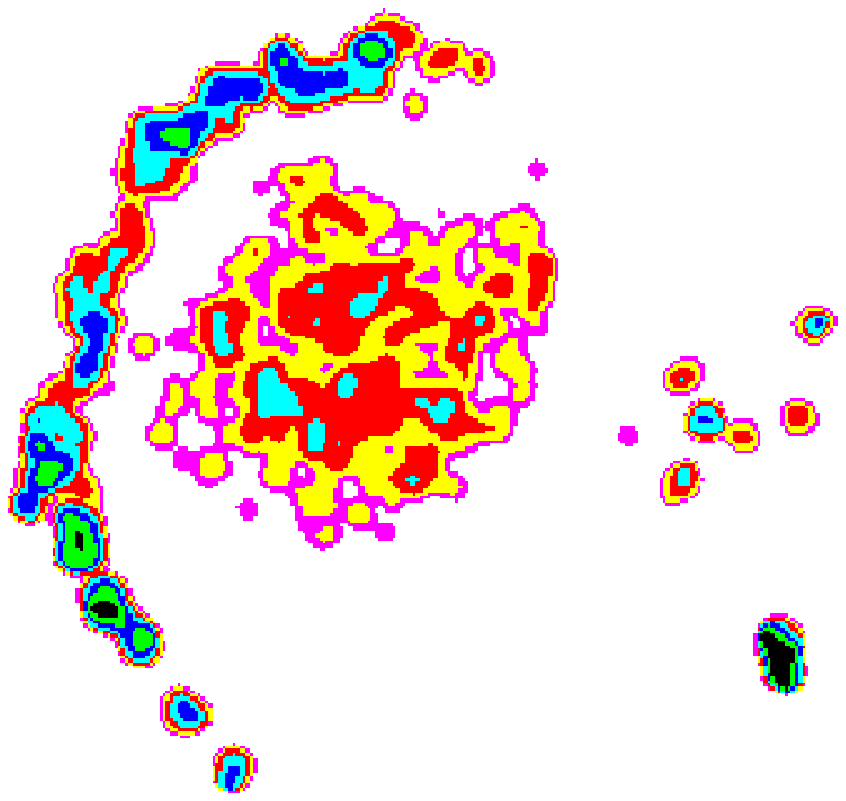}
\includegraphics[%
  scale=0.35]{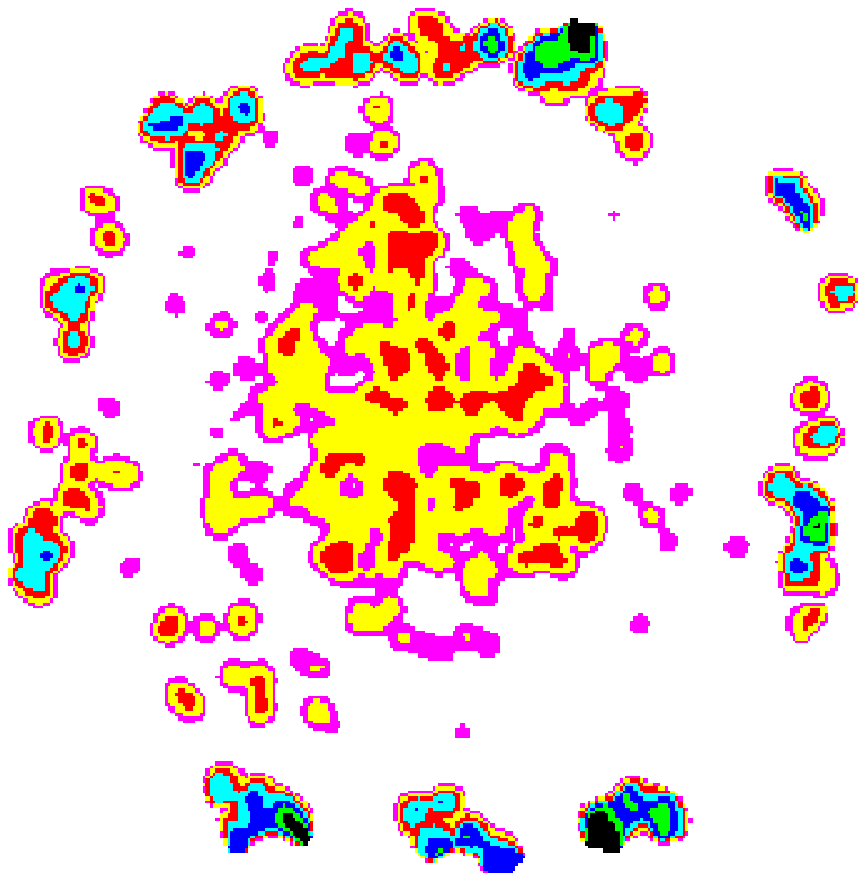}
\includegraphics[%
  scale=0.35]{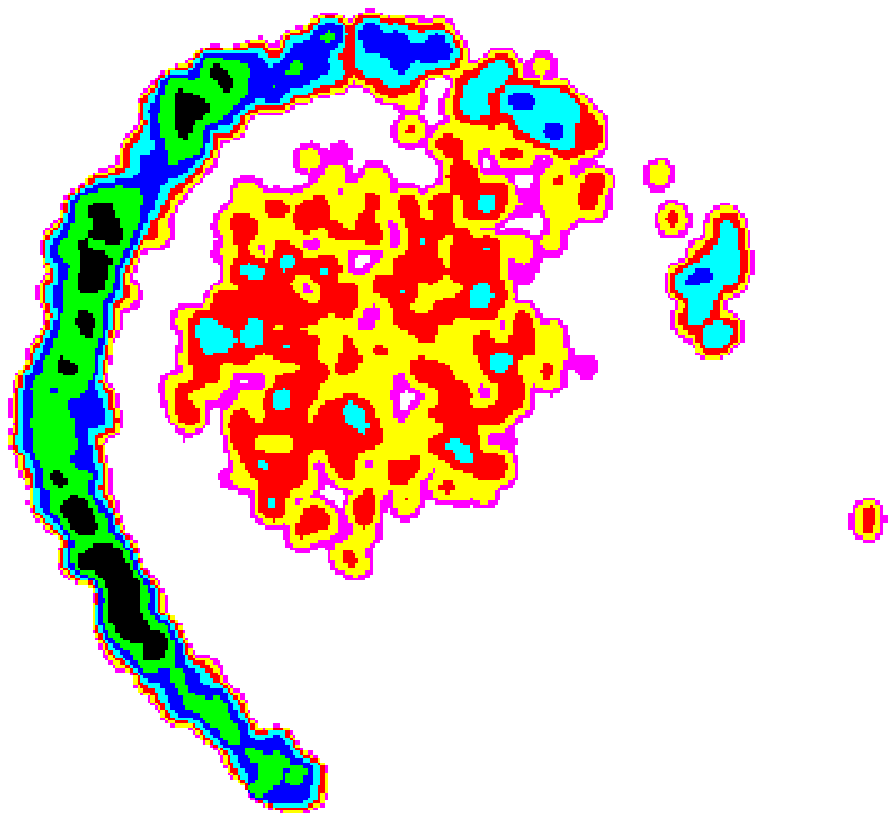}
\includegraphics[%
  scale=0.35]{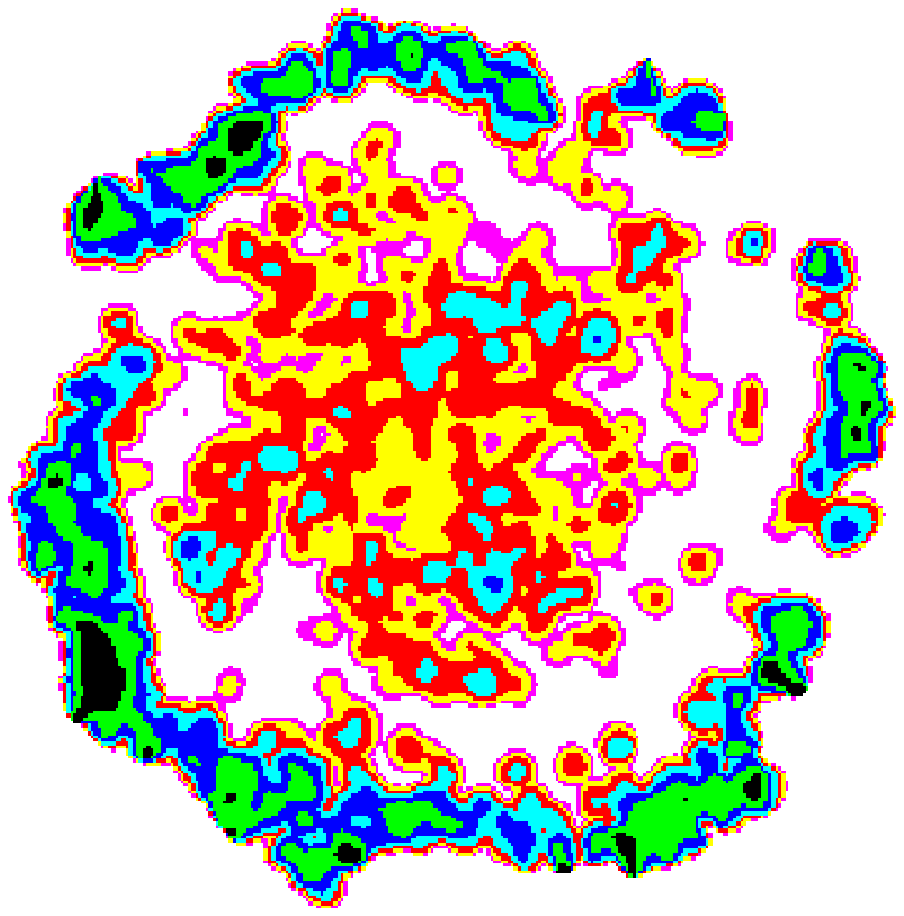}
 \includegraphics[%
  scale=0.35]{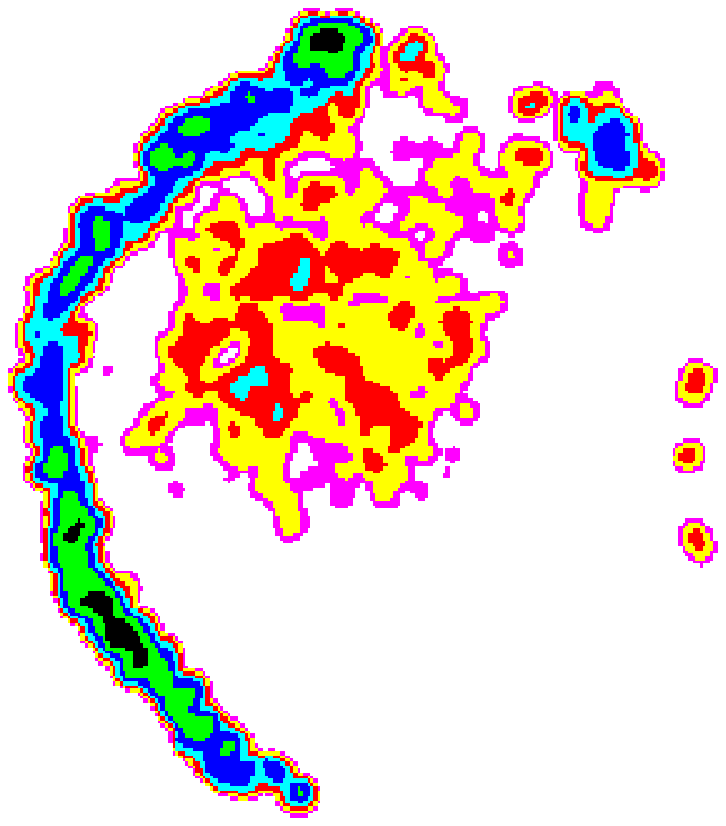}
 \includegraphics[%
  scale=0.35]{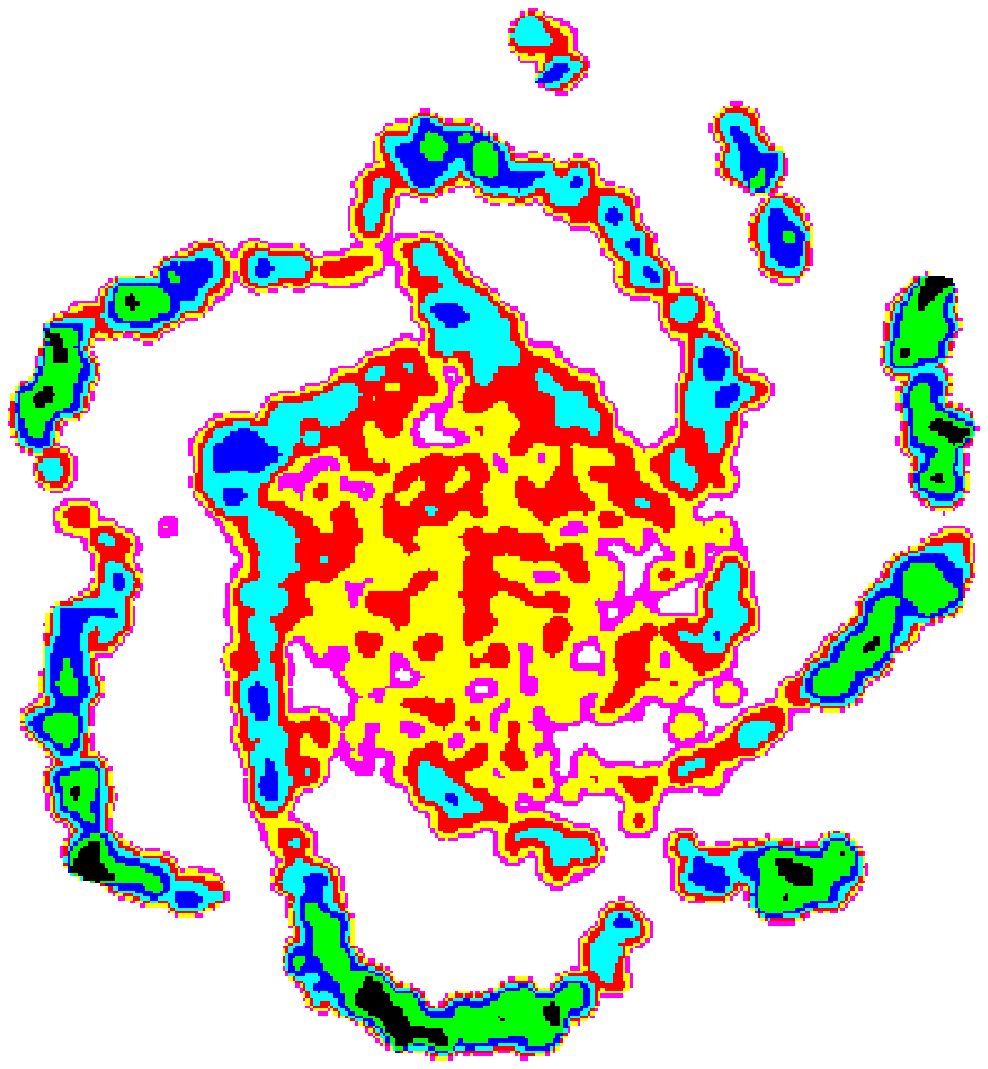} 
\includegraphics[%
  scale=0.35]{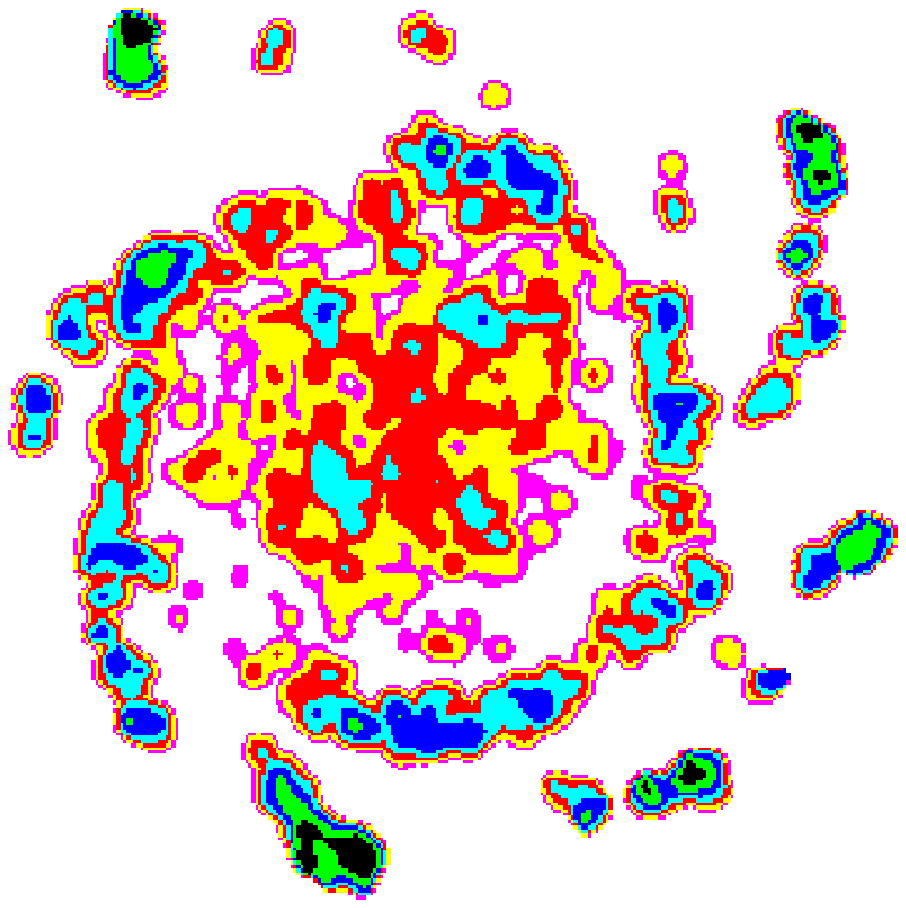}
\includegraphics[%
  scale=0.35]{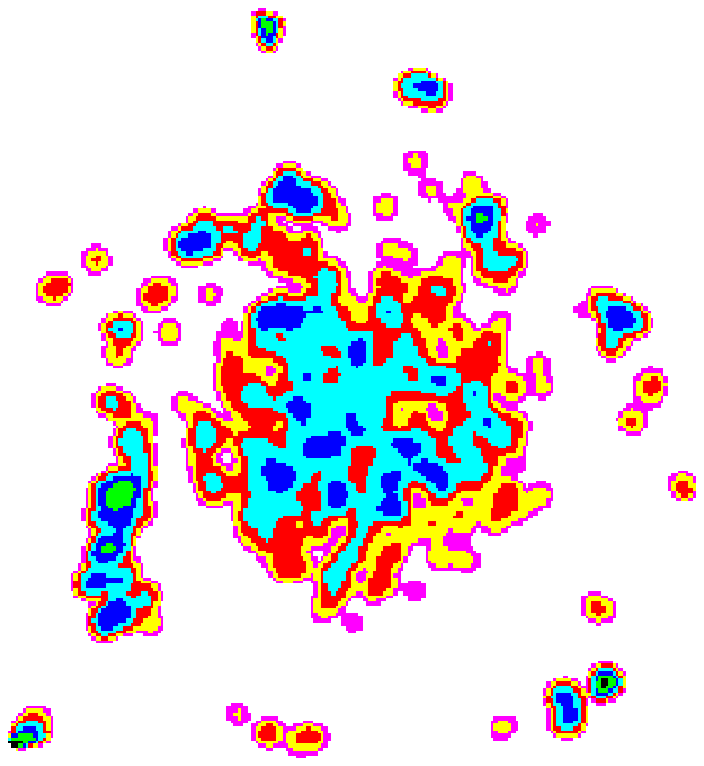}
\includegraphics[%
  scale=0.35]{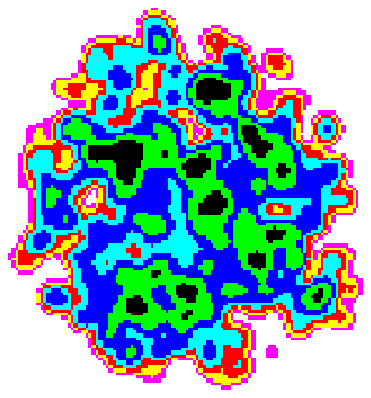}
\includegraphics[%
  scale=0.35]{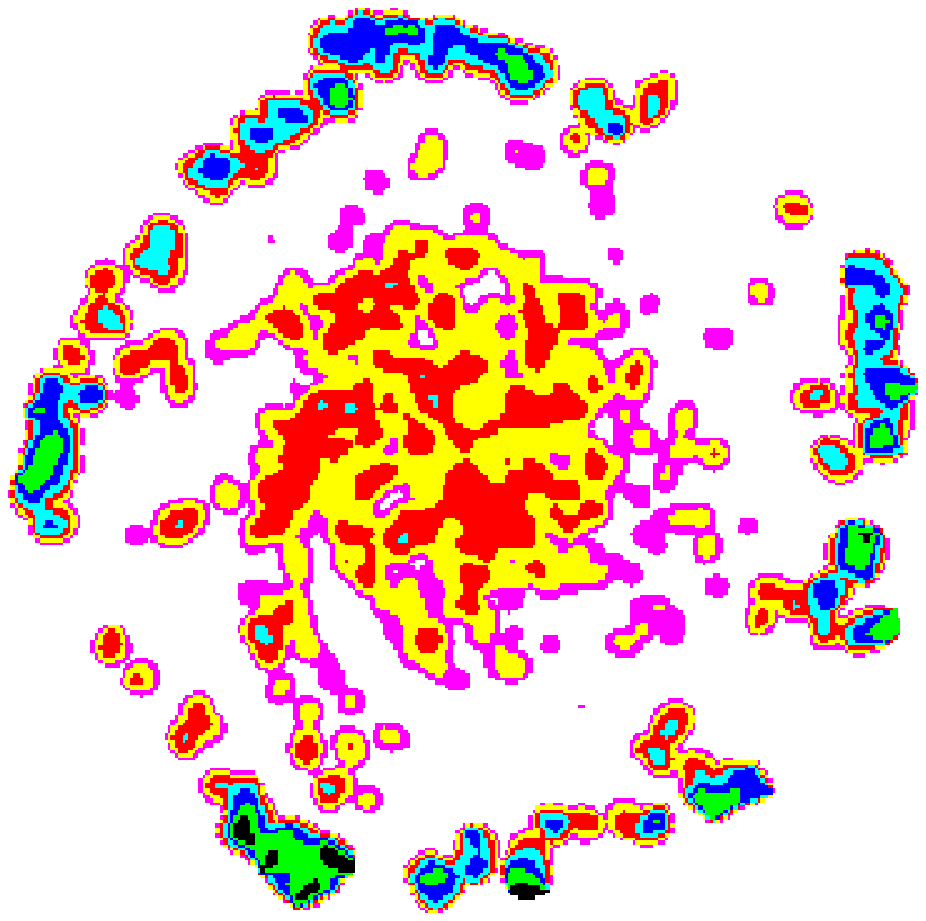}
\includegraphics[%
  scale=0.4]{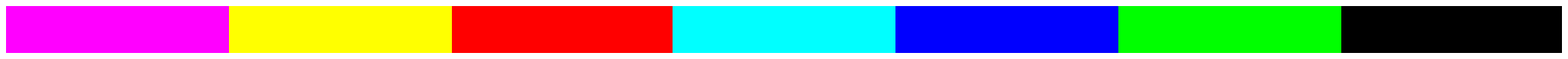}
\caption{H-$\alpha$ maps of the LMC's disk. The different panels
  represent (from top to bottom and from left to right) runs listed in
  Table \ref{sfruns}: SF90, SF45, SF10, SFconv, SF$\epsilon$1, SFconv$\epsilon$1, SF90c0.01, SF10c0.01, SF90c0.05, SF10c0.05, SF90v400, SF10v400, SF10v400t12000, SF10v400t15000, SFld90, SFld10. The color-scale is logarithmic with limits
  $10^{31}$ and $10^{34}$ erg s$^{-1}$ kpc$^{-2}$.} 
\label{lalpha}

\end{figure*}

Converting SFRs to H$\alpha$ luminosities according to
\citet{Kennicutt98}:

\begin{equation}
L(H\alpha)= \frac{\textrm{SFR}(M_{\odot} \textrm{yr}^{-1})}{1.26 \times 10^{41}}, 
\label{kennicuttlow}
\end{equation}

where SFR is the star formation rate averaged over the last 40 Myr
(nearly two times the stellar age of 30 Doradus), 
we obtain the H-$\alpha$ maps illustrated in Fig. \ref{lalpha}.

High emission regions are mainly concentrated in the external disk (with the exception of run SFld90 where the ram-pressure exerted by the low density halo is not enough to induce star formation at the edge of the disk).
The continuous stellar arc forming along the leading side of the disk in edge-on runs breaks 
up into several distinct and very luminous H-$\alpha$ regions that more closely resemble the star-forming complexes observed on the eastern border of the LMC.
The inclusion of stellar and supernovae heating -- which has been neglected in the present simulations -- could prevent further star-formation around highly emitting regions and consequently produce more compact and isolated star-forming complexes.  
Nevertheless, modeling single star-formation complexes -- whose linear extension is smaller than our softening length -- is beyond the scope of this paper. 

The present inclination of the LMC's disk with respect to the orbital motion is about $60^{\circ}$ (according to the convention adopted in this paper). Since the satellite is currently near a perigalactic passage we expect the H-$\alpha$ map at the leading border of the disk to be something in between pure edge-on runs and the run with inclination of $45^{\circ}$. On the other side, it is very likely that the disk inclination during the phase of approach to the pericenter was different.  Indeed in \citet{Mastropietro08} we have simulated the LMC's orbit according to the new proper motion measurements of \citet{Kallivayaliletal06} and  found that the cloud enters the MW halo face-on and moves almost face-on during most of the last 1 Gyr.  It turns nearly edge-on only at the perigalacticon.
This would have a remarkable effect on the star-formation history of the external disk during the last  1 Gyr and some impact also on the H-$\alpha$ maps.
Indeed, although the H-$\alpha$ emission would be mostly concentrated on the eastern side of the disk due to the very recent edge-on motion, we expect to see some luminous clumps  forming a patchy distribution on the entire disk, due to gravitational instabilities and subsequent star-formation induced by a nearly face-on compression of the gaseous disk before 30 Myr ago.

The high velocity edge-on run SF90v400 presents a more elongated and thinner stellar arc along the leading border, with a geometry similar to that obtained increasing the star formation rate parameter to 0.05 (SF90c0.05). 
The H-$\alpha$ map of SF90v400 (third panel of the third raw) shows two distinct luminosity peaks. One is located at the south-east region of the disk, roughly corresponding to the position of 30Doradus and the two compact emission regions N159 and N160.

\begin{figure}
\epsfxsize=9truecm \epsfbox{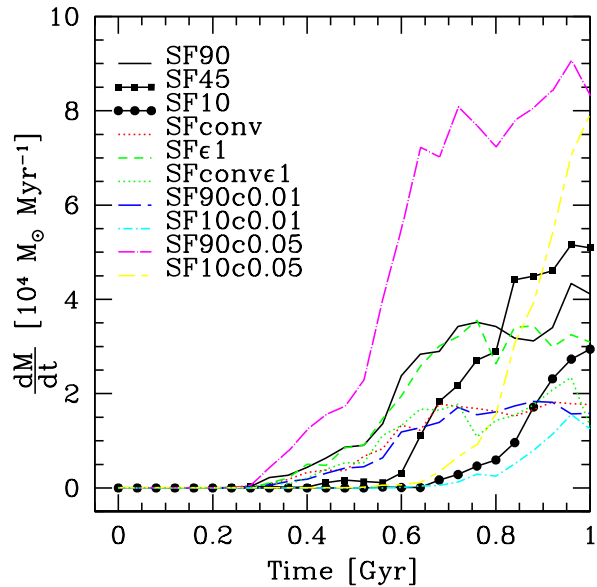} 
\caption{Star formation rate of the external disk 
  $\phi$ for the first ten runs listed in Table \ref{sfruns}. Time increases approaching to the present position.} 
\label{dmdt}
\end{figure}

Fig. \ref{dmdt} illustrates how the star formation rate of the external disk changes with time in the different models. The three black curves refer to the 
standard star formation runs SF, characterized by low orbital velocities. 
The edge-on disk starts forming stars earlier, but for large ram pressure
values the star formation rate of SF10 and SF45 grows faster. At the
perigalacticon passage SF45 has indeed a higher star formation rate
than SF90. 
In the case of an isolated LMC model star formation is almost absent for $r > 7$ kpc.
The remaining curves in Fig. \ref{dmdt} refer to different star formation
recipes (rows 4-9 of Table \ref{sfruns}).      
The location of the star formation events in the external regions of the disk 
does not change significantly choosing different parameters in the star 
formation algorithm. 
The star formation rate parameter $c_{\star}$ determines the amount of new
stars forming but does not affect the minimum threshold in ram-pressure 
neither the evolution of the star formation rate. 
In particular, in the case of edge-on runs after an initial steep increment 
the curve seems to converge to a constant value for increasing external 
pressures.
The consequences of an increased star formation efficiency ($\epsilon 1$) are 
almost negligible (but not in H$\alpha$ maps where only the very recent star formation rate is taken in account: compare the first panels of the first and second row) while including the convergency requirement (SFconv) has 
nearly the same effect than reducing the star formation rate parameter of a
factor two.
 \begin{figure}
\epsfxsize=9truecm \epsfbox{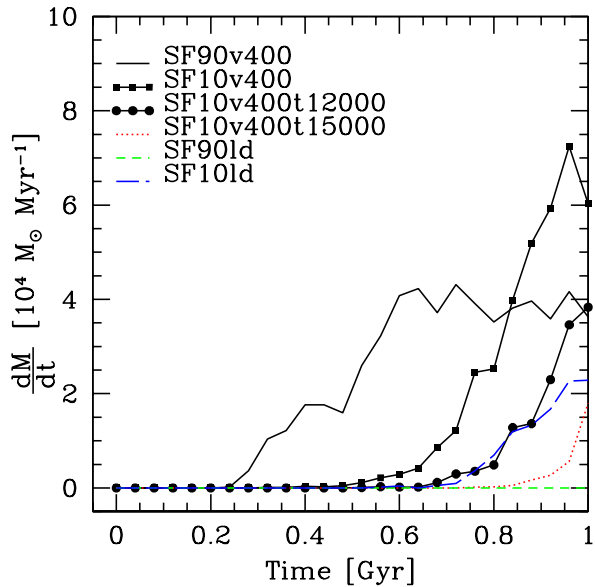} 
\caption{Same as in \ref{dmdt} for the last six runs listed in Table \ref{sfruns}. } 
\label{dmdtalt}
\end{figure}

For convenience, the star formation rate of the last six runs of Table
\ref{sfruns} is plotted separately (Fig. \ref{dmdtalt}).
The high velocity edge-on model SF90v400 is characterized
by a steeper increment in star formation at earlier times but later on
the curve flattens 
and the star formation rate at the perigalacticon is similar to
that of the low velocity case SF90.  
Differences in H-$\alpha$ maps are produced by a difference of $\sim 5 \times 10^3$ M$_{\odot}$ Myr$^{-1}$ about Time = 1 Gyr.  
On the contrary, the star formation generated by compression
perpendicular to the disk increases with increasing ram-pressure
values (it shows a decrement only towards the end of the simulation) and in
the case of the high velocity run SF10v400 reaches a  peak $\sim 1.5$
times higher than in SF10.
The star formation rate of SF10v400 is strongly
affected by the introduction of an artificial temperature threshold.
With a temperature floor of 15000 K we nearly suppress star formation in the external disk. However, a threshold of 12000 K is already very high for a LMC model and more typical of luminous disk galaxies like the MW.
A temperature floor lower than 10000 K would not make sense since below this temperature the cooling function adopted in the present paper drops rapidly to zero. 
If, as pointed out by \citet{Mastropietro08}, the satellite is moving almost face-on until it gets very close to perigalacticon, we would not expect to see star formation before 0.6 Gyr independently of the orbital velocity.
Differences between high and low velocity runs should be marginal also near the pericenter since star formation in edge-on runs -- assuming our standard prescriptions for star formation -- seems to saturate around $\sim 4 \times 10^4$ M$_{\odot}$ Myr$^{-1}$.
Finally, a hot halo ten times less dense than the one assumed in our favorite model would reduce drastically the star formation in a face-on LMC and suppress completely the star formation on the leading edge of the disk.

\begin{figure}
\epsfxsize=9truecm \epsfbox{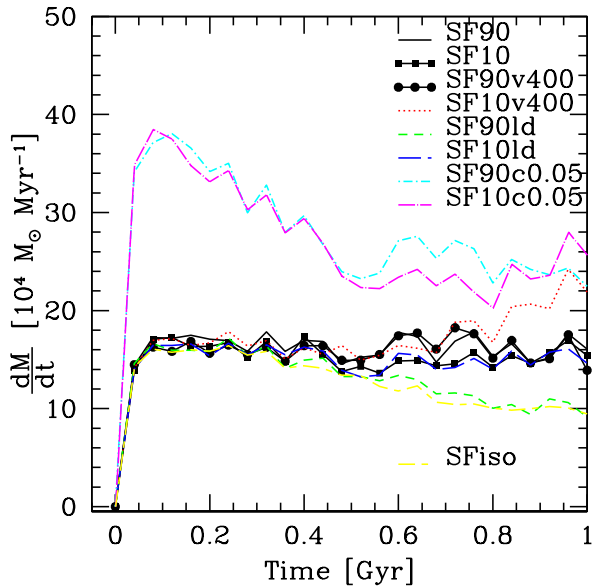} 
\caption{Star formation rate of the entire disk for selected runs in Table \ref{sfruns}. SFiso refers to the LMC model run in isolation.
} 
\label{dmdttot}
\end{figure}

The star formation rate of the entire disk is illustrated in Fig. \ref{dmdttot}. 
This plot is only indicative since we neglect the presence of the bar and its influence on the star formation history of the satellite. The star formation rate of a LMC model evolved in isolation is plotted for comparison.
The largest  contribution to SFR is given by star formation events in the central region of the disk. 
Curves peak between 15 and 35 M$_{\odot}$ Myr$^{-1}$, comparable  with observations of the Magellanic Clouds Photometric Survey (Harris et al. in prep). 
The total star formation rate is clearly not affected by ram-pressure before 0.4  Gyr.
Indeed, after an initial sharp increment curves are rather  flat  despite orientation and intensity of the external pressure. An initial burst in star formation is obtained only by increasing the star formation efficiency parameter to 0.05.
For Time$ >$ 0.5 Gyr the largest deviations from an average star formation rate of $ \sim 15$   M$_{\odot}$ Myr$^{-1}$ are produced in the edge-on low density run SF90ld, where dM/dt drops to 10 M$_{\odot}$ Myr$^{-1}$ at Time = 1 Gyr, and in the high velocity face-on model SF10v400 whose star formation rate increases up to $ \sim 24$   M$_{\odot}$ Myr$^{-1}$ toward the end of the simulation.
Differences among the other models are of the order of few  M$_{\odot}$ Myr$^{-1}$ and vary with time so that it would be quite difficult to use the star formation history of the entire disk to test the Cloud's orbital parameters and the hot halo density.

Fig. \ref{agephi} represents the mean stellar age of the external disk versus the azimuthal angle $\phi$ for the same selected runs of Fig. \ref{dmdttot}. The maximum increment in age in a clockwise direction is associated with edge-on runs, where stars forming at the leading edge move, in time, away to the side, due to the clockwise rotation of the disk.
The youngest stars are located at $30^{\circ}<\phi < 100^{\circ}$.
Clearly the gradient in age is much weaker in models with $i<90^{\circ}$. 
SF10v400 -- with a mean stellar age of $\sim 250$ Myr -- forms stars earlier with respect to the other nearly face-on runs, while the difference between  SF90v400 and the corresponding low velocity run SF90  is about 5 Myr at the leading edge and not significant in the rest of the external disk. 

For the same runs we also plotted the final radial gas density profile (Fig. \ref{densf}). In most of the cases a secondary peak is still present.

\begin{figure*}
\includegraphics[%
  scale=0.35]{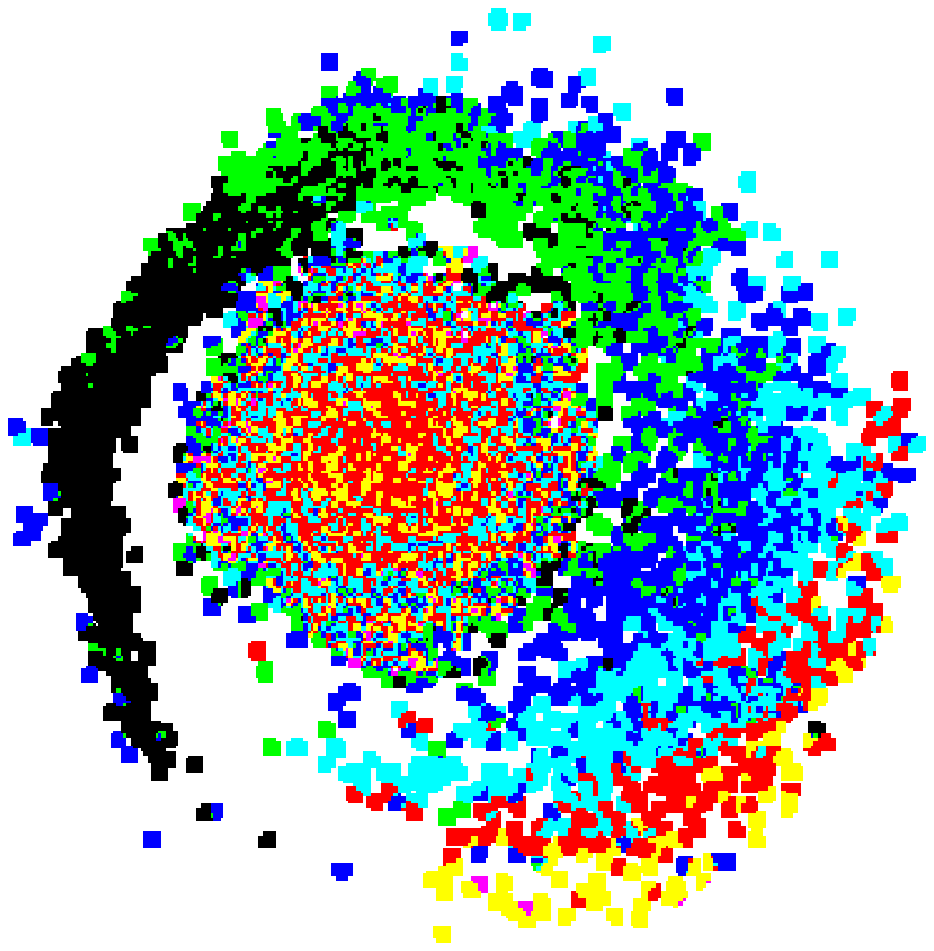}
\includegraphics[%
  scale=0.35]{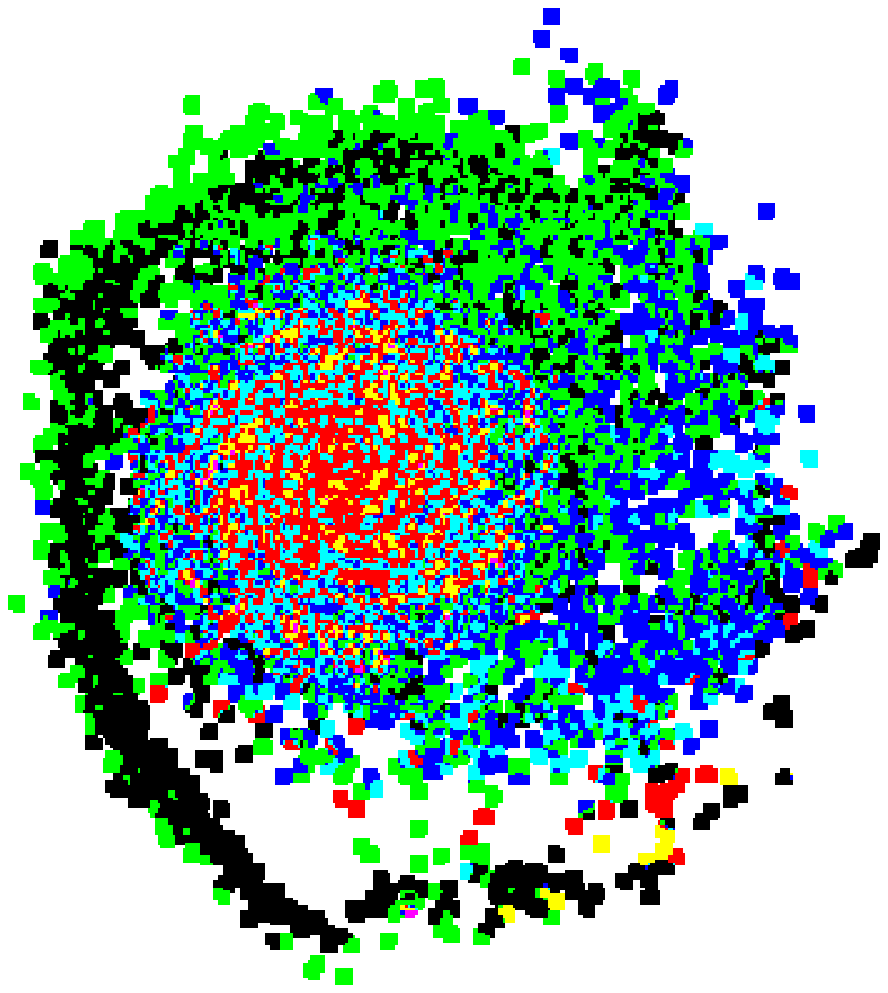}
\includegraphics[%
  scale=0.35]{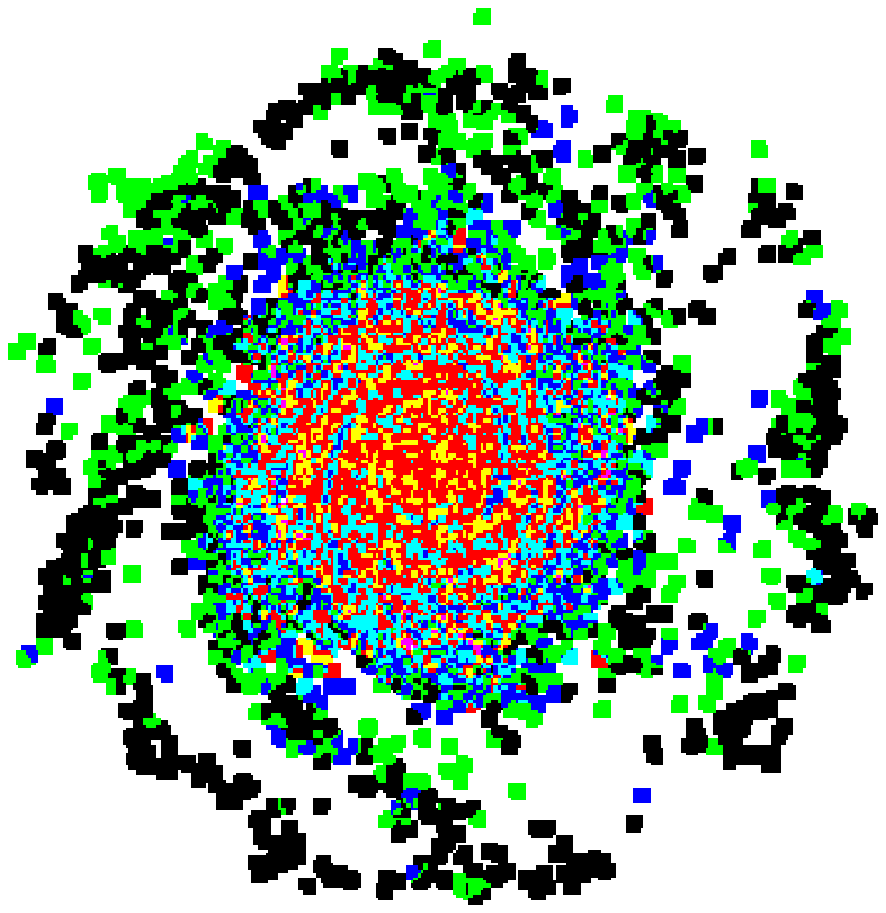}
\includegraphics[%
  scale=0.35]{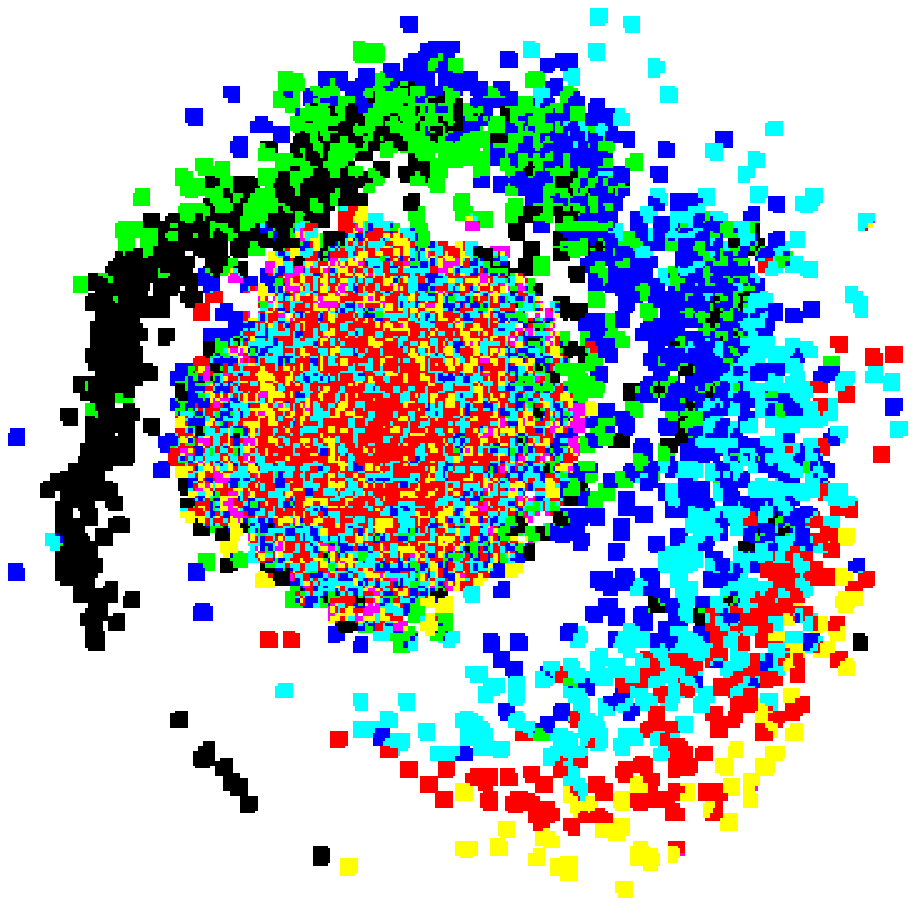}
\includegraphics[%
  scale=0.35]{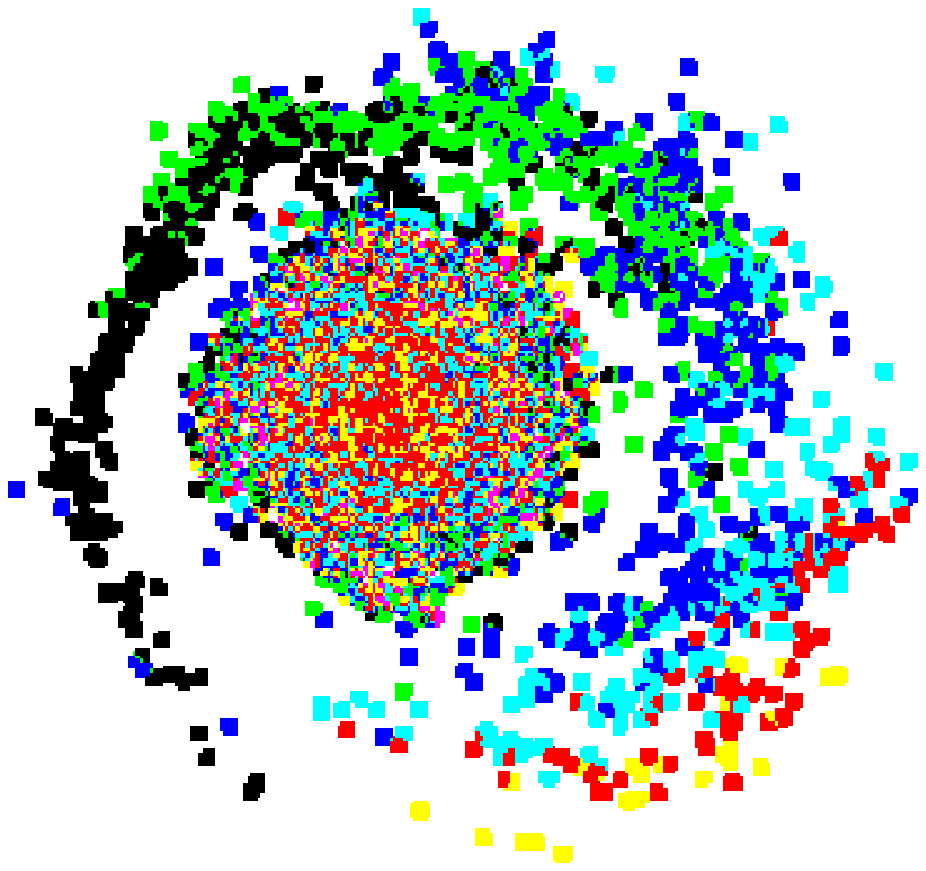}
\includegraphics[%
  scale=0.35]{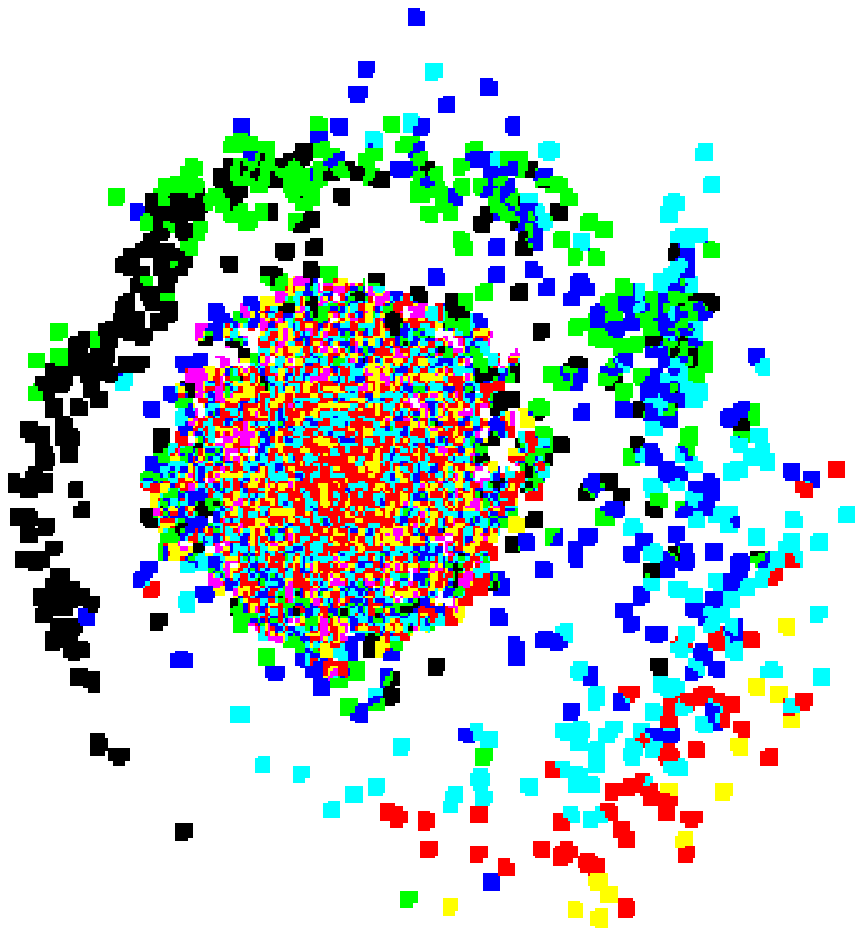}
\includegraphics[%
  scale=0.35]{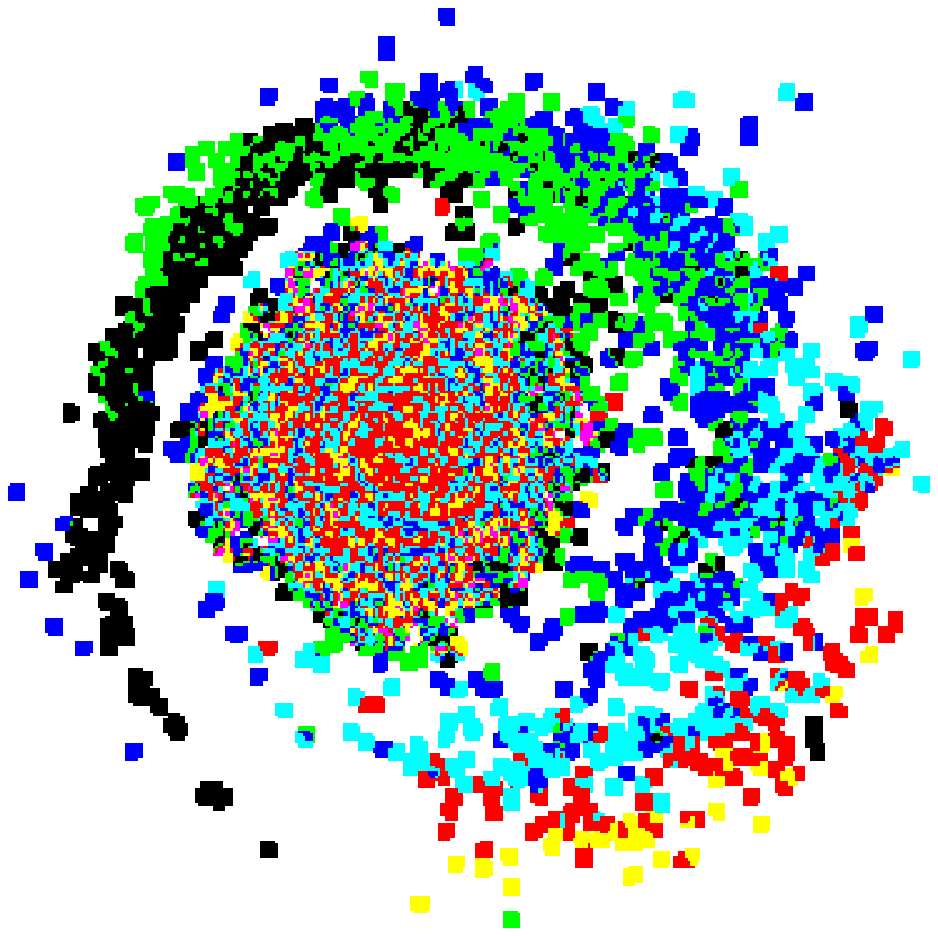}
\includegraphics[%
  scale=0.35]{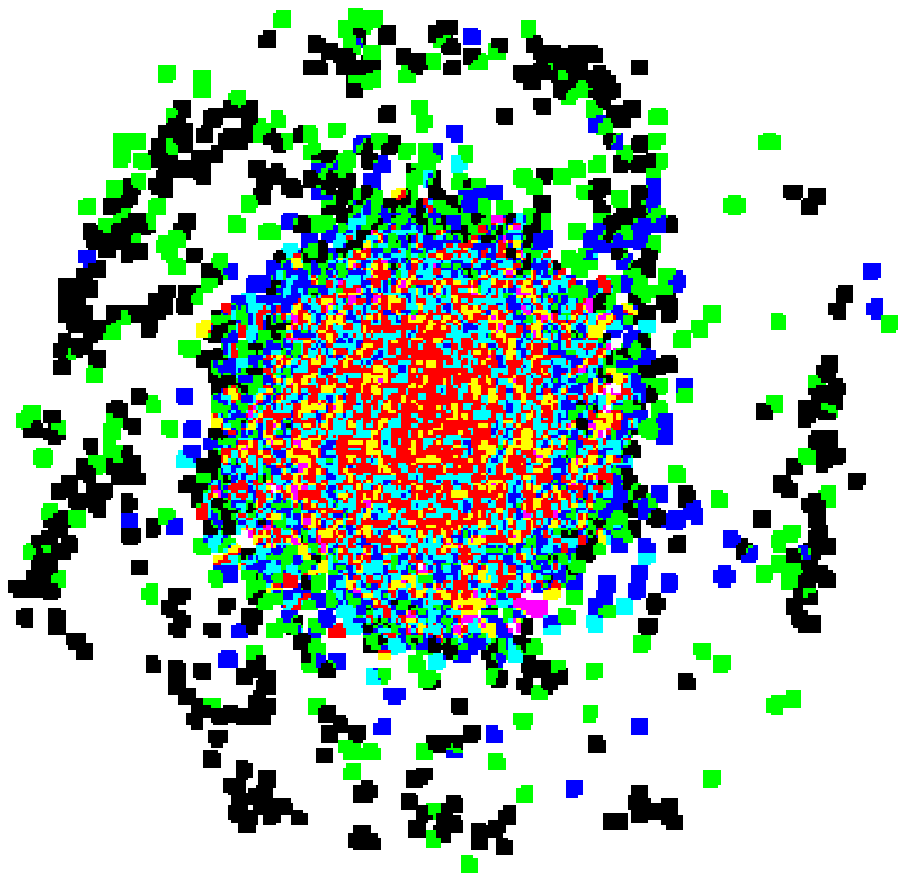}
\includegraphics[%
  scale=0.35]{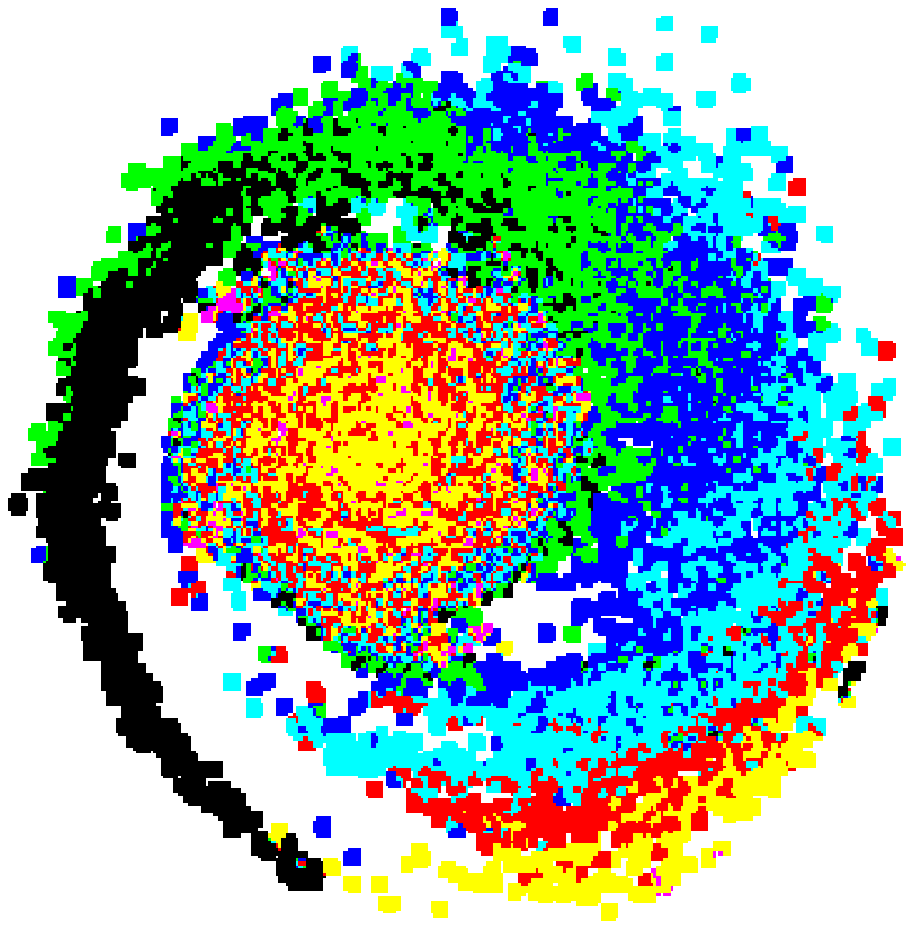}
\includegraphics[%
  scale=0.35]{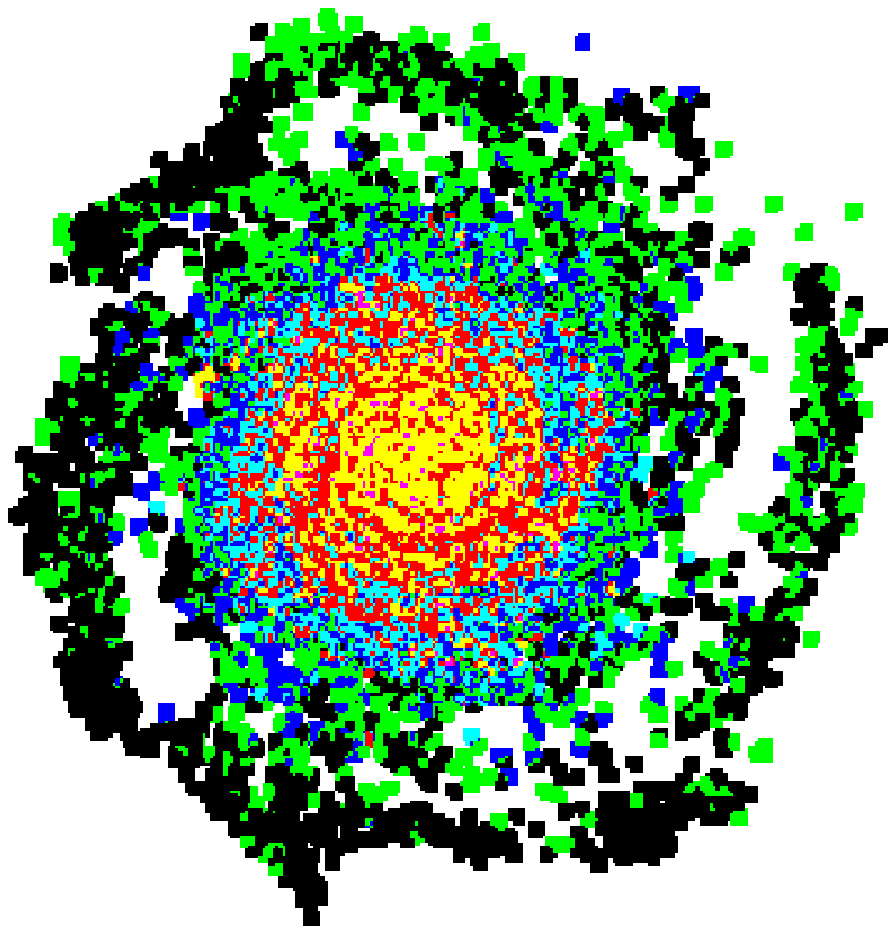}
 \includegraphics[%
  scale=0.35]{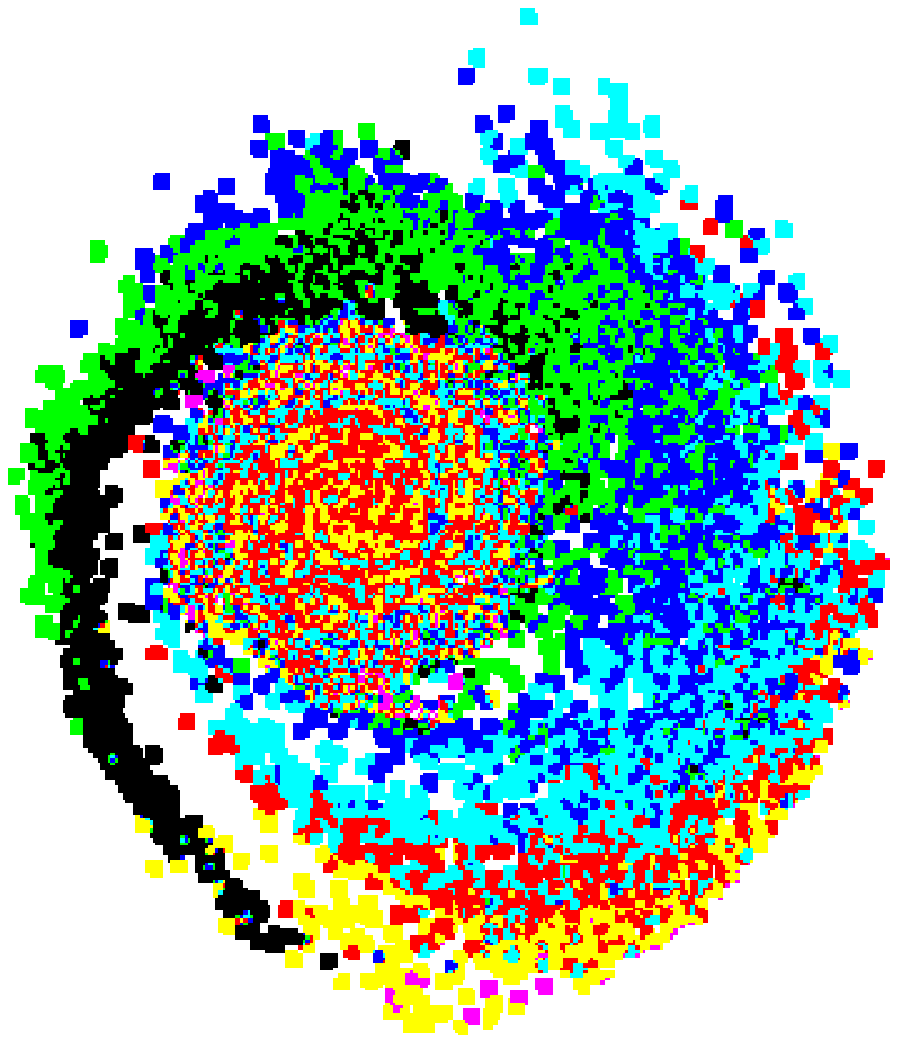}
 \includegraphics[%
  scale=0.35]{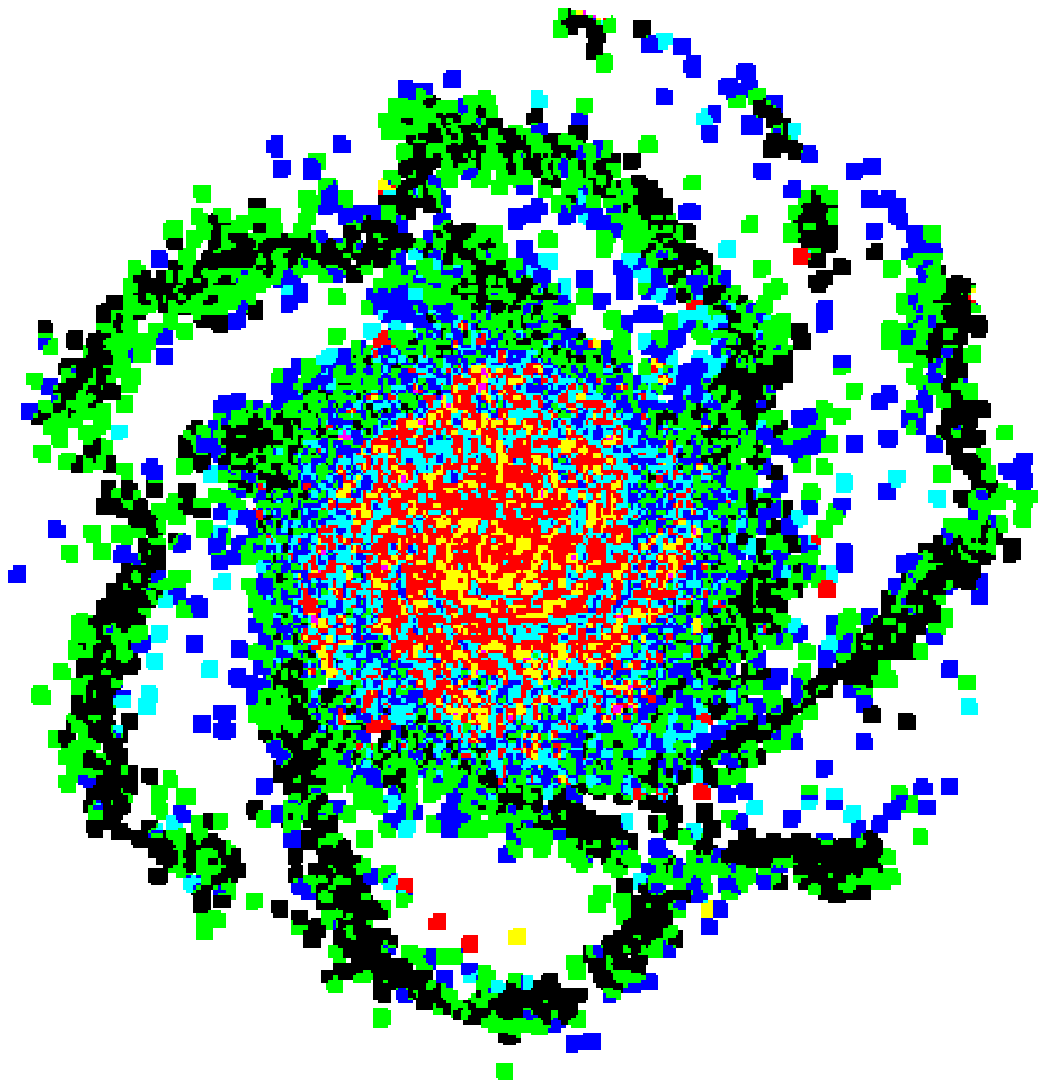} 
\includegraphics[%
  scale=0.35]{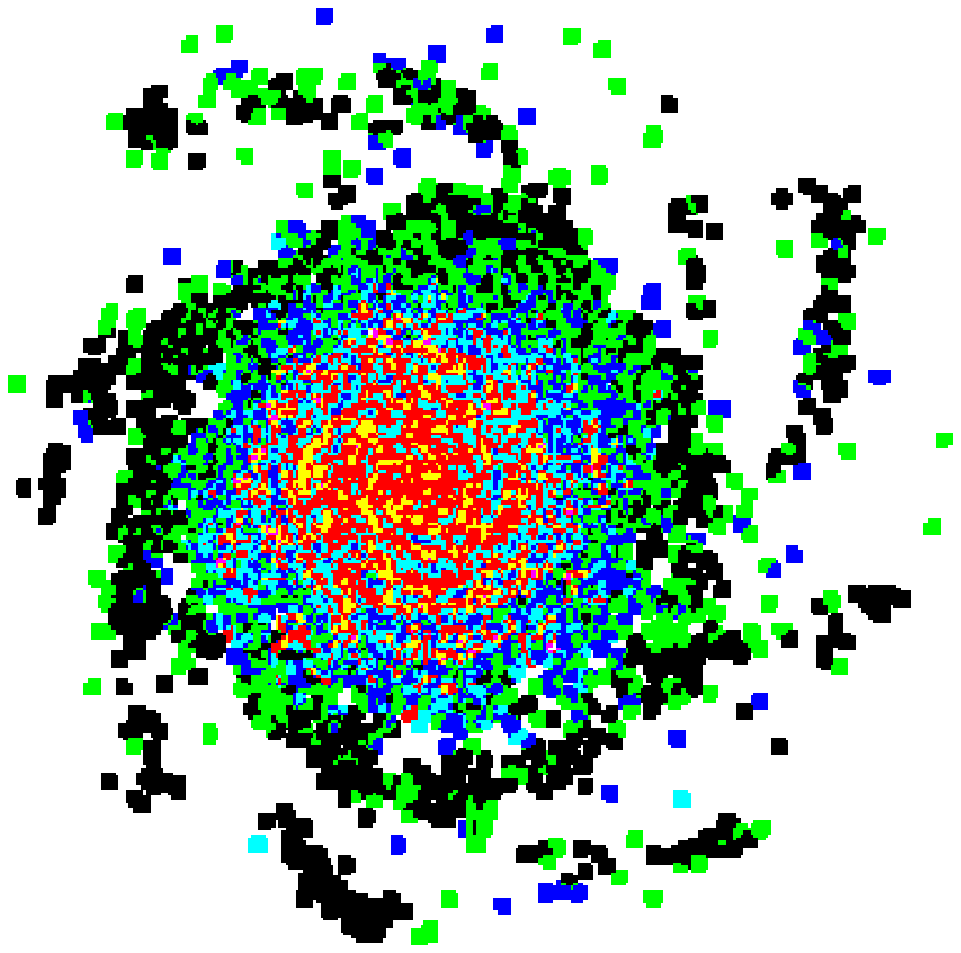}
\includegraphics[%
  scale=0.35]{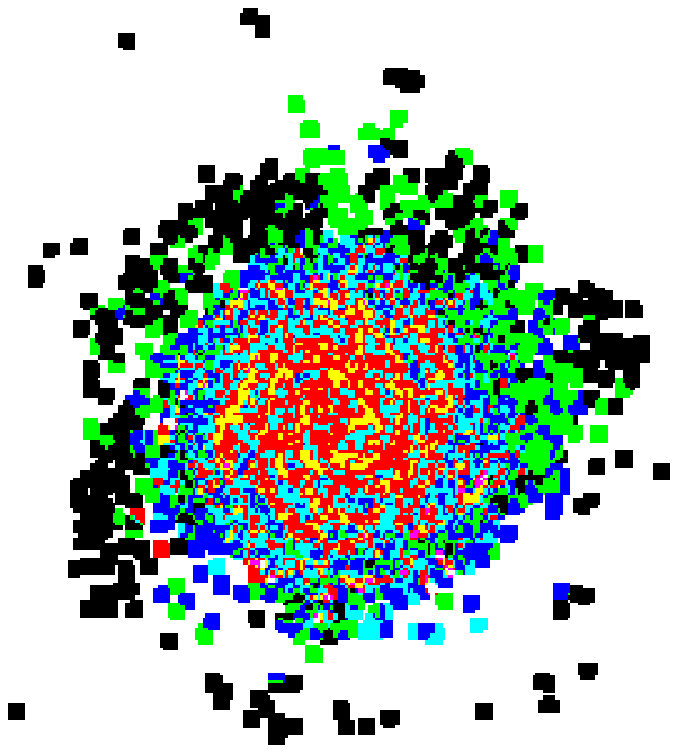}
\includegraphics[%
  scale=0.35]{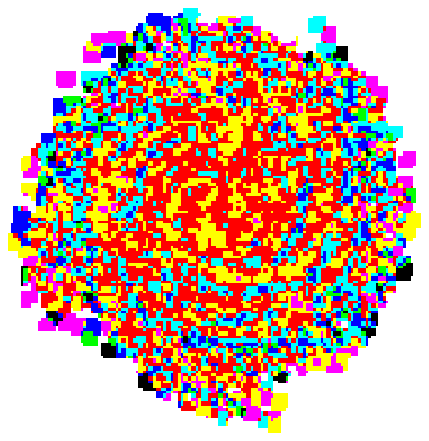}
\includegraphics[%
  scale=0.35]{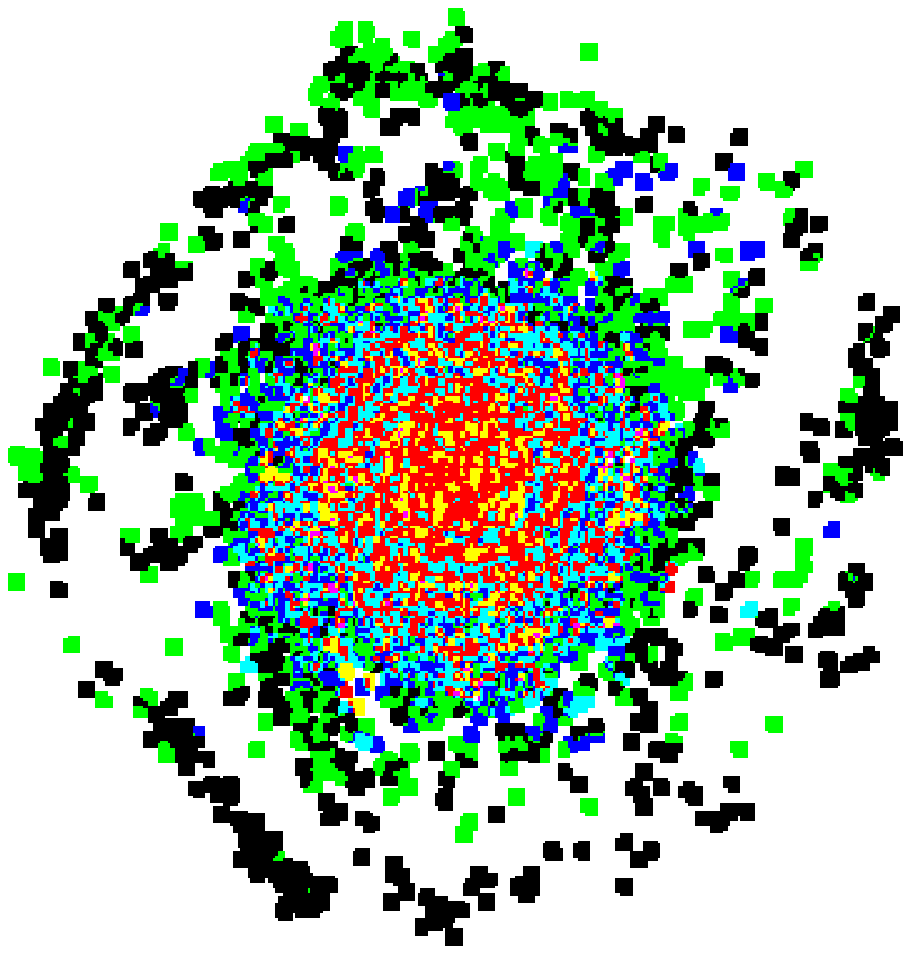}
\includegraphics[%
  scale=0.4]{colorbar2.eps}

\caption{Age distribution of the newly forming stars in the LMC's disk. The different panels
  represent (from top to bottom and from left to right) runs listed in
  Table \ref{sfruns}: : SF90, SF45, SF10, SFconv, SF$\epsilon$1, SFconv$\epsilon$1, SF90c0.01, SF10c0.01, SF90c0.05, SF10c0.05, SF90v400, SF10v400, SF10v400t12000, SF10v400t15000, SFld90, SFld10. Magenta indicates stars older than  725 Myr.} 
\label{agedistribution}

\end{figure*}

\begin{figure}
\epsfxsize=8truecm \epsfbox{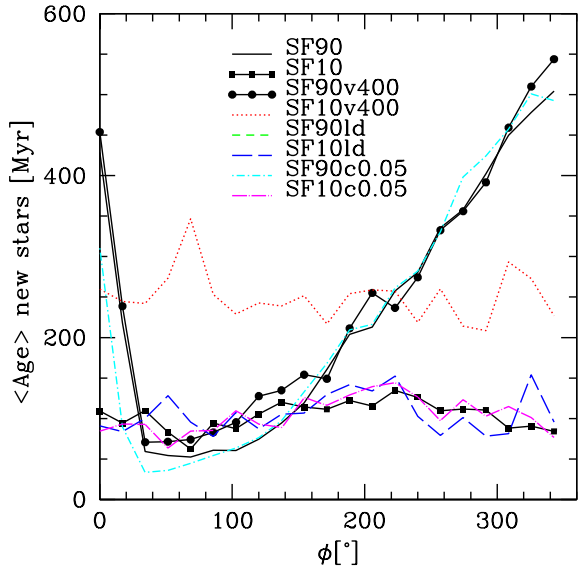} 
\caption{External disk mean stellar age in function of the azimuthal angle $\phi$ for selected runs in Table \ref{sfruns}. }
\label{agephi}
\end{figure}

\begin{figure}
\epsfxsize=8truecm \epsfbox{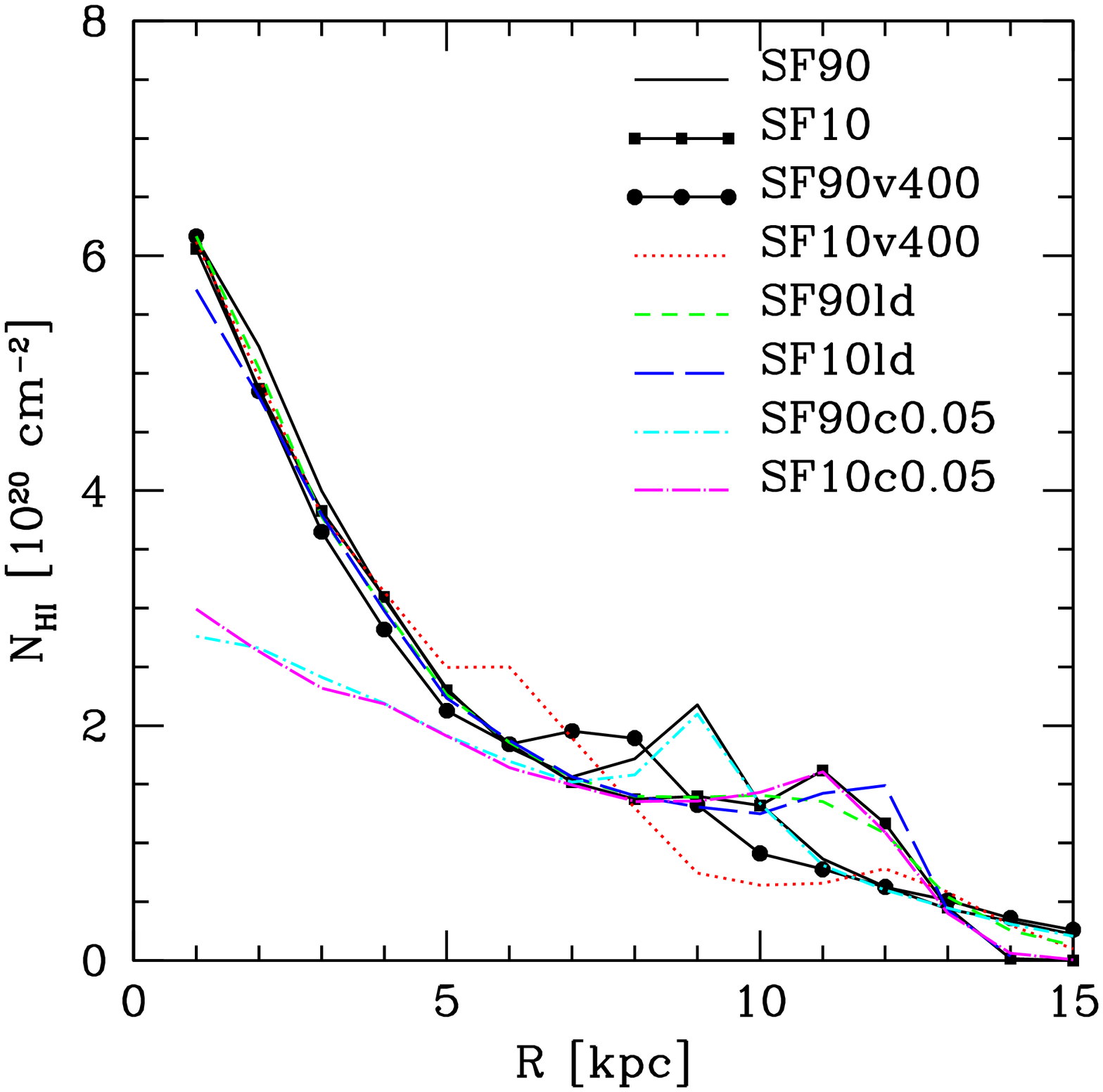} 
\caption{Azimuthally averaged HI column density profile of the final LMC disk for a selection of the runs listed in Fig. \ref{sfruns}. }
\label{densf}
\end{figure}

\section{Conclusions}

We have performed  high resolution ``wind tunnel'' simulations to study the effects of ram-pressure by a tenuous Galactic hot halo on the HI morphology of the LMC's disk, its recent star formation history, and location of the youngest star forming regions.  
We did not focus on  the mass loss produced by ram-pressure stripping since this would also be affected by tidal interactions. For the same reason our galaxies do not form any bar and we actually start with a galaxy model very stable against bar formation so that the pure effects of  external pressure are more clearly visible. 
Our LMC is a multi-component system with a spherical NFW halo, an exponential stellar disk  and a gaseous disk that extends up to 8 times the stellar disk scale length.
In each simulation the external flux of hot particles increases with time as the satellite approaches the perigalacticon in such a way that the pressure experienced by the disk is consistent with the LMC's orbital velocity and an average hot halo density of $2 \times 10^{-5}$ cm$^{-3}$ within 150 kpc from the Galactic center.
Low velocity runs are characterized by a ``classic'' pericentric velocity of 300 km s$^{-1}$ \citep{Mastropietroetal05} while high velocity runs have velocities compatible with the new proper motion measurements of \citet{Kallivayaliletal06}.
We expect the angle between the LMC's disk and its proper motion to vary significantly during the last billion years of orbital interaction.
We have defined the inclination angle $i$ as the angle between the angular momentum 
vector of the disk and the flux of hot particles in the wind tube, so that the observed LMC's disk would have $i\sim 60^{\circ}$.
\citet{Mastropietro08} performed self consistent Nbody/SPH simulations of the interacting system MW/LMC adopting orbital constraints from the last LMC's proper motion measurements and found that the Cloud enters the MW halo face-on ($i\sim 0^{\circ}$), moving nearly face-on for most of the last billion years. It turns edge-on only about 30 Myr ago. 
This means that the LMC is moving nearly edge-on close to the perigalactic passage, corresponding to the maximum ram-pressure values, consistently with the actual disk inclination measured by \citet{Kallivayaliletal06}.

We have performed several simulations varying the inclination angle of the disk, the star formation recipe and  the intensity of the external pressure. We have shown that:

\begin{itemize}

\item {The compression of the leading border of an edge-on LMC disk can account for the high density HI region observed at the south east. In our simulations this high density feature is well defined (with a mean density one order of magnitude higher than the surrounding gas) and localized within 1.5 kpc from the border of the disk.  Its average thickness and velocity dispersion along the line of sight are larger than the average values in the rest of the disk. In cool90 it extends for almost $160^{\circ}$ and could also explain the origin of the spiral arm E described by \citet{StaveleySmithetal03}, which does not have an equivalent in the stellar disk.} 
  
\item{Compression directed perpendicularly to the disk (in runs with $i<90^{\circ}$) produces local instabilities in the gas distribution and a clumpy structure characterized by voids and high density filaments similar to those observed by the Parkes multibeam HI survey (see Fig. 3 of Staveley-Smith et al. 2003). If the satellite was moving nearly face-on in the past -- and according to \citet{Mastropietro08} this is likely to happened during most of the LMC/MW orbital history -- ram-pressure could be responsible for the general mottled appearance of the HI disk. } 

\item{As a result of the increase in density along the edge of the disk the mean HI column density shows a secondary peak at large radius, in agreement with observations. }

\item{The compression of the satellite's IGM is naturally associated with induced star formation activity.
We focussed on the external regions of the disk since  the central parts of the real LMC would be dominated by the bar.
Edge-on disks start forming stars earlier, but for large ram-pressure values the star formation rate of runs with $i<90^{\circ}$ grows much faster.
The high velocity edge-on model SF90v400 is characterized by a steeper increment in star formation at earlier times but later on the curve flattens and the star formation rate at the perigalacticon is similar to  that of the low velocity case SF90.
On the other hand, the star formation generated by a compression perpendicular to the LMC's disk increases with increasing ram-pressure values and in the case of the high velocity run SF10v400 reaches a peak $\sim 1.5$ times higher than in SF10.
If the satellite is moving almost face-on until it gets very close to the perigalacticon we would expect not to see star formation before 0.6 Gyr independently of the orbital velocity.
Differences between high and low velocity runs should be marginal also near the pericenter since star formation in edge-on runs -- assuming our standard prescriptions -- seems to saturate around $\sim 4 \times 10^4$ M$_{\odot}$ Myr$^{-1}$.}

\item{In edge-on models the star formation of the external disk is characterized by a thin stellar arc along the leading border, well distinct from the star formation events in the central disk. If SFR is converted in H$\alpha$ luminosities, this arc breaks in several distinct and very luminous H$\alpha$ regions that more closely resemble the star forming complexes observed on the eastern border of the disk.  
Although the H$\alpha$ emission is mostly concentrated on the eastern side as a consequence of the very recent edge-on motion, we expect to see some luminous clumps forming a patchy distribution on the entire disk, due to gravitational instabilities and subsequent star formation induced by a nearly face-on compression of the disk before 30 Myr ago. 
As observed by \citet{deBoeretal98} stellar complexes on the leading edge show a progression in age in the clockwise direction, but a face-on compression in the recent past of the LMC would circumscribe this trend to the youngest stellar regions, with age $<$ 30-40 Myr.   }

\end{itemize}

\section{Acknowledgements}
We would like to thank M-R. Cioni, T. Kaufmann and  N. Kallivayalil for useful discussions.
The numerical simulations were performed on the zBox1 supercomputer at the University of Zurich and on the local SGI-Altix 3700 Bx2 (partly funded by the cluster of excellence ``Origin and Structure of the Universe'' ). This work was partly supported by the DFG Sonderforschungsbereich 375 ``Astro-Teilchenphysik''.

\label{lastpage}

\end{document}